\documentclass[reprint,pra,twocolumn,preprintnumbers,amsmath,amssymb,superscriptaddress,showpacs,longbibliography]{revtex4-1}
\usepackage{graphicx}
\usepackage{amsmath}
\usepackage{graphics}
\usepackage{amssymb}
\usepackage{amsfonts,epsfig}
\usepackage{dsfont}
\usepackage{natbib}
\usepackage{hyperref}
\usepackage{enumitem}
\usepackage{physics}
\usepackage{color}
\usepackage[mathscr]{eucal}
\usepackage[caption=false]{subfig}
\allowdisplaybreaks

\newcommand{\beq}{\begin{equation}}
\newcommand{\eeq}{\end{equation}}
\newcommand{\bea}{\begin{eqnarray}}
\newcommand{\eea}{\end{eqnarray}}

\newcommand{\mean}[1]{\langle{#1}\rangle{}}

\usepackage{color}

\newcommand{\s}{s}

 \usepackage[usenames,dvipsnames]{pstricks}
 \usepackage{epsfig}
 \usepackage{pst-grad} % For gradients
 \usepackage{pst-plot} % For axes
 \usepackage[space]{grffile} % For spaces in paths
 \usepackage{etoolbox} % For spaces in paths
 \makeatletter % For spaces in paths
 \patchcmd\Gread@eps{\@inputcheck#1 }{\@inputcheck"#1"\relax}{}{}
 \makeatother
\makeatletter
\newcommand*{\rom}[1]{\expandafter\@slowromancap\romannumeral #1@}
\makeatother

\newcommand{\cv}[1]{\left(#1\right)}
\newcommand{\cvb}[1]{\left[#1\right]}

\newcommand{\iden}{\mathbb{I}}

\newcommand{\opS}{\sigma}%{\op{S}}

\newcommand{\prob}{\mathbb{P}}

\begin{document}

\title{Characterization of a quasi-static environment with a qubit}
\author{Fattah Sakuldee}
\email{sakuldee@ifpan.edu.pl}
\affiliation{Institute of Physics, Polish Academy of Sciences, al.~Lotnik{\'o}w 32/46, PL 02-668 Warsaw, Poland}
\author{{\L}ukasz Cywi\'{n}ski}\email{lcyw@ifpan.edu.pl}
\affiliation{Institute of Physics, Polish Academy of Sciences, al.~Lotnik{\'o}w 32/46, PL 02-668 Warsaw, Poland}

\begin{abstract}
We consider a qubit initalized in a superposition of its pointer states, exposed to pure dephasing due to coupling to a quasi-static environment, and subjected to a sequence of single-shot measurements projecting it on chosen superpositions. We show how with a few of such measurements one can significantly diminish one's ignorance about the environmental state, and how this leads to increase of coherence of the qubit interacting with a properly post-selected environmental state. We give theoretical results for the case of a quasi-static environment that is a source of an effective field of Gaussian statistics acting on a qubit, and for a nitrogen-vacancy center qubit coupled to a nuclear spin bath, for which the Gaussian model applies qualitatively provided one excludes from the environment nuclei that are strongly coupled to the qubit. We discuss the reason for which the most probable sequences of measurement results are the ones consisting of identical outcomes, and in this way we shed light on recent experiment (D. D. Bhaktavatsala Rao et al., arXiv:1804.07111) on nitrogen-vacancy centers. 
\end{abstract}

\date{\today}
		
\maketitle

\section{Introduction}
When a quantum system is brought into interaction with an environment, correlations between the two are created, leading to decoherence \cite{Zurek_RMP03,Schlosshauer_book} of the system's state - reduced density matrix of a system correlated with an environment has to be more mixed than the initial density matrix describing the isolated system. 
However, the establishment of such a system-environment correlation also means that performing a projective measurement on the system affects the environment coupled to it \cite{Wiseman,Pfender_NC19,Ma_PRA18}.
Depending on the nature of established correlations and their dynamics, one can either use the measurements on the system to manipulate the environment \cite{Blok_NP14,Muhonen_PRB18}, or to diminish one's ignorance about its state \cite{Klauser_PRB06,Stepanenko_PRL06,Giedke_PRA06}.
%(the two possibilities are not exclusive...)
When the intrinsic dynamics of the environment is slow compared to timescales of system's initialization, evolution, and readout, results of subsequent measurements on the system  become correlated, and each measurement can introduce further modification (dependent on the measurement result) of the state of the environment.

Qubits interacting with a condensed matter environment are typically most strongly affected by dephasing due to exactly such slow fluctuations of nuclear spins \cite{Coish_PSSB09,Cywinski_APPA11,Chekhovich_NM13,Yang_RPP17} or two-level systems associated with charge dipoles \cite{Paladino_RMP14,Szankowski_JPCM17}. The possibility of using a qubit as a {\it probe} tracking the slowest modes of environmental dynamics was first recognized for spin qubits interacting with nuclear spins of the host semiconductor material \cite{Coish_PRB04}, as it was clear that the timescale of intrinsic dynamics of the nuclear bath is orders of magnitude longer than experimentally feasible cycle of qubit's preparation, evolution, and measurement. Various protocols for such a ``narrowing'' of nuclear state (diminishing the spread of values of nuclear fields affecting the spin qubit) were proposed about 10 years ago \cite{Klauser_PRB06,Stepanenko_PRL06,Giedke_PRA06}, and then further refined \cite{Cappellaro_PRA12}. Subsequent  breakthroughs in single-shot readout of quantum dot spin qubits enabled observation of enhancement of qubit's coherence times by gathering data on qubit's precession on timescales shorter than the correlation time of the environment \cite{Barthel_PRL09,Delbecq_PRL16}, and also by using additional feedback from measurement results to the qubit manipulation protocol \cite{Shulman_NC14}.

The effects of environmental state modification induced by even a few measurements on a qubit should be particularly strong for environments consisting of weakly interacting constituents. A nitrogen-vacancy (NV) center in diamond that interacts with nuclear spins of $^{13}$C nuclei (occupying only $1.1$\% of lattice sites for natural diamond) is a natural candidate for investigation of such effects, as dephasing of NV-based spin qubit can be well described by taking into acccount $<\! 100$ nearest nuclei, and the dipolar interactions between the nuclei become relevant only on timescales much longer than that of free evolution dephasing of the NV center qubit \cite{Zhao_PRB12,Yang_RPP17}. Progress in single-shot readout of NV center qubits at low temperatures \cite{Robledo_Nature11} allowed for recent measurement \cite{RaoEAPreprint2018} of probabilities of obtaining various sequences of results of projective measurements on the qubit that were clearly exhibiting history-dependent (non-Markovian) behavior.

In this paper we consider a recently experimentally implemented protocol \cite{RaoEAPreprint2018}, in which a qubit is repeatedly initialized in a superposition state, evolves under an influence of an environment that causes its pure dephasing, and is then subjected to a projective measurement. We focus on the case of an environment that is essentially static during the series of repetitions of preparation-evolution-measurement cycle. We formulate the general theory for calculation of probabilities of obtaining all the possible sequences of measurement results, we derive approximate analytical formulas in the case in which the environment is large enough to be treated in a coarse-grained manner - when the spectrum of environment-induced qubit energy shifts is very dense -  and when the initial distribution of these shifts can be assumed to be Gaussian. 
We apply both the exact and approximate approach to the case of NV center spin qubit interacting with the  environment of $^{13}$C nuclear spins. We reproduce the main observation of Ref.~\cite{RaoEAPreprint2018} that the sequences of identical measurement results are the most probable ones when the evolution time is longer than qubit dephasing time. Derivation of this result shows clearly that such a behavior can be explained by treating the environment as a classical object, and the measurement sequence as a means to decrease the amount of uncertainty in our classical probabilistic description of the state of this object. In other words, in the regime in which the quasi-static environment approximation holds, one can view the qubit as a probe that reveals the pre-existing state of the environment. Finally, we calculate the dephasing the qubit will experience after a sequence of measurements yielded a particular sequence of results. We focus on previously mentioned most probable sequences of identical results, and discuss to what degree they lead to the narrowing of the environmental state, and the resulting enhancement of coherence time of the qubit interacting with an environment post-selected on the basis of obtaining one of such sequences.

The paper is organized in the following way. In Section \ref{sec:general} we give a general theory for changes of the environmental state due to a cycle of multiple preparations, evolutions, and measurements of a qubit that is coupled to the environment via pure dephasing Hamiltonian. This Section also contain a careful definition of the quasi-static bath approximation, and a discussion of somewhat nontrivial conditions that need to be fulfilled for this approximation to apply to real-life experiments with qubits. Application of this general theory to the NV centers in diamond (and other kinds of spin qubits interacting with nuclear spin environments) is given in Section \ref{sec:application}. Then in Section \ref{sec:Gaussian} we introduce the coarse-graining and Gaussian approximations to the description of quasi-static environment and its influence on the qubit. We give there approximate analytical results for probabilities of obtaining of various sequences of measurement results, and show how the distribution of qubit energy shifts caused by interaction with the environment changes after a particular sequence was registered. We compare these predictions results of exact calculations for the case of NV center interacting with the nuclear bath, and show that the Gaussian approximation accounts for all the qualitative features of these results, provided that we consider nuclear environments in which there are no nuclei very close to the qubit. 
Finally, in Section \ref{sec:dephasing} we present exact and approximate results for coherence decay after post-selecting the environmental state based on previously obtained sequence of qubit measurement results.

%%%%%%%%%%%%%%%%%%%%%%%%%%%%%%%%%%%%%%%%%%%%%%%%%%%%%
%%% GENERAL MODEL FOR QUBIT UNDERGOING PURE DEPHASING
%%%%%%%%%%%%%%%%%%%%%%%%%%%%%%%%%%%%%%%%%%%%%%%%%%%%%
\section{Multiple measurements of a qubit initialized in superposition state}  \label{sec:general}
\subsection{Change in state of environment induced by measurement on a qubit}
We consider a qubit ($Q$) coupled to its environment ($E$) in such a way that E induces only dephasing of superpositions of qubit's pointer states denotes as $\ket{\uparrow}$ and $\ket{\downarrow}$. The Hamiltonian of the total Q$+$E system is given by
\beq
\hat{H} = \frac{\Omega}{2}\hat{\sigma}_z\otimes\iden^E + \iden^{Q}\otimes \hat{H}_{E} + \ket{\uparrow}\bra{\uparrow}\otimes \hat{V}_{\uparrow} +  \ket{\downarrow}\bra{\downarrow}\otimes \hat{V}_{\downarrow} \,\, ,\label{eq:Hamiltonian_gen}
\eeq
where $\hat{\sigma}_{z}\ket{\uparrow/\downarrow} \! =\! \pm\ket{\uparrow/\downarrow}$, $\Omega$ is the energy splitting of the qubit, $\hat{H}_{E}$ is the Hamiltonian of the environment, and $\hat{V}_{\uparrow/\downarrow}$ describe the qubit-environment coupling.

We will discuss qubit's dynamics in rotating frame, as this is the natural frame for discussion of qubit's manipulation and readout when electron spin resonance techniques are used for qubit control. We can then remove $\Omega$ splitting from qubit Hamiltonian. One should keep in mind that when we discuss measurement in $\ket{\pm x}$ basis at time $t$ after initialization of the superposition state of the qubit, we refer to $\ket{\pm x}$ states in this rotating frame.

We consider the situation in which the qubit is initialized in a superposition state, taken as $\ket{+x}\! \equiv \! \frac{1}{\sqrt{2}}(\ket{\uparrow}+\ket{\downarrow})$ here without any loss of generality. It is then brought into contact with $E$ described by density matrix $\rho_{0}^{E}$. This contact lasts for time $\tau$, after which the qubit is subjected to measurement.  Since only its coherence (the off-diagonal elements of qubit's density matrix in basis of pointer states) evolves due to interaction with $E$, we consider measurements of transverse (with respect to quantization axis $z$) component of qubit's Bloch vector. 

It will be convenient to define the environmental evolution operator conditioned on the state of the qubit:
\beq
\hat{U}_{\s}(\tau) \equiv e^{-i\hat{H}_{\s}\tau} \,\, ,
\eeq
where the label $\s\! =\! \uparrow$, $\downarrow$ and
\beq 
\hat{H}_{\uparrow/\downarrow} \equiv \hat{H}_{E} + \hat{V}_{\uparrow/\downarrow} \,\, . 
\eeq

We will be interested now in statistics of obtaining a specific string of measurement results. We assume that measurements are projective, described by operators $\hat{P}_{q}\! =\! \ket{q}\bra{q}$ corresponding to orthonormal basis $\ket{q}$ of qubit states. The unnormalized state of $Q+E$ system after obtaining result $q$ of the first measurement is thus given by
\beq
\tilde{\rho}_{1,q}^{S+E} =  \hat{P}_{q} e^{-i\hat{H}\tau} \left( \ket{+x}\bra{+x}\otimes \hat{\rho}_{0}^{E} \right) e^{i\hat{H}\tau} \hat{P}_{q} \,\, ,
\eeq
and the probability of obtaining this result is given by $\mathrm{Tr}(\tilde{\rho}_{1}^{S+E}$). As $\tilde{\rho}_{1,q}^{S+E} \! =\! \ket{q}\bra{q}\otimes \tilde{\rho}_{1,q}^{E}$ we focus now on post-measurement unnormalized state of the environment $\tilde{\rho}_{1,q}^{E}$.

Let us focus now on the protocols in which all the measurements are done in the same basis, chosen here to be $\ket{\pm x}$. For the main purpose of this paper, which is the analysis of the quasi-static environment case, employing measurements along distinct axes, e.g.~both $x$ and $y$, is not necessary, but let us note that for a general environment using multi-axis protocols is clearly advantageous \cite{Wang_arXiv19}.

For measurements in $\ket{\pm x}$ basis we have
\beq
\tilde{\rho}_{1,\pm}^{E}(\tau) = \frac{1}{4} \left( \hat{U}_{\uparrow}(\tau) \pm \hat{U}_{\downarrow}(\tau) \right) \hat{\rho}_{0}^{E} \left( \hat{U}_{\uparrow}^{\dagger}(\tau) \pm \hat{U}_{\downarrow}^{\dagger}(\tau) \right) \,\, , \label{eq:tilderho1}
\eeq
and the probability of obtaining this result is $\prob\cv{\pm\vert\rho^B_{0}} = \tr{\tilde{\rho}^E_{1,\pm}}$, so that the normalized state at time $\tau$ is $\hat{\rho}_{1,\pm}^{E}(\tau) \! =\! \tilde{\rho}_{1,\pm}^{E}(\tau) / \prob\cv{\pm\vert\rho^E_{0}}$. 
%\add{Note that in this work, both preparation and measurement are of projective type in order to incorporate the conventional description of single shot measurement, in which both of them are always sharp and do commute. In this framework, for pure-dephasing, it is clear that the change in probability of obtaining subsequent measurement results will come from the evolution of the bath state rather than that of another degrees of freedom. (???????)}

%PUT HERE THE FORMULATION WITH TWO SETS OF PROJECTION OPERATORS
For further discussion it will be useful to write the $\hat{U}^{\uparrow/\downarrow}(\tau)$ evolution operators as
\begin{equation}
\hat{U}_{\s}(\tau) = \sum_{\alpha} e^{-i\omega{\alpha_\s}\tau} \hat{P}_{\alpha_\s} \,\, , \label{eq:Palphas}
\end{equation}
in which $\hat{P}_{\alpha_\s}$ are projectors on eigenstates of $\hat{H}_{\s}$:
\beq
\hat{H}_{\s} \ket{\alpha_\s} = \omega_{\alpha_\s} \ket{\alpha_\s} \,\, . \label{eq:wa_general}
\eeq
Using the above we can rewrite Eq.~(\ref{eq:tilderho1}) as
\begin{align}
\tilde{\rho}_{1,\pm}^{E}(\tau) &= \frac{1}{4} \sum_{\alpha,\beta} \Big( e^{-i(\omega_{\alpha_\uparrow}-\omega_{\beta_\uparrow})\tau} \hat{P}_{\alpha_\uparrow}\hat{\rho}_{0}^{E} \hat{P}_{\beta_\uparrow} + \nonumber \\
 & e^{-i(\omega_{\alpha_\downarrow}-\omega_{\beta_\downarrow})\tau} \hat{P}_{\alpha_\downarrow}\hat{\rho}_{0}^{E} \hat{P}_{\beta_\downarrow} \pm e^{-i(\omega_{\alpha_\uparrow}-\omega_{\beta_\downarrow})\tau} \hat{P}_{\alpha_\uparrow}\hat{\rho}_{0}^{E} \hat{P}_{\beta_\downarrow} \nonumber  \\
& \pm e^{-i(\omega_{\alpha_\downarrow}-\omega_{\beta_\uparrow})\tau} \hat{P}_{\alpha_\downarrow}\hat{\rho}_{0}^{E} \hat{P}_{\beta_\uparrow} \Big) \,\, , \label{eq:tilderho1P}
\end{align}

If after the measurement of the qubit the environment is then allowed to evolve for time $\Delta t$ in the absence of the qubit, its state changes to
\beq
\hat{\rho}_{1,\pm}^{E} (\tau,\Delta t) = e^{-i\hat{H}_{E}\Delta t} \hat{\rho}_{1,\pm}^{E}(\tau) e^{i\hat{H}_{E}\Delta t} \,\, . \label{eq:rhotaut}
\eeq
Using the projectors on eigenstates $\ket{\epsilon}$ of $\hat{H}_{E}$ we have then
\beq
\hat{\rho}_{1,\pm}^{E} (\tau,\Delta t) = \sum_{\epsilon,\epsilon'} e^{-i(\epsilon-\epsilon')\Delta t} \hat{P}_{\epsilon}\hat{\rho}_{1,\pm}^{E}(\tau) \hat{P}_{\epsilon'} \,\, .  \label{eq:finalrho1general}
\eeq
Plugging the form of $\hat{\rho}_{1,\pm}^{E}(\tau)$ from Eq.~(\ref{eq:tilderho1P}) into Eq.~(\ref{eq:finalrho1general}) clearly results in a rather complicated expression in the general case, in which the bases $\ket{\alpha_\uparrow}$, $\ket{\alpha_\downarrow}$, and $\ket{\epsilon}$ are unrelated.
If we then reinitialize the qubit in $\ket{+x}$ state, let $Q$ and $E$ evolve for time $\tau$, and perform measurement again in $\ket{\pm}$ basis, we will obtain a new state of $E$ given by Eq.~(\ref{eq:tilderho1}), in which $\hat{\rho}_{0}^{E}$ is replaced by $\hat{\rho}_{1,\pm}^{E}(\tau,\Delta t)$.

Let us remark here that if one replaces the projective measurements $\hat{P}_q$ by binary outcomes POVMs that are unsharp but do commute as in Ref. \cite{Ma_PRA18}, one will be able to achieve a similar behaviour as in Eq. \eqref{eq:tilderho1}. For illustration, we consider a measurement operator $\hat{M}_{q}^\theta=\dfrac{1}{\sqrt{2}}\cv{\iden^Q\cos\theta + q\opS_x\sin\theta}$ where $\theta\in\cvb{0,\pi/4}$ denote the strength parameter associate with sharpness of the measurement for the outcome $q \! =\! \pm 1$, see Ref.~\cite{Ma_PRA18}. Note that for $\theta=\pi/4$ one can recover the projection $\hat{P}_q$, i.e.~$\hat{M}_q$ is a weak version of $\hat{P}_q.$ The state of the composite system, with qubit initialized in $\ket{+x}$ state, after first measurement is then given by 
\beq
\tilde{\rho}_{1,\pm}^{S+E} =  \hat{M}_{\pm}^\theta e^{-i\hat{H}\tau} \left( \ket{+x}\bra{+x}\otimes \hat{\rho}_{0}^{E} \right) e^{i\hat{H}\tau} \hat{M}_{\pm}^\theta \,\, ,
\eeq	
and the state of the bath is of the form 
	\begin{align}
		\tilde{\rho}_{1,\pm}^{E}(\tau) &= \frac{1}{4} \left[ \hat{U}_{\uparrow}(\tau)\hat{\rho}_{0}^{E}\hat{U}_{\uparrow}^{\dagger}(\tau) + \hat{U}_{\uparrow}(\tau)\hat{\rho}_{0}^{E}\hat{U}_{\uparrow}^{\dagger}(\tau)\right.\nonumber\\
		&\left.\hspace{0.5cm} \pm \sin2\theta \cv{\hat{U}_{\downarrow}(\tau)\hat{\rho}_{0}^{E}\hat{U}_{\uparrow}^{\dagger}(\tau) + \hat{U}_{\uparrow}(\tau)\hat{\rho}_{0}^{E}\hat{U}_{\downarrow}^{\dagger}(\tau)}\right]\label{eq:tilderho1-POVM}
	\end{align}
where $q=\pm 1$ are measurement outcomes, as in Eq.~\eqref{eq:tilderho1}. From this form, one can see that the operation structure following from Eq. \eqref{eq:tilderho1-POVM} will be similar to that obtained from Eq. \eqref{eq:tilderho1}, only with a modification of coefficients. Consequently, in order to see the effect of the measurement on the change of environment state, it is sufficient to consider the projective type measurement instead of commuting POVMs.

%%%%%%%%%%%%%%%
%% QUASI-STATIC
%%%%%%%%%%%%%%
\subsection{Quasi-static environment case} \label{sec:QSE}
This procedure can be iterated to obtain state of the environment after a sequence of $n$ measurements on the qubit. However, the number of distinct contributions to the final $\hat{\rho}^{E}$, each corresponding to a distinct sets of products of projectors $\hat{P}_{\alpha_s}$ and $\hat{P}_{\epsilon}$ acting on the original $\hat{\rho}^{E}_{0}$ will grow exponentially with $n$. Without making any additional assumptions about $\hat{\rho}_{0}^E$, and the Hamiltonian of $Q$ and $E$, no simple structure arises, and states of $E$ after multiple measurements have to be obtained by brute force iteration of the formulas from the previous section.

Let us focus on the case of a quasi-static environment. It is defined by the following conditions:
\begin{enumerate}
\item $\hat{\rho}_{0}^{E} \! =\! f(\hat{H}_{E})$, with $f(\hat{H}) \!=\! e^{-\beta\hat{H}}/\mathrm{Tr}{e^{-\beta\hat{H}}}$ corresponding to the case of $E$ in thermal equilibrium at inverse temperature $\beta$,
\item $[\hat{H}_{E},\hat{V}_{\uparrow/\downarrow}]\! =\! 0$ and $[\hat{V}_{\uparrow},\hat{V}_{\downarrow}]\!= \! =0$,  \label{cond:quasi-cond_comm}
\end{enumerate}
that together mean that there is a common eigenbasis for $\rho_{0}^{E}$, $\hat{H}_{E}$, and $\hat{V}_{\uparrow/\downarrow}$:
\begin{align}
\hat{\rho}_{0}^{E} \ket{\alpha} & = p^{(0)}_{\alpha}\ket{\alpha} \,\, , \\
\hat{H}_{E} \ket{\alpha} & = \epsilon_{\alpha}\ket{\alpha} \,\, , \\
e^{-i\hat{H}_{\uparrow/\downarrow}\tau}\ket{\alpha} & = e^{-i\omega^{\uparrow/\downarrow}_{\alpha} \tau} \ket{\alpha} \,\, . \label{eq:wa}
\end{align}
Under these assumptions we can express both $\hat{\rho}_{0}^{E}$ and $\hat{U}_{\uparrow/\downarrow}(\tau)$ using the common set of projection operators $\hat{P}_{\alpha}\! \equiv \! \ket{\alpha}\bra{\alpha}$,
\begin{align}
\hat{\rho}_{0}^{E} & = \sum_\alpha p^{(0)}_{\alpha} \hat{P}_{\alpha} \,\, , \label{eq:rho0} \\
\hat{U}_{\uparrow/\downarrow}(\tau) & = \sum_{\alpha} e^{-i\omega_\alpha^{\uparrow/\downarrow}\tau}\hat{P}_{\alpha} \,\, . 
\end{align}

The $E$ state after the first measurement can now be written as
\begin{align}
\tilde{\rho}_{1,\pm}^{E}(\tau) & = \frac{1}{4} \sum_{\alpha,\alpha'} \left( e^{-i\omega^{\uparrow}_{\alpha}\tau} \pm e^{-i\omega^{\downarrow}_{\alpha}\tau} \right ) \hat{P}_{\alpha}\hat{\rho}_{0}^{E} \hat{P}_{\alpha'} \left( e^{i\omega^{\uparrow}_{\alpha'}\tau} \pm e^{i\omega^{\downarrow}_{\alpha'}\tau}  \right ) \,\, \nonumber \\
%& = \frac{1}{4}\sum_{\alpha} p^{(0)}_{\alpha} \left( e^{-i\omega^{\uparrow}_{\alpha}\tau} \pm e^{-i\omega^{\downarrow}_{\alpha}\tau} \right )\left( e^{i\omega^{\uparrow}_{\alpha}\tau} \pm e^{-i\omega^{\downarrow}_{\alpha}\tau} \right ) \hat{P}_{\alpha} \,\, , \\
& = \sum_{\alpha}A^{\pm}_{\alpha}(\tau) p^{(0)}_{\alpha}  \hat{P}_{\alpha} \,\, , \label{eq:rho1pm}
\end{align}
where
\begin{align}
A^{\pm}_{\alpha}(\tau)  & =   \frac{1}{4}\left( e^{-i\omega^{\uparrow}_{\alpha}\tau} \pm e^{-i\omega^{\downarrow}_{\alpha}\tau} \right )\left( e^{i\omega^{\uparrow}_{\alpha}\tau} \pm e^{i\omega^{\downarrow}_{\alpha}\tau} \right ) \,\, , \nonumber\\
& = \frac{1}{2}[1\pm \cos(\omega^{\uparrow}_{\alpha}-\omega^{\downarrow}_{\alpha})\tau ] \,\, . \label{eq:Apm}
\end{align}
Defining now
\beq
\Delta \omega_{\alpha} = \omega^{\uparrow}_{\alpha}-\omega^{\downarrow}_{\alpha} \,\, ,
\eeq
we have 
\begin{align}
A^{+}_{\alpha}(\tau) & = \cos^2(\Delta\omega_{\alpha}\tau/2) \,\, ,  \\
A^{-}_{\alpha}(\tau) & =\! \sin^2(\Delta\omega_{\alpha}\tau/2) \,\, .
\end{align}
The above formulas shows that the post-measurement density operator of $E$ is still diagonal in $\ket{\alpha}$ basis. This means that it is invariant under evolution due to $\hat{H}_{E}$ in the absence of the qubit, so that $\tilde{\rho}^{E}_{1,\pm}(\tau,t) \! =\! \tilde{\rho}^{E}_{1,\pm}(\tau)$. One could ask now, why we call the environment characterized by the above-introduced conditions ``quasi-static'', and not simply ``static''. This is a valid question, to which we will come back in Sec.~\ref{sec:QSBAdecoh}.

Let us also note that if we average the state of $E$ over the results of measurements of the qubit we obtain
\begin{align}
\hat{\rho}^{E}_{1}  &\equiv \prob\cv{+\vert\rho^E_{0}}\hat{\rho}^{E}_{1,+}  +  \prob\cv{+\vert\rho^E_{0}}\hat{\rho}^{E}_{1,+} = \tilde{\rho}^{E}_{1,+} + \tilde{\rho}^{E}_{1,-}   \nonumber\\
& = \sum_{\alpha} [A^{+}_{\alpha}(\tau) + A^{-}_{\alpha}(\tau)]p^{(0)}_{\alpha}\hat{P}_{\alpha} = \hat{\rho}^{E}_{0} \,\, \label{eq:invariant}
\end{align}
where we have used Eqs.~(\ref{eq:rho1pm})  and (\ref{eq:Apm}). This means that the normalized state of $E$ after making a measurement on the qubit and discarding its result is the same as the pre-measurement state. This is the essence of the behavior in the quasi-static environment case when multiple measurements on the qubit are considered: making the measurements allows one to learn about the state of $E$ based on obtained results, as $\hat{\rho}^{E}_{1,x}$ does depend on the result $x\! =\! \pm 1$, but simply making the measurements on the qubit while discarding their results does not change the environmental state. 

We can now easily see that if the unnormalized state of $E$ after $n-1$ measurements in given by $\tilde{\rho}^{E}_{n-1}$, after obtaining $\pm$ result in the $n$-th measurement we obtain the state of $E$ given by
\beq
\tilde{\rho}^{E}_{n,\pm} = \sum_{\alpha} A^{\pm}_{\alpha}(\tau) \hat{P}_{\alpha} \tilde{\rho}^{E}_{n-1}\hat{P}_{\alpha} \,\, . \label{eq:iterative}
\eeq
Note that for the aforementioned case of commuting POVMs $\hat{M}_\pm^\theta,$ the coefficients $A^\pm_\alpha\cv{\tau}$ take the form 
	\begin{equation}
		A^{\theta,\pm}_{\alpha}(\tau)   =   \frac{1}{2}[1\pm \sin2\theta\cos(\omega^{\uparrow}_{\alpha}-\omega^{\downarrow}_{\alpha})\tau ]
	\end{equation}
 and Eqs. \eqref{eq:invariant}-\eqref{eq:iterative} are still valid in this case. From now on, for simplicity, we will concentrate on the results for conventional projective measurement.

%%%%%%%%%%%%%%%%%%%%%%%%%
% SEQUENTIAL MEASUREMENTS
%%%%%%%%%%%%%%%%%%%%%%%%%
\subsection{Sequential Measurement Protocol}
Let us consider $n$ successive measurements on the qubit, each of which takes an outcome from dichotomous alternatives $\{-x,+x\}$ (projective measurements of $\ket{\pm x}$ states). Let $M_j= {'}\cv{j}{'}_{binary}$ denotes a particular measurement sequences labelled by an integer $j=1,\ldots, 2^{n}-1$ in a binary form where the order of measurements runs from last digit to the first one; each digit will be denoted as $'0'$ if the outcome is $+x$ and denoted by $'1'$ otherwise. For example, in the case of $4$ measurements $M_0='0000'$ represents the sequence of all measurement outcomes are all $+x$, and  $M_3='0011'$ represents the sequence of which the first and second measurements' outcomes are $-x$ and the results turn $+x$ at the later steps.

Apart from the description via measurement sequences, which are not naturally ordered, one can adopt a parametrisation using a path length $k$. For a measurement outcome $M_j$ we define its path length $k(j)$ as a summation of $'1'$ inside the measurement sequence's labelling number $j$. In this representation, using again the example of $4$ measurements, $M_{15} \! =\! '1111'$ corresponds to $k(15)\! =\! 4$, while $M_1$, $M_2$, $M_4$, and $M_8$ all correspond to $k\! =\! 1$. 

We use now Eq.~(\ref{eq:iterative}) together with (\ref{eq:Apm}) for $A^{\pm}_\alpha(\tau)$ and Eq.~(\ref{eq:rho0}) for the initial state of $E$ to obtain the (unnormalized) state of $E$ conditioned on $n$ measurements giving $M_j$ sequence of results:
\beq
\tilde{\rho}^{E}_{n,M_j} = \sum_{\alpha} p^{n,M_j}_\alpha \hat{P}_{\alpha} \,\, ,  \label{eq:rhoMj}
\eeq
with
\beq
 p^{n,M_j}_\alpha = [A^{+}_{\alpha}(\tau)]^{n-k(j)}[A^{-}_{\alpha}(\tau)]^{k(j)}   \label{eq:pMjalpha}
\eeq
with the normalized state given by $\hat{\rho}^{E}_{n,M_j} = \tilde{\rho}^{E}_{n,M_j} / \mathrm{Tr}[\tilde{\rho}^{E}_{n,M_j}]$.	
%For convenience, since the unnormalised state $\tilde{\rho}$ reflects useful information about measurement sequence - the probability of obtaining it -  we will employ it in our further formulation.
The probability of obtaining such a sequence $M_j$ is given by
\beq
\prob_n (M_j \vert\rho^E_{0}) = \mathrm{Tr} \tilde{\rho}_{n,M_j} =  \sum_{\alpha} p^{(0)}_{\alpha} [A^{+}_{\alpha}(\tau)]^{n-k(j)}[A^{-}_{\alpha}(\tau)]^{k(j)} \,\,  \label{eq:pMj}
\eeq
and the probability of observing a sequence with path length $k$ is given by
$$\prob_n (k \vert\rho^E_{0}) = \displaystyle\sum_{M_j,k(j)=k}\prob_n (M_j \vert\rho^E_{0}) \,\, .$$ 

It is now crucial to note the following fact. The above expressions for $\prob_n (M_{j})$ can be obtained using a completely classical model of environment's influence on the qubit. Let us identify the result of measurement of $\pm x$  with a coin toss, and the influence of the environmental state $\alpha$ on the qubit with a degree to which the coin is biased. At the beginning of the experiment we draw a biased coin from an ensemble of coins: the only things that we know is {\it a priori} probability $p^{(0)}_{\alpha}$ of drawing a coin biased in a particular way. Then we throw the coin $n$ times. Learning the number $k$ of times in which we got the $-x$ result increases our knowledge about which $\alpha$ coin we are dealing with. Formula \eqref{eq:pMjalpha} corresponds exactly to classical Bayesian expression for {\it a posteriori} probability of having the $\alpha$ biased coin. 
The only nontrivial feature of the quantum formulation of this problem is that we can have multiple states $\ket{\alpha}$ corresponding to the same ``coin bias'', i.e.~the same values of $A^{\pm}_{\alpha}(\tau)$. This happens when $\Delta \omega_{\alpha}\tau$ and $\Delta\omega_{\alpha'}\tau$ differ by a multiple of $2\pi$ for states $\ket{\alpha}$ and $\ket{\alpha'}$.

%%% SHORT TIMES
\subsubsection{Short evolution times}  \label{sec:short_times}
When $\tau$ is shorter than the inverse bandwidth of the environment, i.e.~when $\Delta \omega_{\alpha}\tau \! \ll \! 1$ for every $\alpha$, we easily see that probability of obtaining sequence $M_j$ with $j\! > \! 0$ is $\sim x^{2k(j)}$, where $x \! \ll \! 1$ is the maximal value of $\Delta \omega_{\alpha}\tau $. Consequently, $\prob (M_{j>0}) \! \ll \! 1$.
On the other hand, $\prob (M_0) \! \! \approx \! 1$. 
This fact is not surprising since the unitary evolution of the composite system always approaches identity for very short evolution times.  
 
%%% LONG TIMES
\subsubsection{Long evolution times} \label{sec:long_times}
\begin{figure}
	\begin{center}
		\includegraphics[width=\linewidth]{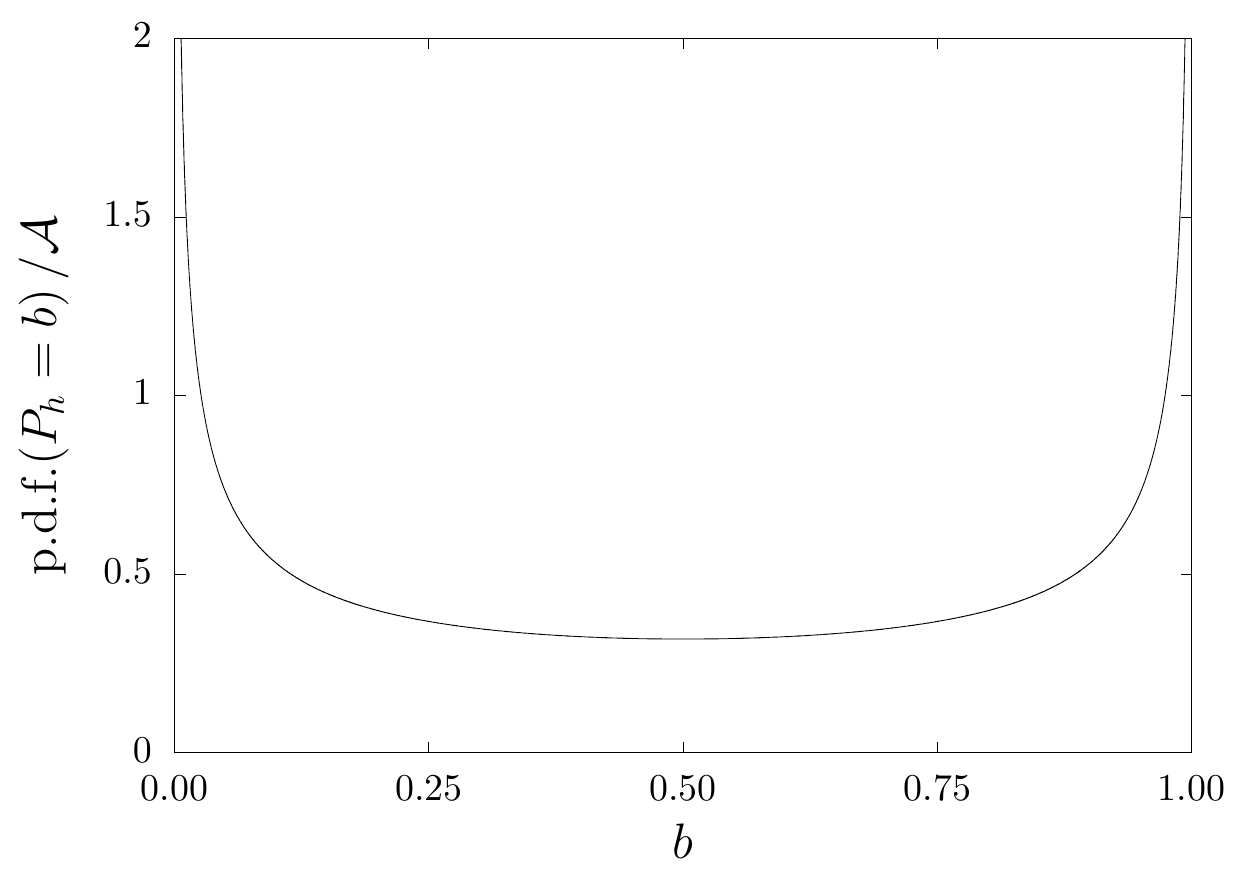}
	\end{center}
	\caption{(Non-normalised) probability density function of the value $P_h$ over the interval $\cv{0,1}$}\label{fig:Pb}
\end{figure}

Let us assume now that the evolution time $\tau$ is much longer than the inverse bandwidth of distribution of $\Delta \omega_\alpha$ energy differences. For a large environment, the number of distinct $\Delta \omega_\alpha$ should be very large, and it should be feasible to work rather work with smooth ``density of states'' corresponding to a given narrow range of $\Delta \omega$: a coarse-grained distribution of $\Delta \omega$, discussed in more detail in Sec.~\ref{sec:coarse_grained}. For a reasonably smooth distribution of this kind, and for large $\tau$ considered now, the distribution of $\theta \! \equiv \! \Delta \omega_\alpha \tau$ modulo $2\pi$ should be approximately flat in $\theta \in [0,2\pi]$ range. In the classical model discussed above, $\theta$ is a parameter that controls the degree to which the coin is biased, as the probability of a coin toss to give a ``heads'' ($+x$) result is $P_h\! =\! \cos^2 \theta/2$, and for the ``tails'' probability we have $P_t \! =\! \sin^2 \theta/2$. For flat distribution of $\theta$ the probability distribution density of the coin being characterized with $b$ value of $P_h$ on the interval $\cv{0,1}$ is
\begin{equation}
\text{p.d.f.}\cv{P_h =b} \! =\! \frac{1}{2\pi}\frac{1}{\sqrt{b(1-b)}}  \,\, .
\end{equation}
The above function is shown in Fig.~\ref{fig:Pb}. Clearly, heavily biased coins, tossing of which gives predominantly mostly heads or tails, are the most probable to be picked. This means that for long enough $\tau$, a sequence of measurements on the qubit will give results $M_{0}$ (all heads, or $+x$) or $M_{2^{n}-1}$ (all tails, or $-x$), with probabilities larger than for any other result. Note that this reproduces qualitatively the main result of Ref.~\cite{RaoEAPreprint2018}.

%%%%%%%%%%%%%%%
%%% DECOHERENCE
%%%%%%%%%%%%%%%
\subsection{Qubit decoherence without and with postselection} \label{sec:QSBAdecoh}
Let us now take a closer look at how the  {\it decoherence} of the qubit is actually measured. The universally accepted procedure is the following: the qubit is initialized, it evolves under influence of the environment for certain time $\tau$, and then a measurement of, say, its $\hat{\sigma}_x$ is performed, and the result is recorded. This has to be repeated $n\! \gg \! 1$ times for each value of $\tau$. The $\tau$ dependence of expectation value of $\hat{\sigma}_x$ is then widely identified with the real part of decoherence function $W(\tau)$:
\beq
\mean{\hat{\sigma}_{x}(\tau)}  = \mathrm{Re}W(\tau) \,\, ,
\eeq
where
\begin{align}
W(\tau) & = 2\rho^{Q}_{\uparrow\downarrow}(\tau) = 2\mathrm{Tr}_{E}\left( \bra{\uparrow} e^{-i\hat{H}\tau}\ket{+x}\bra{+x}\otimes\hat{\rho}_{0}^{E} e^{i\hat{H}\tau}   \ket{\downarrow} \right ) \,\, , \nonumber\\
& = \mathrm{Tr}_{E}\left( \rho_{0}^{E}\hat{U}_{\downarrow}^{\dagger}(\tau) \hat{U}_{\uparrow}(\tau) \right) \,\, , \label{eq:Wgeneral}\\
& = \sum_{\alpha} p^{(0)}_{\alpha} e^{-i\Delta\omega_{\alpha}\tau} \,\, .\label{eq:W}
\end{align}
in which $\hat{\rho}^{Q}(\tau)$ is the reduced density matrix of the qubit at time $\tau$, and the last formula in the equation holds for the quasi-static bath that is our focus here.
 This equation corresponds to dephasing by random unitary (RU) channel, describing the situation in which with probability $p^{(0)}_{\alpha}$ rotation by angle $\Delta\omega_{\alpha}\tau$ about $z$ axis is applied to the qubit.
In this expression, the qubit's coherence, obtained by averaging over many repetitions of the initialization-evolution-measurement cycle, is completely determined by $\hat{H}_{\uparrow/\downarrow}$ and the initial state of the environment $\rho_0^E$. It has to stressed that this is by no means obvious that the above expression applies in situation in which measurements on the qubit influence the state of the environment. While a general discussion of conditions under which Eq.~(\ref{eq:Wgeneral}) describes decoherence is beyond the scope of this paper, we do have to explain when Eq.~(\ref{eq:W}) holds for the quasi-static environment case that is our main focus here. 

In the light of results from previous sections that showed how subsequent measurements on the qubit modify the state of the quasi-static environment, the applicability of Eq.~(\ref{eq:W}) should appear doubtful. However, this  equation has been widely used to describe decoherence caused by the quasi-static environment (see e.g.~\cite{Paladino_RMP14,Cywinski_APPA11,Szankowski_JPCM17} and references therein) - and it has been done for physically well-motivated reasons. The resolution of the problem lies in precise understanding of why we talk about {\it quasi}-static and not simply static environment. 

In reality, no environment can strictly fulfill the two conditions given in Sec.~\ref{sec:QSE}. In order for the state of the environment to be thermalized (or, more generally, correspond to a probability distribution over $\hat{H}_{E}$ eigenstates), the {\it exact} Hamiltonian of the environment, $\hat{H}^{\mathrm{exact}}_{E}$ has to contain terms that do not commute with $\hat{H}_{E}$ introduced previously. These could describe coupling with the thermal reservoir (e.g.~with the crystal lattice in the case of nuclei), or interactions pertaining to $E$ alone - but in both cases the rationale for their omission from $\hat{H}_{E}$ is that dynamics caused by them is so much slower than the one caused by terms retained in $\hat{H}_{E}$ that it can be {\it completely} neglected on timescale $\tau$ of evolution of qubit coupled to $E$. 

The conceptually cleanest way to justify the use of Eq.~(\ref{eq:W}) is to assume that qubit evolutions lasting for $\tau$ are separated by $\Delta t$ waiting times (between measurement and re-initialization) that are long enough for evolution under $\hat{H}^{\mathrm{exact}}_{E}$ to bring back $E$ to its previous state. Practically, however, using such a long $\Delta t$ is apparently not necessary in many cases. 
For example, in \cite{Barthel_PRL09}, where single-shot measurement of transverse component of an electron spin interacting with a nuclear bath in a quantum dot were considered, $\Delta t \! \sim \! 10$ $\mu$s was used, while the nuclear environment autocorrelation time for such quantum dots is of the order of seconds \cite{Reilly_PRL08,Reilly_PRL10,Malinowski_PRL17}. However, averaging of single-shot results for time $\sim \! 10$ s leads to decay of coherence signal that is in good agreement with theoretical predictions for inhomogeneous broadening due to averaging over equlibrium ensemble of environmental states \cite{Barthel_PRL09}.
It thus appears that it is enough for total experiment time ($\approx M \Delta t$ where $M$ is the number of measurement and $\Delta t \! \gg \! \tau$ is assumed) to be long enough to guarantee that the dynamics caused by $\hat{H}^{\mathrm{exact}}_{E}$ makes $E$ ergodic, and thus the average over the sequence of measurements corresponding to averaging over $\hat{\rho}^{E}_{0}$ ensemble, {\it even in the presence of modifications of state of $E$ induced by projective measurements on the qubit}. 
Let us note that in experiment on NV centers described in \cite{RaoEAPreprint2018} the $T_{2}^{*} \!=\! 1.2$ $\mu$s coherence decay time observed after averaging a large dataset of results had a value typical for NV centers, supporting our assumption that environment dynamics can be considered ergodic on timescales of acquistion of all the measurement results used for averaging and obtaining expectation values of all the relevant observables.

While we cannot precisely delineate here the sufficient conditions for applicability of this approximation, we can state what is necessary for our discussion of sequential measurement protocol to describe a real world scenario. 
The theory for environmental state modification discussed in this paper applies to the situation, in which the influence of additional terms from $\hat{H}^{\mathrm{exact}}_{E}$ is negligible on timescale $\approx n(\tau+\Delta t)$ of the whole sequence of $n$ measurements. On this timescale the environment is really static, with its state changing only due to our acquisition of new information from measurements on the qubit. 

On the other hand, when we discuss dephasing of the qubit, or probability of obtaining a given measurement sequence, we assume that the acquisition of all the data necessary for reconstruction of $W(\tau)$, or $\prob (M_j)$, takes time much longer than the time on which the additional terms in $\hat{H}^{\mathrm{exact}}_{E}$ overcome the effect of measurements of the qubit. Then, for purposes of averaging over its influence, the environment can be assumed to be independent of the qubit, and described by its initial thermal state $\hat{\rho}^{E}_{0}$. This will also apply to discussion of dephasing due to a post-selection of environmental state in Sec.~\ref{sec:dephasing}. We can think about first making $n$ consecutive measurements, characterized by fixed $\tau$ and $\Delta t$, and then taking one datapoint for qubit's coherence after time delay $\tau'$ only after a certain string of measurement results $M_j$, was obtained. The total data acquisition time for such post-selected coherence signal, $W_{M_j}(\tau')$, will be $\sim \! 1/\prob (M_j)$ times large than in the previously discussed case of measurement of $W(\tau')$. For such a measurement protocol, the decoherence will be theoretically described by
\begin{align}
W_{M_j}(\tau') & = \mathrm{Tr}_{E}\left( \rho^{E}_{n,M_j}\hat{U}_{\downarrow}^{\dagger}(\tau') \hat{U}_{\uparrow}(\tau') \right) \,\, , \nonumber\\
& = \sum_{\alpha} p_{\alpha}^{(n,M_j)} e^{-i\Delta\omega_{\alpha}\tau}  \,\, . \label{eq:Wpost}
\end{align}

Let us also note that the above reasoning concerning the ergodicity of dynamics of $E$ during the whole experiment, means that even if the environment was actually in a random pure state before the beginning of the whole experiment, any observable, such as probability of obtaining a given sequence $M_j$ of measurement results, or time-dependence of post-selected qubit coherence $W_{M_j}$, corresponds to previously given formulae in which $\rho_{0}^{E}$ is taken as a thermal state.

%%%%%%%%%%%%%%%%%%%%%%%%%%%%%%%%%%%%%%%%
%%% MEASUREMENT STATISTICS FOR NV CENTER
%%%%%%%%%%%%%%%%%%%%%%%%%%%%%%%%%%%%%%%%
\section{Application to a spin qubit interacting with nuclear spin bath}  \label{sec:application}
\subsection{NV center spin qubit and its environment}
Now we consider an example of a qubit interacting with a quasi-static bath. For concreteness we focus on the case of nitrogen vacancy (NV) center surrounded by an environment consisting of $N$ nuclear spins of $^{13}$C, but, as we discuss below, the qualitative results will be the same for a broad class of spin qubits interacting with nuclear baths. A general Hamiltonian of the system \cite{Doherty_PR13} reads
\begin{equation}
	\begin{split}
		\hat{H} &= \cv{\Delta_0\hat{S}^2_z + \Omega\hat{S}_z}\otimes\iden^E\\
		 &\hspace{0.5cm} + \iden^Q\otimes\cv{\omega\sum_{k=1}^N\hat{I}_z^k+ \sum_{k<l} \sum_{i,j=x,y,z} \hat{I}^k_i\mathbb{B}^{i,j}_{k,l}\hat{I}^l_j} \\
		 &\hspace{0.5cm} + \sum_{k=1}^N \sum_{j=x,y,z} \hat{S}_z\otimes\cv{\mathbb{A}^{z,j}_{k}\hat{I}^k_j}\label{eq:NV-Hamiltonian}
	\end{split}
\end{equation}
where $\hat{S}_{z}$ is the operator of spin 1, $\hat{I}^{k}_{j}$ is the operator of the $j$-th component of nuclear spin $1/2$, $\Delta_0$ is the splitting between $m_s\! =\! 0$ and $m_s\! =\! \pm 1$ states, $\Omega \! =\! -\gamma_e B_z$ is the Zeeman splitting of the qubit ($\gamma_e \! =\! 28.02$ GHz/T is the electron gyromagnetic factor and $B_z$ is the magnetic field directed along the $z$ axis connecting the $N$ impurity to carbon vacancy), $\omega=-\gamma_{C^{13}}B_z$ with $\gamma_{C^{13}}=10.71$ MHz/T is the Larmor precession frequency of the nuclei, inter-nuclear dipolar interactions are parametrized by $\mathbb{B}_{k,l}^{i,j}$ couplings, and $\mathbb{A}_{k}^{z,j}$ is the hyperfine interaction between the qubit and $k$-th spin.

We focus now on the qubit based on $\ket{m_s=\pm 1}$ states of the NV center. This choice will make the subsequent calculations applicable to a wider class of single-electron spin qubits (for which $S_z$ is spin-$1/2$ operator) that interact with a nuclear bath, e.g.~those based on III-V compound quantum dots \cite{Koppens_PRL08,Bechtold_NP15,Bechtold_PRL16}, silicon quantum dots \cite{Kawakami_PNAS16}, phosphorous \cite{Tyryshkin_NM12,Pla_Nature12} and bismuth \cite{Wolfowicz_NN13} donors in silicon, and other color centers in diamond \cite{Rogers_PRL14} and SiC \cite{Widmann_NM15,Carter_PRB15}. The NV system Hamiltonian is then brought in the form of Eq.~(\ref{eq:Hamiltonian_gen}), with $\hat{H}_{E}$ given by terms describing Larmor precession of nuclei and their mutual dipolar interactions, and 
\beq
\hat{V}_{\uparrow} = -\hat{V}_{\downarrow} = \sum_{k=1}^N \sum_{j=x,y,z} \mathbb{A}^{z,j}_{k}\hat{I}^k_j \,\, .  \label{eq:V}
\eeq
We choose now to rotate about the $z$ axis the coordinate system for each nucleus in such a way that out of two transverse coupling terms, $\mathbb{A}^{z,x/y}_{k}$, only one is nonzero. We have then $\hat{V}_{\uparrow} = -\hat{V}_{\downarrow} = \sum_{k=1}^{N} \hat{V}_{k}$ with
\beq
\hat{V}_{k} = A^{x}_{k} \hat{I}_{k}^x +  A^{z}_{k} \hat{I}_{k}^z \,\, ,  \label{eq:Vk}
\eeq
where the longitudinal, $A^{z}_{k}$, and transverse, $A^{x}_{k}$ couplings are given by
 \begin{align}
		A_{k}^{z} &= \frac{\mu_0\gamma_e \gamma_{C^{13}}}{4\pi R_{k}^3}\cv{1-3\cos^2\cv{\theta_k}}\\
		A_{k}^{x} &= \frac{\mu_0\gamma_e \gamma_{C^{13}}}{4\pi R_{k}^3}\cv{1-3\sin\cv{\theta_k}\cos\cv{\theta_k}}
	\end{align}
with $R_{k}$ being the distance between $k^{th}$ spin and the qubit, $\theta_k$ thr angle between the vector connecting the $k^{th}$ spin with the qubit and the $z-$axis, and $\mu_0=4\pi\cdot 10^{-7} \text{N}\cdot\text{A}^2$  a magnetic permeability in vacuum.

The last approximation that we will now make is to disregard the dipolar interactions between the nuclear spins. In a rarely encountered (for natural concentration of $^{13}$C) case of nearest-neighbor nuclei, the timescale of nuclear dynamics is $\approx \! 1$ ms, and for most nuclei the timescale of their precession due to dipolar interactions with the remaining bath spins is much longer. In the following we fill focus on dynamics of the system on much shorter  timescales of $\sim \! 10-100$ $\mu$s, and we will approximate the environment Hamiltonian by $\hat{H}_{E} = \omega \sum_{k} \hat{I}_{k}^{z}$.

Consequently, using the notation from Section \ref{sec:general}, in the rotating frame we have $\hat{H}_{E} = \sum_{k} \omega\hat{I}_z^k$, and $\hat{V}_\s$ operators given by Eq.~(\ref{eq:V}).

%%%%%%%%%%%%%%%%%%%%%%%%%
%%% APPLICABILITY OF QSBA
%%%%%%%%%%%%%%%%%%%%%%%%%
\subsection{Regimes of applicability of quasi-static bath approach}
We will assume that the initial state of the environment is a thermal one. Furthermore, for experimentally relevant temperatures and magnetic fields smaller than $\sim \! 1$ Tesla, one can safely assume that the initial nuclear state is completely mixed: $\hat{\rho}^{E}_{0} \! =\! \iden^E/2^{N}$ for environment consisting of $N$ nuclei.

While it is obvious that $[\hat{V}_\uparrow,\hat{V}_\downarrow]\! = \! 0 $ in the considered case, the condition $[\hat{H}_{E},\hat{V}_{\uparrow/\downarrow}]\! =\! 0$ requires a more careful discussion. 

%%%%%%%%%%%%%%
%%% ZERO FIELD
%%%%%%%%%%%%%%
\subsubsection{Zero magnetic field}
One situation in which all the conditions for quasi-static character of the bath are fulfilled is the case of zero magnetic field, when $\omega \! =\! 0$ makes $\hat{H}_{E}$, and consequently the above commutator, vanish. 

In this case the basis $\ket{\alpha}$ consists of products of eigenstates of $\hat{V}_k,$ from Eq.~(\ref{eq:Vk}) i.e. 
\beq
\ket{\alpha}\in\displaystyle\bigotimes_{k=1}^N\left\{\ket{v^\uparrow_k},\ket{v^\downarrow_k}\right\} \,\, , \label{eq:zero_field_basis}
\eeq
where $\hat{V}_k\ket{v^{\uparrow/\downarrow}_k} = \pm v_k\ket{v^\uparrow_k}$, and $v_k=\sqrt{\cv{A_k^x}^2 + \cv{A_k^z}^2}$. Hence the frequencies defined in Eq.~(\ref{eq:wa}) are given by 
\beq
\omega_\alpha^\uparrow = -\omega^\downarrow_\alpha = \frac{\Delta\omega_\alpha}{2} = \frac{1}{2}\displaystyle\sum_{k=1}^N \sigma_\alpha\cv{k} v_k \,\, ,  \label{eq:dw_low}
\eeq
with $\sigma_\alpha\cv{k} = +1$ if the $k^{th}$ element in $\ket{\alpha}$ is $\ket{v_k^\uparrow}$ and is equal to $-1$ otherwise. 

%%%%%%%%%%%%%%%
%%% HIGH FIELDS
%%%%%%%%%%%%%%%
\subsubsection{Purely longitudinal couplings or high magnetic fields}
Another case in which the quasi-static character of the environmental influence on the qubit is obvious is that of purely longitudinal coupling of nuclear spins to the central spin, i.e.~$A^{x}_{k} \! =\! 0$, which leads to $[\hat{H}_{E},\hat{V}_{\uparrow/\downarrow}]\! =\! 0$. 

This applies to systems in which the qubit-nuclear coupling is dominantly of contact hyperfine character, i.e.~only  $\mathbb{A}^{j,j}$ are nonzero, while at high magnetic fields $\mathbb{A}^{x,x}$ and  $\mathbb{A}^{y,y}$ couplings are too small compared to qubit's splitting $\Omega$ to visibly affect the dynamics of the qubit and the environment (for discussion of the effect of these couplings on qubit dephasing at magnetic fields that are not high enough to completely suppress them, see e.g.~\cite{Cywinski_PRB09,Neder_PRB11,Bluhm_NP10,Malinowski_PRL17}).
We encounter such a situation for qubits based on electrons and holes in quantum dots \cite{Chekhovich_NM13}, and for electrons bound to phosphorous \cite{Tyryshkin_NM12,Pla_Nature12} and bismuth \cite{Wolfowicz_NN13} donors in silicon (the effects of transverse couplings are visible in Si:P system only for a few nuclei closest to the donor \cite{Witzel_AHF_PRB07}).

The common basis $\ket{\alpha}$ in this case is given by 
\beq
\ket{\alpha} \in \bigotimes_{k=1}^N\left\{\ket{\uparrow}_k,\ket{\downarrow}_k\right\} \label{eq:high_field_basis}
\eeq
 where $\ket{\uparrow}_k$ and $\ket{\downarrow}_k$ are eigenstates of $\hat{I}^k_z$. The eigenvalues of $\hat{H}_{\uparrow/\downarrow}$ are
\beq
\omega_\alpha^{\uparrow/\downarrow} = %\pm \Omega + 
\frac{1}{2}\displaystyle\sum_{k=1}^N \sigma_\alpha\cv{k} (\omega \pm A_k^z ) \,\, , 
\eeq
so that $\Delta \omega_{\alpha} \! =\! \displaystyle\sum_{k=1}^N \sigma_\alpha\cv{k}  A_k^z$.

{% HIGH MAGNETIC FIELD APPROXIMATION
When the transverse components of interactions are nonzero, the projection operators $\hat{P}_{\alpha_s}$ from Eq.~(\ref{eq:Palphas}) are distinct for $s\! =\! \uparrow$, $\downarrow$, and neither of them projects on the $\ket{\alpha}$ states from Eq.~(\ref{eq:high_field_basis}). However, for high magnetic fields, for which  $\omega\gg v_k$ for all $k$, we can still use the common ($s$-independent) basis $\ket{\alpha}$, with eigenvalues	
\beq
\omega_\alpha^{\uparrow/\downarrow} = %\pm \Omega + 
\frac{1}{2}\displaystyle\sum_{k=1}^N \sigma_\alpha\cv{k}\sqrt{\cv{A_k^x}^2 + \cv{\omega \pm A_k^z}^2} \,\, .
\eeq
Here one can see that the purely longitudinal interaction gives the leading term in the approximation 
	\begin{equation}
		\Delta \omega_{\alpha} \! \approx\! \displaystyle\sum_{k=1}^N \sigma_\alpha\cv{k}  A_k^z +\mathcal{O}\cv{\displaystyle\sum_{k=1}^N\cv{A_k^x/\omega}^2} \,\, . \label{eq:dw_high}
	\end{equation} 

Let us note that in both of the above cases the intermediate evolution (between the measurement and re-initialization of the qubit) can be omitted, since there is simply no evolution due to $\hat{H}_{E}$ in  zero field case, and the evolution operator $e^{-i\hat{H}_E t}$ in the high field case commutes with the post-measurement density operator of the environment.

%%%%%%%%%%%%%%%%
%%% SMALL FIELDS
%%%%%%%%%%%%%%%%
\subsubsection{Small magnetic fields}
Apart from zero field and high field cases, let will briefly discuss the case of nonzero but small fields field. First we observe that for magnetic field such that $v_k\gg\omega$ for all $k$, we can repeat the reasoning from the previous section (only with $\omega/v_{k}$ being now the small parameter ), and using the states  $\ket{\alpha}$ from Eq.~(\ref{eq:zero_field_basis}), defined by coupling operators in zero field case, in expression (\ref{eq:rho1pm}) for the $E$ state after the first measurement, should be a good approximation. However, since the projections onto $\ket{\alpha}$ states do not commute with free evolution generated by $\hat{H}_E$, in order for the intermediate evolution to be negligible we have to assume that the delay between measurement and re-initialization, $\Delta t$, is short compared to characteristic timescales of dynamics generated by $\hat{H}_{E}$. In the case at hand this means $\omega \Delta t\! \ll \! 1$.

% THIS COULD ALSO GO THE END OF THE PAPER, TO THE CONLCUSION AND OUTLOOK SECTION:
For larger magnetic fields, for which it is no longer reasonable to maintain that $\hat{P}_{\alpha_\uparrow} \! \approx \! \hat{P}_{\alpha_\downarrow}$, or for longer delays $\Delta t$, there is no preferred basis that is  invariant under both the conditional evolution map and the free evolution map. 
One will have then to deal with nontrivial post-measurement (and conditioned on the outcome if this measurement) evolution of the environment. Collecting a sequence of results of projective measurements does not correspond then to a characterization of a quasi-static environmental state, but it should give information about the dynamics of the environment, and also to the way in which these dynamics are influenced by measurements on the qubit.  
This interesting topic is however out of our scope of this paper, and in the following we will consider only the zero field and high field cases. 

%%%%%%%%%%%%
%%% GAUSSIAN
%%%%%%%%%%%%
\section{Gaussian approximation and its application to the NV center qubits}  \label{sec:Gaussian}
In this section we will first take a detour from discussion of general model of quasi-static environment, and focus on a Gaussian approximation to the bath state: the case in which tracing over the environmental states $\ket{\alpha}$ can be replaced by performing an average over parameter $\Delta \omega$ having a Gaussian distribution. We will compare the analytical results obtained within this approximation to results of an exact calculation for an NV center interacting with $N\! =\! 20$ nuclear spins, showing that the two approaches agree very well, provided that we look only at the centers that do not have nuclei very close to them.
Furthermore, such an approach could be used to described other kinds of environments characterized by slow dynamics that can be neglected on timescale of both free evolution $\tau$ and measurement-reinitialization delay $\Delta t$. An example distinct from the nuclear environment case discussed in detail here is an environment that is a source of $1/f$ charge noise \cite{Paladino_RMP14} that dephases a qubit endowed with finite electric dipole moment.

%%% GAUSSIAN DISTRIBUTION OF FREQS
\subsection{Gaussian approximation for a spin environment} \label{sec:coarse_grained}
With number of nuclei $N$ as small as $20$ the cardinality of $\left\{\alpha\right\}$, and consequently the number of possible values of $\Delta \omega_\alpha$, is  of the order $10^6$. In expressions such as (\ref{eq:rhoMj}), (\ref{eq:pMj}), and (\ref{eq:Wpost}) we are summinng functions of $\Delta \omega_\alpha$ over all possible values of $\alpha$. Due to the fact that $A^{x/z}_{k}$ couplings quickly decay with nucleus-qubit distance, for any spatial arrangement of the nuclei most of these couplings are much smaller than the value of the largest one of them, and the possible values of $\Delta \omega_\alpha$ are expected to form a rather dense set. This suggests that summation over $\alpha$ could be replaced by integral over a smooth, coarse-grained distribution of $\Delta \omega$ splittings, even in the case of a rather small nuclear environment.

Let us define equivalent classes of similar values of $\Delta \omega$ as
	\begin{equation}
		\cvb{\Delta\omega} \cong \left\{\Delta\omega_\alpha : \Delta\omega_\alpha \in\left(\Delta\omega - w/2, \Delta\omega + w/2\right]\right\}
	\end{equation}
where $w$ is a bin-width of appropriate size. For any probability $p_\alpha$ that can be written as a function of random variable $\Delta\omega_\alpha,$ one can the practically replace the probability space $\cv{\left\{\alpha\right\},p\cv{\Delta\omega_\alpha}}$ by $\cv{\mathbb{B},n\cv{\Delta\omega} p\cv{\Delta\omega}}$ where $\mathbb{B}=\cvb{ \min\Delta\omega_\alpha,\max\Delta\omega_\alpha }$ is a compact subset of $\mathbb{R}$ and $n\cv{\Delta\omega}$ is the cardinality of $\cvb{\Delta\omega}.$ 
For the case of completely mixed state of the environment, we have $p_\alpha=p\cv{\Delta\omega_\alpha}=\frac{1}{2^N}\chi_{\cvb{\Delta\omega_\alpha}}$ being a characteristic function of the set $\cvb{\Delta\omega_\alpha}$, and  one can replace direct counting on $\left\{\alpha\right\}$ by a density $n\cv{\Delta\omega}$ on the interval $\mathbb{B}$. In Figures Fig.~\ref{fig:coarse_freq} and \ref{fig:landscape} we show that the the coarse-grained distribution follows a smooth curve for $N\! = \! 20$ nuclei, in both zero and high field ($B_z\! =\! 100$ mT) cases.

%%% GAUSSIAN DISTRIBUTION FIGURE
\begin{figure}[tbh]
\centering
%\subfloat[{direct counting}\label{fig:freq_20}]{%
  \includegraphics[width=0.8\linewidth]{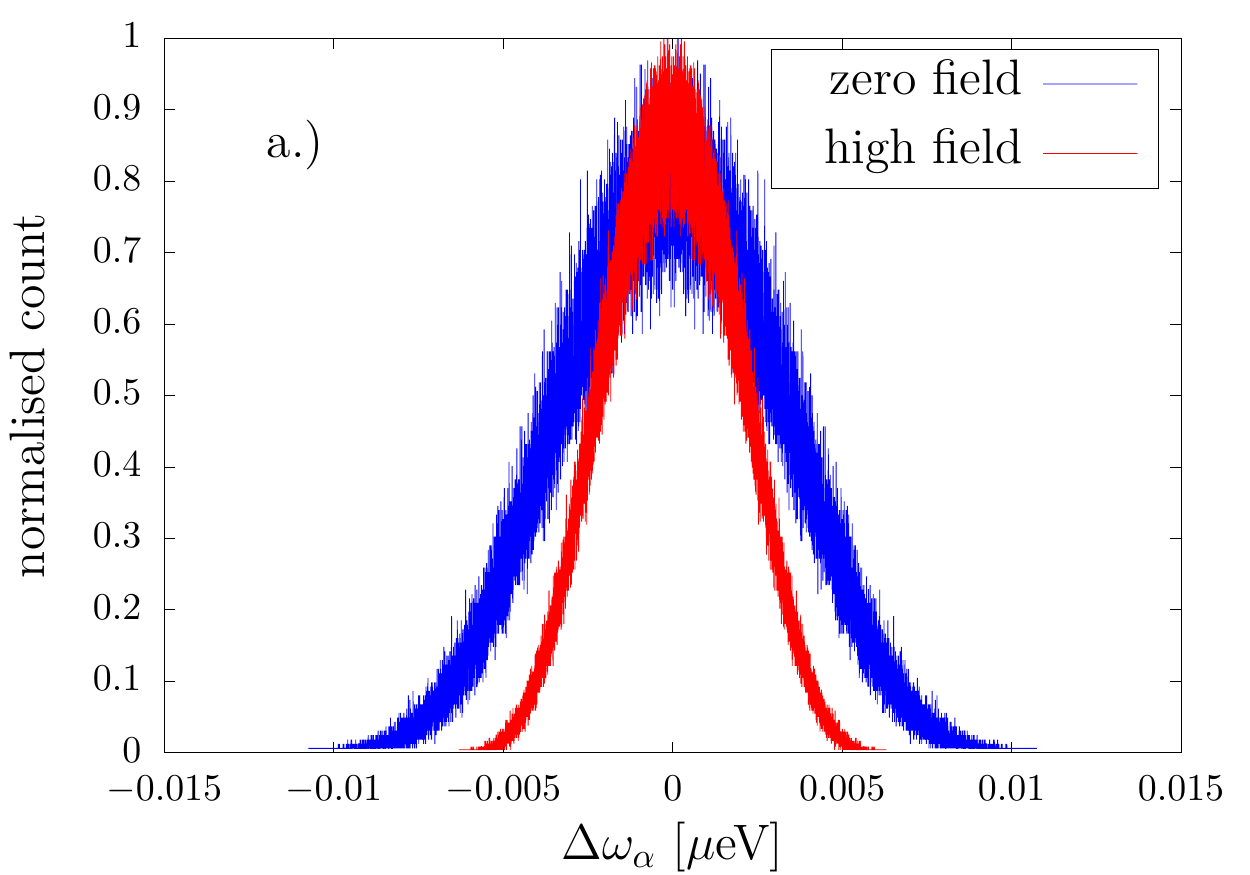}%
%}
\hfill
%\subfloat[{coarse graining calculation}]{%
  \includegraphics[width=0.8\linewidth]{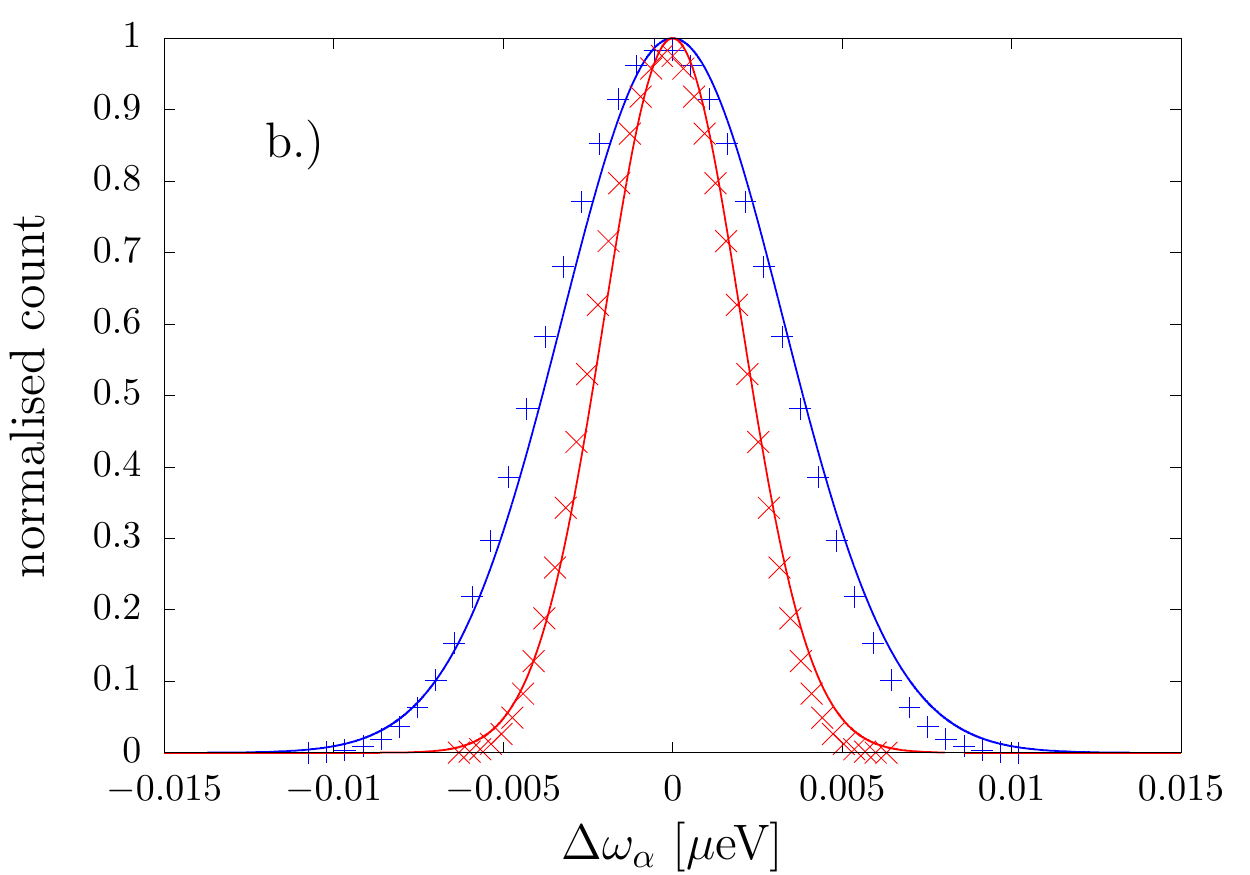}%
%}
\caption{Distributions of the effective frequency $\Delta\omega_\alpha,$ generated for $N=20$ nuclear spins located farther than $0.57$ nm from the qubit, with `high field'' results corresponding to $B_z \! =\! 100$ mT. (a) Direct counting with resolution of $5\times 10^{-5}\mu$eV. (b) Coarse-graining with $40$ bins (symbols) and Gaussian fits (lines). These results are qualitatively representative for NV center not having nuclei closer than $\approx \! 0.5$ nm from it, i.e.~for majority of spatial arrangements of the nuclei the distributions are well fit by Gaussians. An example of an outlier spatial arrangement that does not lead to such distribution is shown in Fig.~\ref{fig:landscape}. }\label{fig:coarse_freq}
\end{figure}
%%% STRONG COUPLING FINGERPRINT FIGURE
\begin{figure}[tbh]
\subfloat{\label{fig:landscape-pos}
  \includegraphics[width=0.8\linewidth]{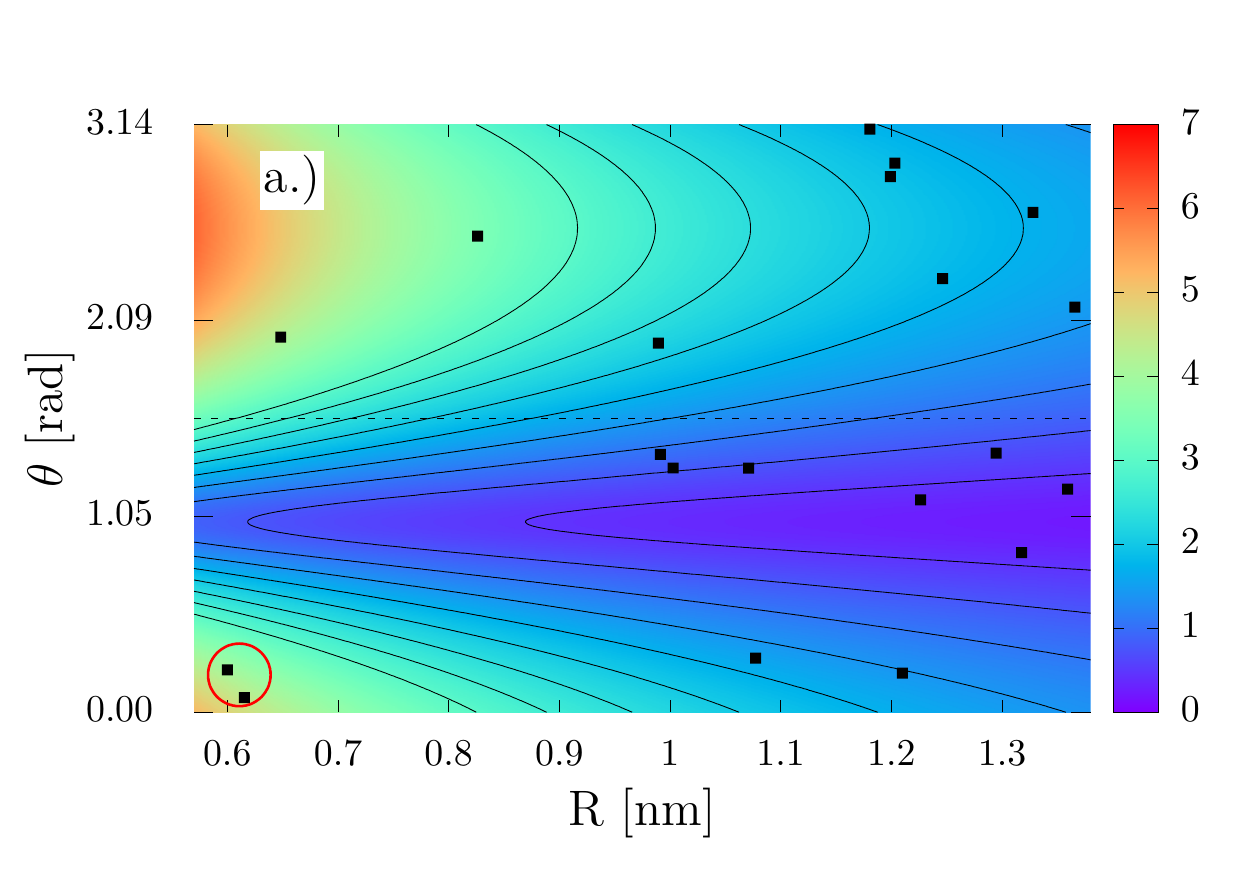}%
}\hfill
\subfloat{\label{fig:Gaussianised}
  \includegraphics[width=0.8\linewidth]{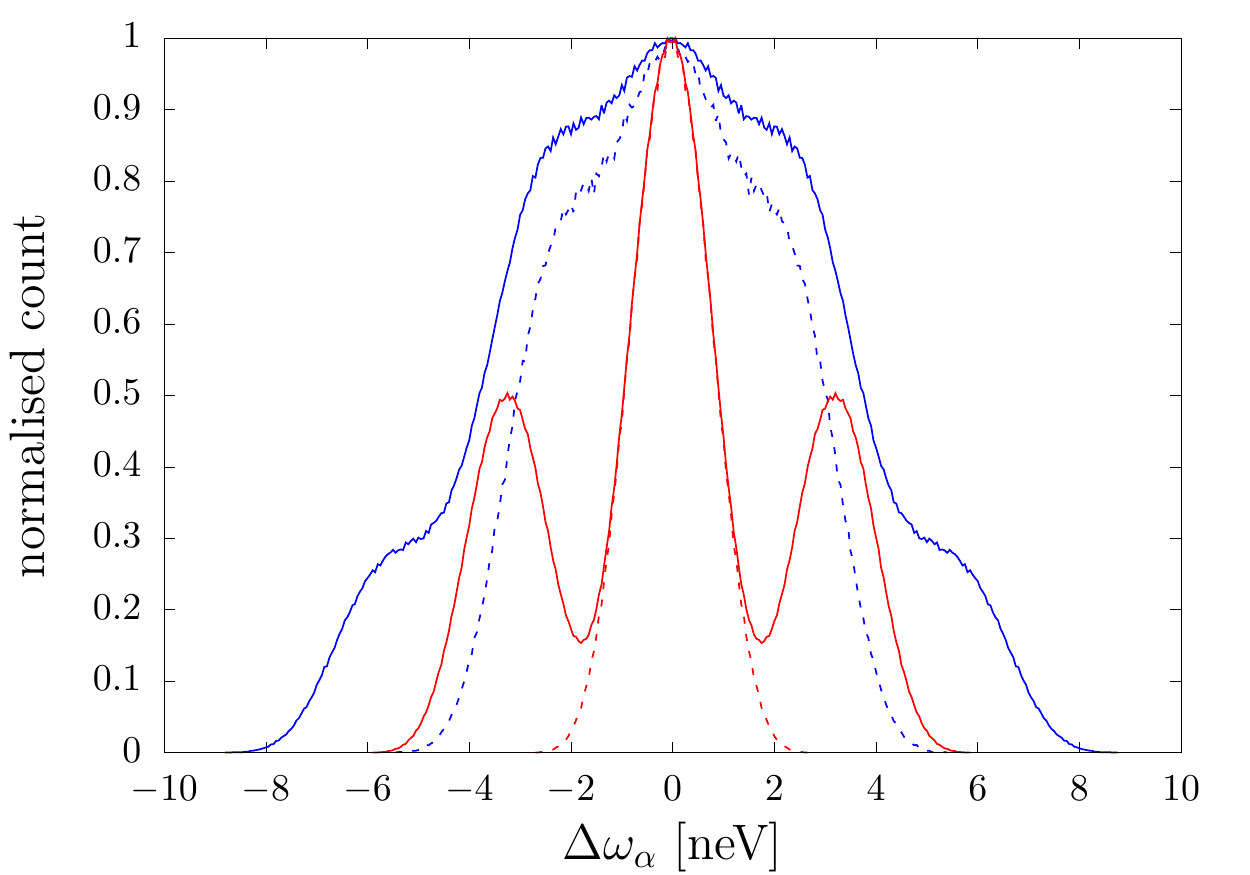}%
}
\caption{(a) An example of spatial realization of environment consisting of $N\! = \! 20$ nuclear spins located at least $0.57$ nm from the qubit that leads to non-Gaussian character of distribution of $\Delta \omega$. In panel (a) $R$ is the distance from the qubit, and $\theta$ is the angle between the vector connecting the nucleus to the qubit and the $z$ axis. The color scale corresponds to the qubit-nucleus coupling $v_k$ (in arbitrary units). The contour lines represent isolevels of magnitude of $v_k$ separated by steps equal to $10\%$ of the maximum value of $v_k$ visible in the figure.
The presence of two nuclei located one close to another, and both about $0.6$ nm from the qubit, marked by red circle in the lower left part of the panel, leads to a prominently non-Gaussian shape of the coarse-grained distributions of $\Delta \omega$ shown in panel (b) (blue solid line for zero field, red dash-dotted line for $B_{0}\! =\! 100$ mT and bin width of $1\times 10^{-4}\mu$eV). Coarse-grained distributions with influence of these two spins removed are shown as dashed and dotted lines. 
}\label{fig:landscape}
\end{figure}

In both Fig.~\ref{fig:coarse_freq} and \ref{fig:landscape} we are showing distribution of $\Delta \omega$ obtained for spatial arrangements of the nuclear spins in which no nuclei are located closer than $0.57$ nm from the center. This distance was chosen by requesting that no $A^{x/z}_k$ coupling is larger than $10\%$ of nuclear Zeeman splitting $\omega$ for $B_z \! = \! 100$ mT, which we use in calculations for the ``high field'' regime. 
After such an exclusion of nuclei most strongly coupled to the qubit, for the majority of spatial arrangements of the remaining nuclei, the coarse-grained distribution of $\Delta \omega$ can be well fit by a Gaussian:  
\begin{equation}\begin{split}
		n\cv{\Delta\omega} p\cv{\Delta\omega} &=\frac{1}{\sqrt{2\pi\sigma^2}}\exp\cvb{-\frac{\cv{\Delta\omega}^2}{2\sigma^2}} \,\, , \label{eq:coarse-distribution}
		\end{split}
	\end{equation}
	with $\sigma \! = \! 2.0$ neV in zero field case and $\sigma \! =\! 1.4$ neV in the high field case for the environment realization illustrated in Fig.~\ref{fig:coarse_freq}. The difference in width of the Gaussians at zero and high fields is due to the fact that while at high fields the contribution of transverse couplings $A^{x}_{k}$ to $\Delta \omega$ is suppressed, see Eq.~(\ref{eq:dw_high}), at zero field both longitudinal and transverse couplings enter on equal footing the formula for $v_k$ and thus for $\Delta \omega$, see Eq.~(\ref{eq:dw_low}).

However, the above observation does not hold for all the spatial arrangements of nuclei located farther than $0.57$ nm from the center. In Figure \ref{fig:landscape} we present an example of such a spatial arrangement for which, due to the presence of only two nuclei with similar $A_k$ located $\approx 0.6$ nm from the qubit, the coarse-grained disctribution of $\Delta \omega$ splittings is clearly of non-Gaussian shape. Such a result becomes typical when at least one nucleus is present within the radius of $\approx 0.6$ nm from the qubit. 

For natural concentration of $^{13}C$, the expected number of nuclei inside a ball of $\approx \! 0.6$ nm radius from the qubit is only about one. Consequently, the probability of having an NV center without such strongly coupled nuclei in its vicinity is sizable, so it is reasonable to focus on such a  class of qubits, the dynamics of which is not dominated by strong effects specific to one or a few proximal nuclei. In fact, when centers strongly interacting with a few proximal nuclei are identified in experiment, for example by observation of prominent oscillations in their free evolution coherence decay, it is more practical to treat the few most strongly coupled nuclei as additional qubits in a multi-qubit register. These spins can be then controlled with rf waves tuned to their precession frequencies that strongly depend on the state of the qubit \cite{Dutt_Science07,Jiang_Science09,Robledo_Nature11,Taminiau_NN14,Waldherr_Nature14}.

Within the approximation discussed above, in all the expressions involving summation over $\alpha$ states of quantities that depend on $\Delta \omega_{\alpha}$, such as Eq.~(\ref{eq:pMj}) for probability of obtaining $M_j$ sequence of results, and Eq.~(\ref{eq:W}) for coherence decay, the summation should be replaced by inregral over $\Delta \omega$ of  a function of $\Delta \omega$ multiplied by weighing factor from Eq.~(\ref{eq:coarse-distribution}). For example, for decoherence function of qubit interacting with the initial equilibrium state of the environment we have
\beq
W(\tau) = \int_{\-\infty}^{\infty}  \frac{1}{\sqrt{2\pi\sigma^2}}\exp\cvb{-\frac{\cv{\Delta\omega}^2}{2\sigma^2}} e^{-i\Delta\omega \tau} = e^{-(\tau/T_{2}^{*})^2} \,\, ,\label{eq:Wcoarse}
\eeq
in which the decay time $T_{2}^{*} \! = \! \sqrt{2}/\sigma$. 
In the following throughout the paper, unless stated otherwise, the example of realization of environment with $N\! =\! 20$ nuclei shown in Fig. \ref{fig:coarse_freq} will be adopted numerically to illustrate the Gaussian approximation, while the $A^{x,z}_{k}$ couplings defining this environment will be used in {\it exact} calculations based on formulas from Section \ref{sec:general} in which summations over all $2^{N}$ states $\ket{\alpha}$ will be carried out.

%%%%%%%%%%%%%%%
% PROBABILITIES
%%%%%%%%%%%%%%%
\subsection{Probabilities of obtaining various sequences of measurement results} \label{sec:probabilities}
%\subsubsection{Results within Gaussian approximation} \label{sec:prob_Gaussian}
Within Gaussian approximation for the initial state of the environment, the expression for probability of obtaining a sequence of measurement results $M_j$, given by general form from Eq.~(\ref{eq:pMj}), reads 
	\begin{equation}
		\begin{split}
		\prob_n\cv{M_j} &=  \int_{-\infty}^{\infty} \mathrm{d} \Delta\omega\frac{e^{-\frac{\cv{\Delta\omega}^2}{2\sigma^2}}}{\sqrt{2\pi\sigma^2}}\\
		&\hspace{0.5cm}\times\cos^{2\cv{n-k\cv{j}}}\cv{\frac{\tau\Delta\omega}{2}}\sin^{2k\cv{j}}\cv{\frac{\tau\Delta\omega}{2}}  \,\, , \label{eq:pMj_Gaussian}
		\end{split}
	\end{equation}
where $n$ is the number of measurements, and $k(j)$ is the number of ``failures'' (measurements giving $-x$ result). 
It is easy to convince oneself that  $\cos^{2n}\cv{\frac{\tau\Delta\omega}{2}}$ (for $j=0$) and $\sin^{2n}\cv{\frac{\tau\Delta\omega}{2}}$ (for $j=2^n-1$) have narrower widths and higher amplitudes than these of products  $\cos^{2\cv{n-k\cv{j}}}\cv{\frac{\tau\Delta\omega}{2}}\sin^{2k\cv{j}}\cv{\frac{\tau\Delta\omega}{2}}$ with $k \! \neq \! 0$, $2^n-1$. 

By straightforward calculation, the probability to obtain the identical sequences $M_0$ and $M_{2^n-1}$ can be written as
		\begin{align}
			\prob_n\cv{M_0} &= \dfrac{1}{2^{2n}}\displaystyle\sum_{r=0}^{2n} \cv{\begin{array}{c}2n\\r\end{array}}e^{-\frac{1}{2}\sigma^2\cv{n-r}^2\tau^2},\\
			\prob_n\cv{M_{2^n-1}} &= \dfrac{1}{2^{2n}}\displaystyle\sum_{r=0}^{2n} \cv{\begin{array}{c}2n\\r\end{array}}\cv{-1}^{n-r}e^{-\frac{1}{2}\sigma^2\cv{n-r}^2\tau^2}. \label{eq:pM0G}
		\end{align}
%Then one can see that, for 
For arbitrary sequence $M_j$ one can derive a relation
\begin{equation}
	\begin{split}
		\prob_n\cv{M_j} &= \sum_{r=0}^{k(j)} \cv{-1}^{k\cv{j}-r}\cv{\begin{array}{c}k(j)\\r\end{array}}\prob_{n-r}\cv{M^{n-r}_0}\label{eq:pMj_inpM0}\\
			&=\sum_{r=0}^{n-k(j)} \cv{-1}^{r}\cv{\begin{array}{c}n-k(j)\\r\end{array}}\prob_{k(j)+r}\cv{M^{k(j)+r}_{2^{k(j)+r}-1}}
	\end{split}
\end{equation}
where $M_j^m$ stands for a subsequence of $m$ measurements. These relations confirm again that only the path length $k(j)$ of the measurement sequence affects the probability of obtaining it. An example of probability distribution for $n=4$ measurements with $\tau=2.7~\mu$s and $\tau=13.5~\mu$s is shown in Fig \ref{fig:prob-all-post}. 
One can see that for short $\tau$, $M_{0}$ is the most probable result, as discussed in Sec.~\ref{sec:short_times}, while for longer $\tau$ both $M_{0}$ and $M_{15}$ are equally probable, as predicted in Section \ref{sec:long_times}.

\begin{figure}
\begin{center}
	\includegraphics[width=0.9\linewidth]{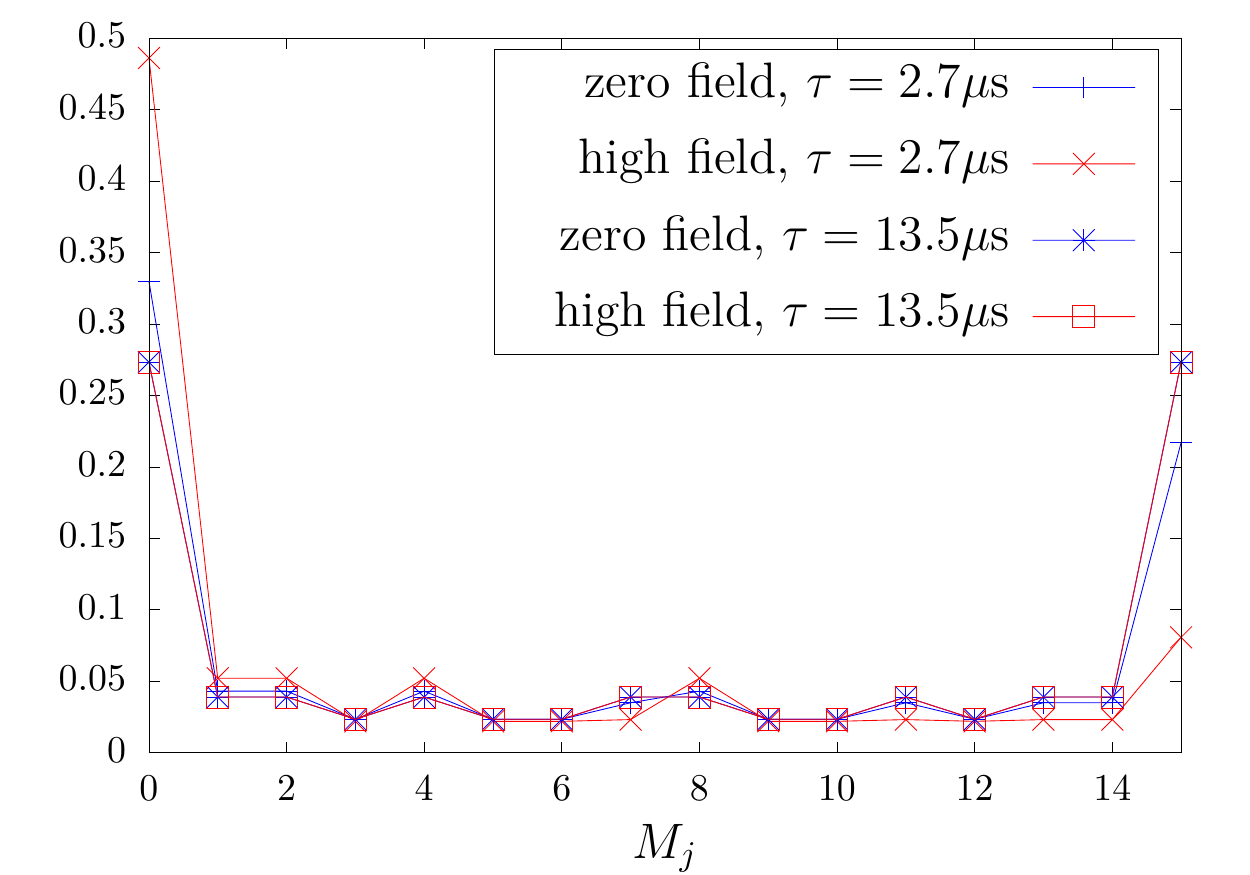}
	\caption{Probability of obtaining $16$ measurement sequences $M_j$ for $n=4$ given by Eq. \eqref{eq:pMj_inpM0} in zero field and high field for  $\tau=2.7~\mu$s and $\tau=13.5~\mu$s, obtained using Gaussian approximation from Eq.~(\ref{eq:pMj_Gaussian}), and parameters $\sigma$ of Gaussian distributions fit to data from Fig.~\ref{fig:coarse_freq}. }\label{fig:prob-all-post}
\end{center}
\end{figure}

% FIGURE: N=20 results
\begin{figure*}[t]
\subfloat{\label{fig:dist-meas-seq-20-zero}
  \includegraphics[width=0.5\linewidth]{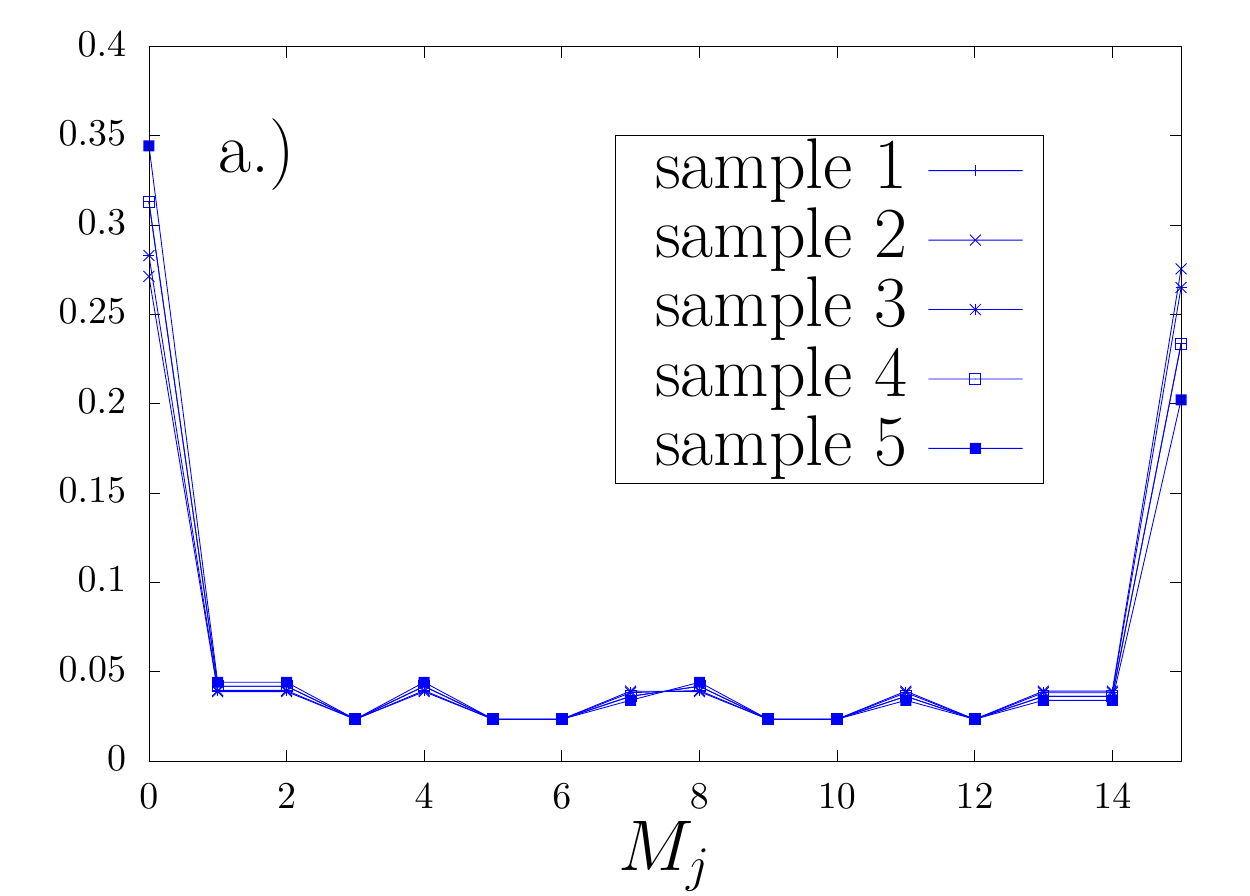}%
}
\subfloat{\label{fig:dist-meas-seq-20-zero-long}
  \includegraphics[width=0.5\linewidth]{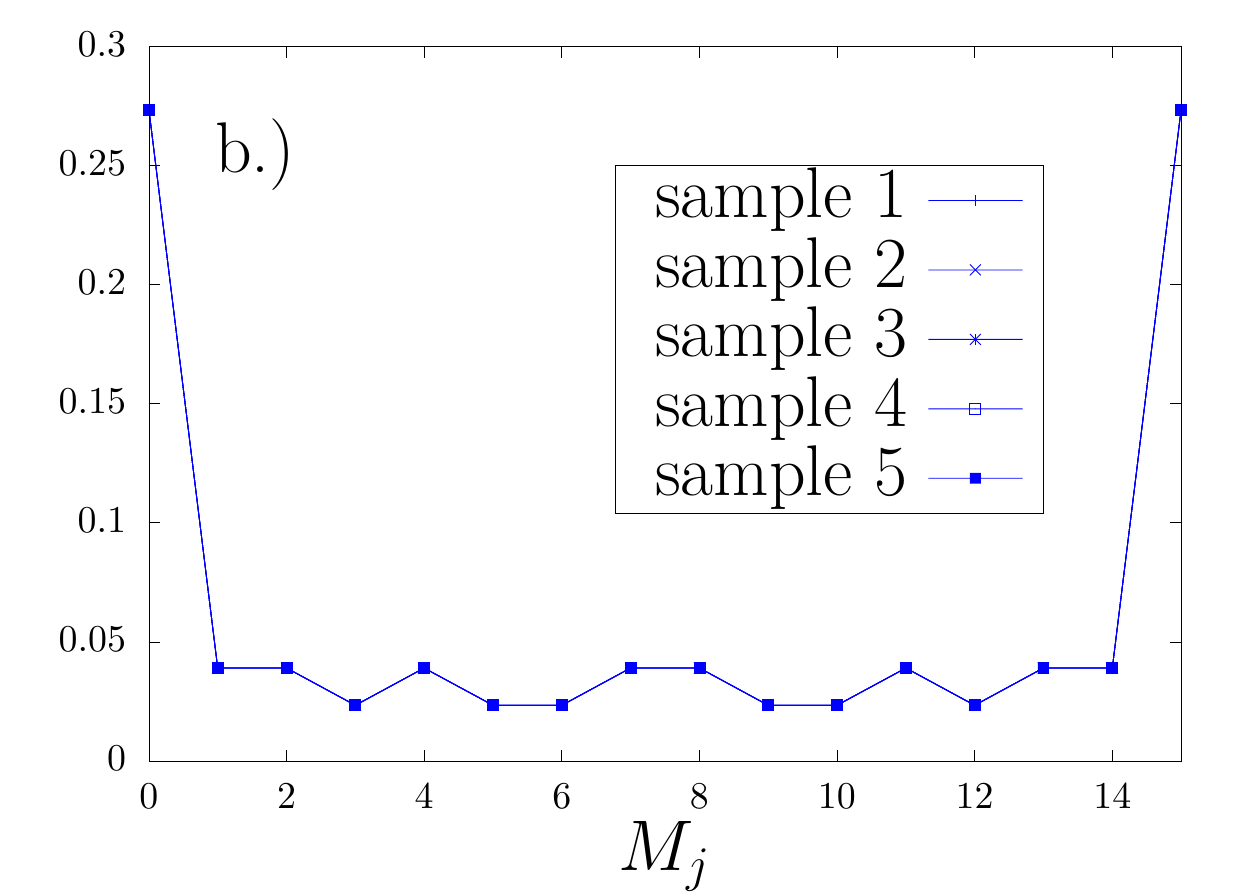}
}\hfill
\subfloat{\label{fig:dist-meas-seq-20-high}
  \includegraphics[width=0.5\linewidth]{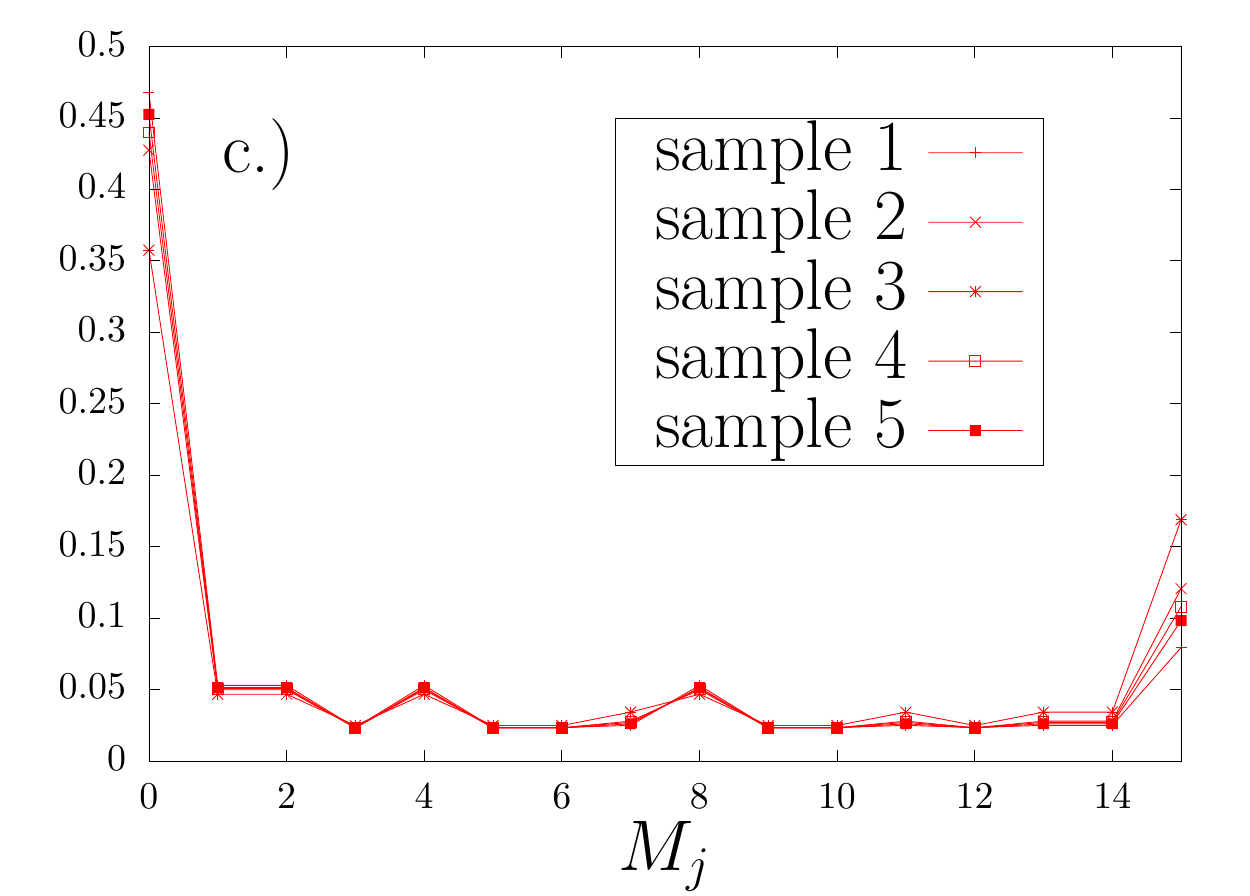}%
}
\subfloat{\label{fig:dist-meas-seq-20-high-long}
  \includegraphics[width=0.5\linewidth]{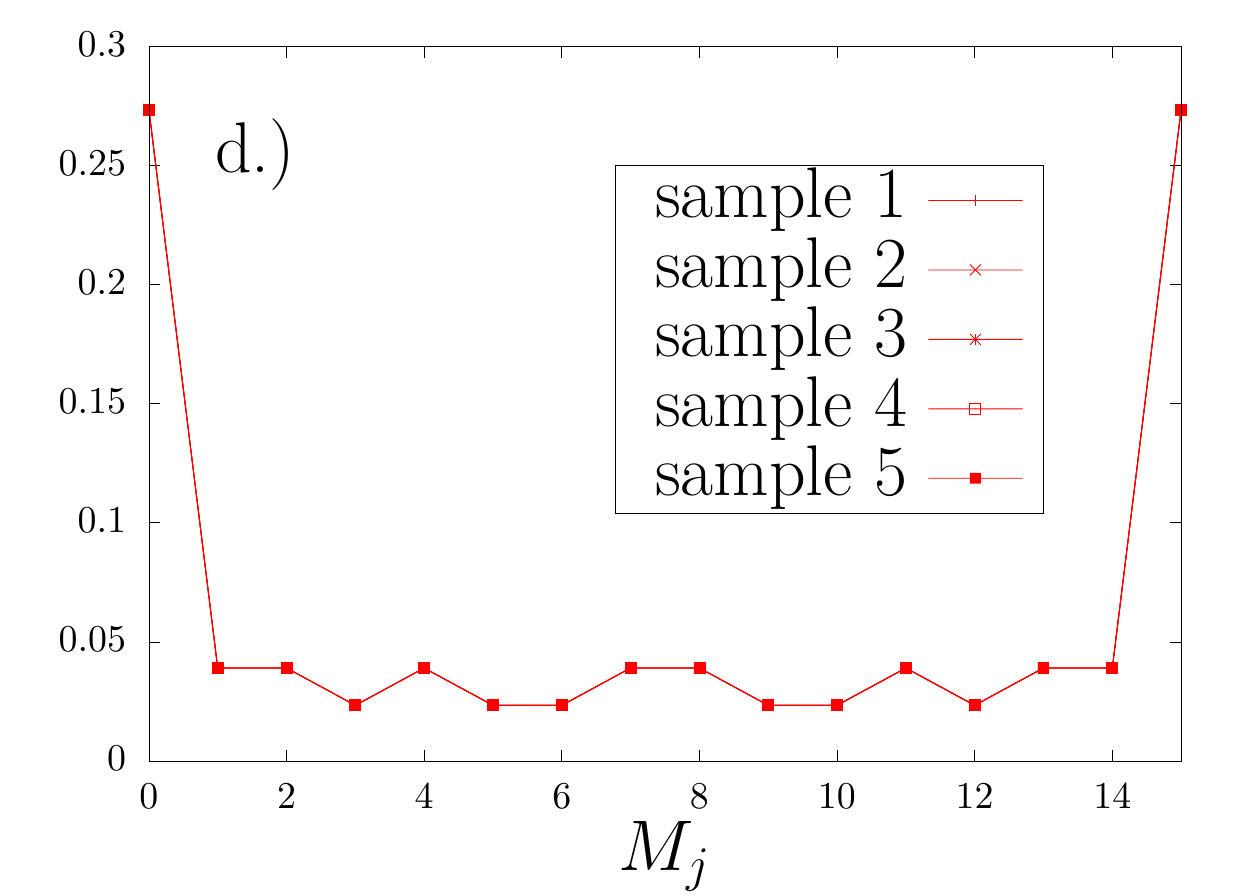}
}
\caption{Probabilities of obtaining all possible measurement sequences of $4$ consecutive measurements for zero field (blue line, upper panels) and high field (red line, lower panels) cases for $N=20$ spins in the environment with $\tau=2.7\mu$s in the left panels, and $\tau=13.5\mu$s in the right panels. Different spatial arrangements of spins in the environment are labeled as sample $1\ldots 5$.
For longer $\tau=13.5\mu$s the symmetry is more prominent and all the spatial arrangements of the nuclei give essentially the same result. }\label{fig:dist-meas-seq-20}
\end{figure*}

% N=20 results
Let us now compare the results obtained with Gaussian approximation with the exact calculations for an environment consisting of $N\!= \! 20$ nuclei located at least $0.57$ nm from the qubit. In addition to the nuclear configuration used to obtain the coarse-grained distribution of $\Delta \omega$ in Fig.~\ref{fig:coarse_freq}, we have investigated four other spatial arrangements of the environment (with nuclear positions randomly generated). For all of them the coherence decay time $T_{2}^{*} \! \approx \! 2$ $\mu$s.
We have calculated $\prob_n\cv{M_j}$ corresponding to $n\!= \! 4$ for each one of them using Eq.~(\ref{eq:pMj}). The results for zero and high field, and for two values of qubit-environment interaction time $\tau$, are shown in Fig.~\ref{fig:dist-meas-seq-20}. The first thing to note is that the differences between results for different environment realizations are almost invisible, especially for longer $\tau$.
This shows that for environment consisting of $20$ randomly positioned spins, its influence on the qubit is nearly self-averaging: all the ``typical'' spatial realizations of the environment lead to very similar  $\prob_n\cv{M_j}$ that are well-represented by a Gaussian approximation calculation shown in Fig.~\ref{fig:prob-all-post}.

All the features of the results in zero field case are in agreement with experimental study from Ref.~\cite{RaoEAPreprint2018}. Results in high fields are qualitatively the same. However, as shown in the Fig.~\ref{fig:dist-meas-seq-20}, precise characteristics of the behaviour (i.e.~probabilities of obtaining each specific $M_k$ sequence of measurements)  depend on the value of magnetic field, especially for times $\tau$ not much longer than $T_{2}^{*}$. Let us also remind that for the described above behavior to be seen in experiment, the measured NV center has to be surrounded by an environment in which there are no nuclear spins with couplings to the qubit that are significantly larger than the maximal couplings of the remaining nuclei (this also means that there should be no strong oscillatory ``fingerprints'' of such exceptionally strongly coupled nuclei in decoherence signal of the qubit). In the presence of a strongly coupled proximal nucleus (or a few of them), the results for $\prob_n\cv{M_j}$ would exhibit more visible differences between distinct spatial realizations of the environment. Independence of $\prob_n\cv{M_j}$ from the exact positions of the spins in the bath, clearly visible in results for longer $\tau$ in Fig.~\ref{fig:dist-meas-seq-20}, arises only when we exclude environment containing nuclei in very close vicinity to the qubit.

% P(M0) FIGURE
	\begin{figure}[tbh]
	\begin{center}
			\includegraphics[width=0.9\linewidth]{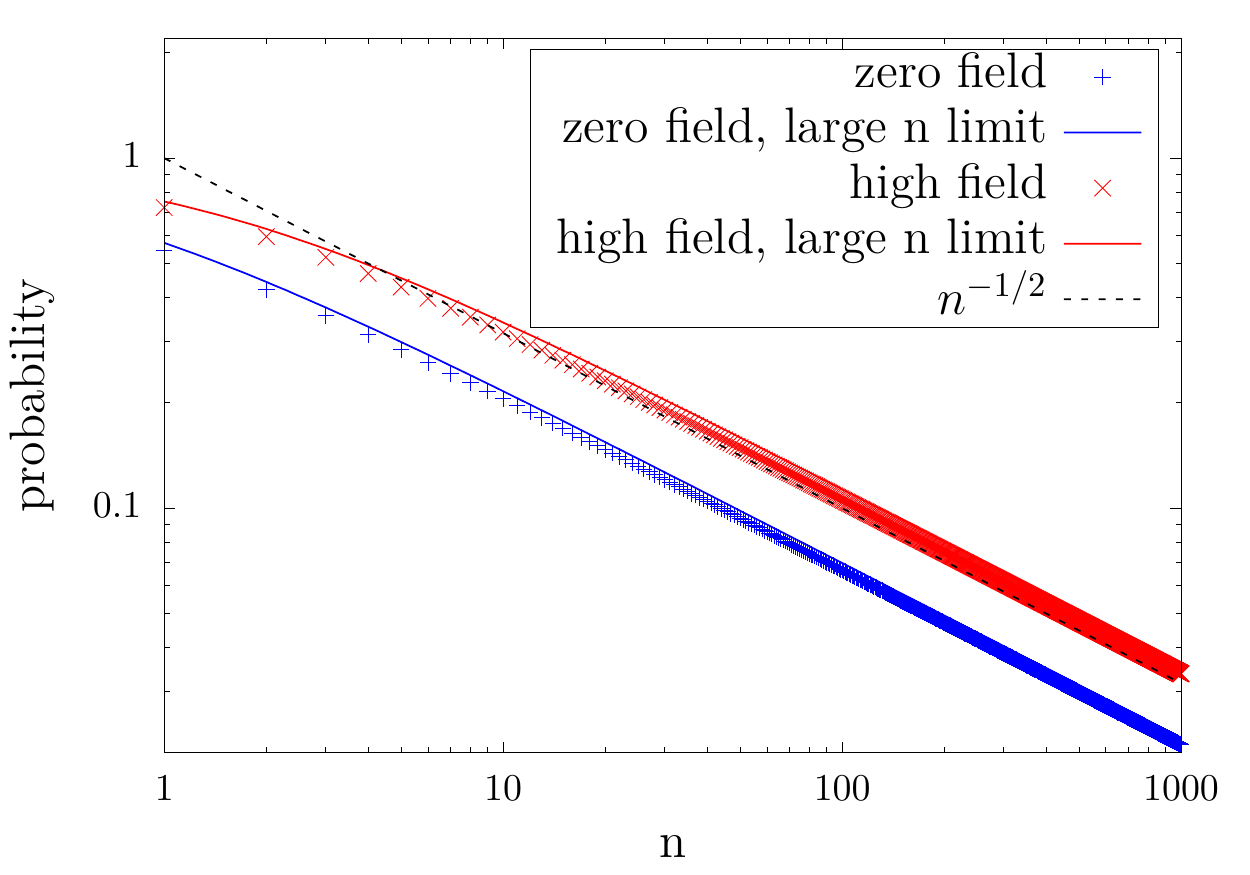}
		\caption{%\cz{MENTION THIS IN THE TEXT HERE!} 
		Probabilities of obtaining sequences $M_0$ of identical measurement results (all projections on $\ket{+x}$) versus number $n$ of measurements for zero field (blue) and high field (red) cases. The symbols correspond to an exact calculation for spatial arrangement of the nuclei used to generate Figs.~\ref{fig:coarse_freq} and \ref{fig:prob-all-post}, and the solid lines are the results of Gaussian approximation in large $n$ limit, Eq.~(\ref{eq:pM0large}).
		The dashed line shows the prediction of $\prob_n\cv{M_j}\! \propto \! 1/\sqrt{n}$ from Eq.~(\ref{eq:sqrtn}) for the high field case.   }\label{fig:ends_vs_n}	
	\end{center}
	\end{figure}

% ADAPTIVE PROBABILITY FIGURE
\begin{figure}[tbh]
		\includegraphics[width=0.9\linewidth]{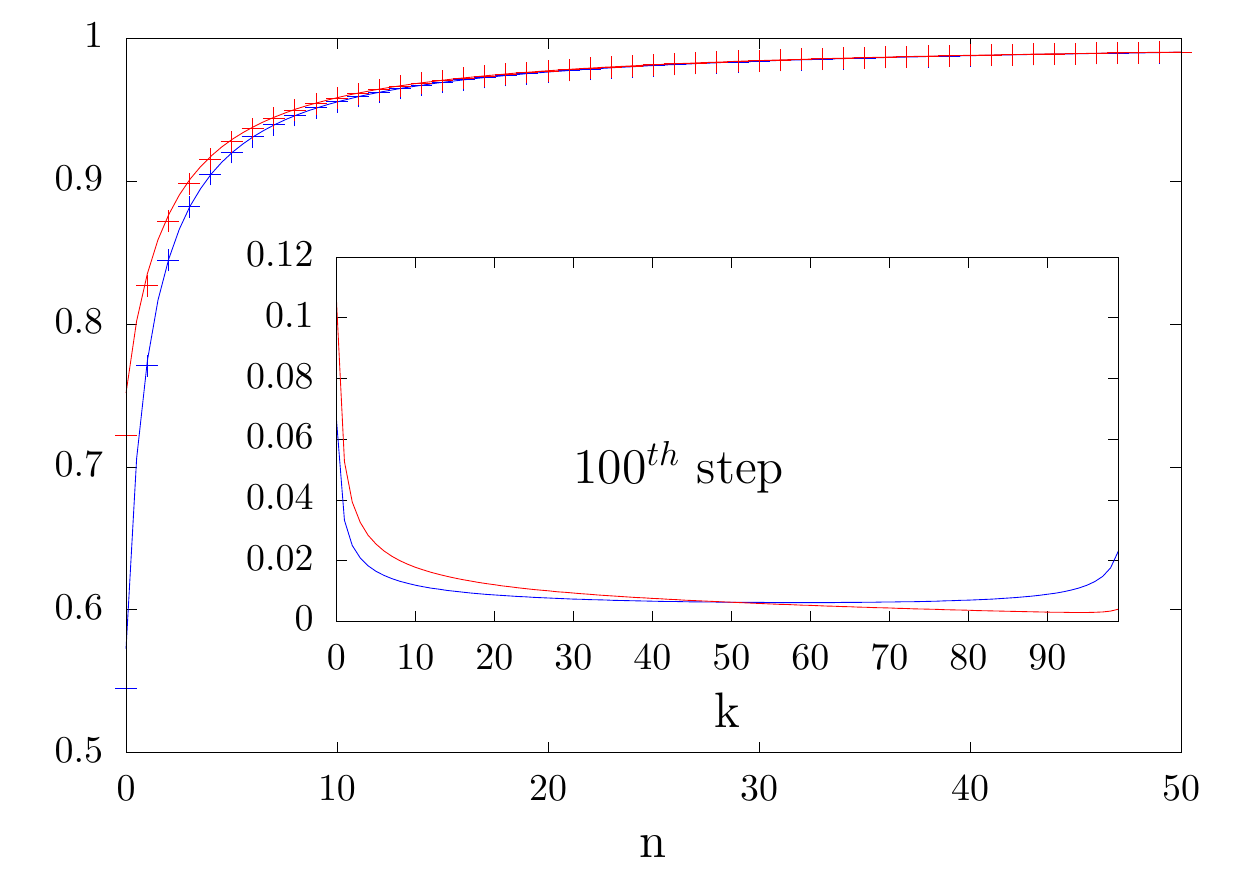}
\caption{Adaptive probability of getting an identical sequence $M_0$ in zero field (blue) and high field (red) regimes. The cross signs represent an exact calculation for spatial arrangement of the nuclei used to generate Figs.~\ref{fig:coarse_freq} and \ref{fig:prob-all-post} and the solid lines are the results of Gaussian approximation in large $n$ limit, Eq.~(\ref{eq:P_adaptive_M0}). In the inset, the probabilities over pathlength (the number of $\ket{-x}$ (failures) of measuring $100$ with $\hat{P}_x$) is displayed. The results are plotted for $\tau=13.5~\mu$s corresponding to long time regime, in which both $M_{0}$ and $M_{2^n-1}$ sequences are the most probable ones (see Eq.~(\ref{eq:sqrtn})), as one can see in the inset.}\label{fig:adaptive} 
\end{figure}

In the limit of large measurement number $n,$ where the binomial distribution with Bernoulli's probability $p=0.5$ above can be approximated by Gaussian distribution, the probability of obtaining  $M_0$ sequence from Eq.~(\ref{eq:pM0G}) can be written as 
		\begin{eqnarray}
			\prob_{\text{large }n}\cv{M_0} 
				 &\approx  \dfrac{1}{\sqrt{n\pi}}\displaystyle\int_{-\infty}^{\infty} e^{-\cv{\frac{1}{n}+\frac{1}{2}\sigma^2\tau^2}\cv{n-x}^2} \mathrm{d}x \nonumber\\		
			&\approx \sqrt{\dfrac{2}{2+n\sigma^2\tau^2}} \equiv f\cv{n,\tau} \,\, ,  \label{eq:pM0large}
		\end{eqnarray}
		while for $M_{2^{n}-1}$ sequence, defining $\sigma''^2 \! \equiv\!  \frac{1}{2}(\frac{1}{n} + \frac{1}{2}\sigma^2\tau^2 )^{-1},$ in the same large $n$ limit we get
\begin{align}
		\prob_{\text{large }n}\cv{M_{2^n-1}} &\approx  \dfrac{1}{\sqrt{n\pi}}\cvb{\int_{-\infty}^{\infty} e^{i p \cv{x-n}}e^{-\frac{\cv{x-n}^2}{2\sigma''^2}} dx}\bigg\vert_{p=\pi} \,\, \nonumber\\
		 & = \prob_{\text{large }n}\cv{M_0}\exp\cv{-\dfrac{\pi^2}{4} f^{2}(n,\tau)} \,\, .			
		\end{align}	
The latter probability vanishes as $\tau\rightarrow 0$, and becomes identical to the probability of $M_0$ result when $\tau \! \gg \! 1/\sigma$. In this limit, we obtain 
\beq
	\prob_{\text{large }n}\cv{M_0}  \approx \prob_{\text{large }n}\cv{M_{2^n-1}} \approx \sqrt{\frac{2}{\sigma\tau}} \frac{1}{\sqrt{n}} \,\, . \label{eq:sqrtn}
\eeq
The predictions of the above formulas for $M_0$ is compared in Fig.~\ref{fig:ends_vs_n} with an exact calculation for one of the previously used spatial arrangements of $20$ nuclei. 

% ADAPTIVE PROBABILITIES
%\subsection{Adaptive probabilities after many measurements}
Using the above expressions for large $n$ asymptotics of $\prob_{n}\cv{M_{0}}$ and $\prob_{n}\cv{M_{2^n-1}}$, the adaptive probabilities of obtaining $\ket{+x}$ result provided that the previous $n$ measurements all gave $\ket{+x}$ or $\ket{-x}$ are
		\begin{align}
				\dfrac{\prob_{n+1}\cv{M_0}}{\prob_{n}\cv{M_0}} &\approx \sqrt{1-\dfrac{\sigma^2\tau^2}{2+\cv{n+1}\sigma^2\tau^2}}
				%\longrightarrow 1,  
				\,\, , \label{eq:P_adaptive_M0}\\
				\dfrac{\prob_{n+1}\cv{M_{2^n-1}}}{\prob_{n}\cv{M_{2^n-1}}} &\approx \cv{\dfrac{\prob_{n+1}\cv{M_0}}{\prob_{n}\cv{M_0}}}\nonumber\\
				&\times\exp{-\dfrac{\cv{n+1}\pi^2}{4+2\cv{n+1}\sigma^2\tau^2}+\dfrac{n\pi^2}{4+2n\sigma^2\tau^2}} \label{eq:P_adaptive_M2-1}
				%\\
				%&\longrightarrow 1,  
			\end{align}
		respectively. Note that for large $n$ both these conditional probabilities approach $1$.
		The results for $M_0$ are shown in Fig.~\ref{fig:adaptive} for both an exact calculation and the Gaussian approximation prediction of Eq.~(\ref{eq:P_adaptive_M0}). This also lead to the preference of length $0$ and length $n$ over all possible path length $k,$ i.e. the identical results $\ket{+x}$ or $\ket{-x}$ are more preferable as in the inset of Fig. \ref{fig:adaptive}.

%%%%%%%%%%%%%%%%%%%%%%%%%%%%%
%%% POST-SELECTED STATES OF E
%%%%%%%%%%%%%%%%%%%%%%%%%%%%%
\subsection{Post-Selected State of Environment} \label{sec:postselected_state}
\begin{figure}[tbh]
\subfloat{\label{fig:dist-post-short}
		\includegraphics[width=\linewidth]{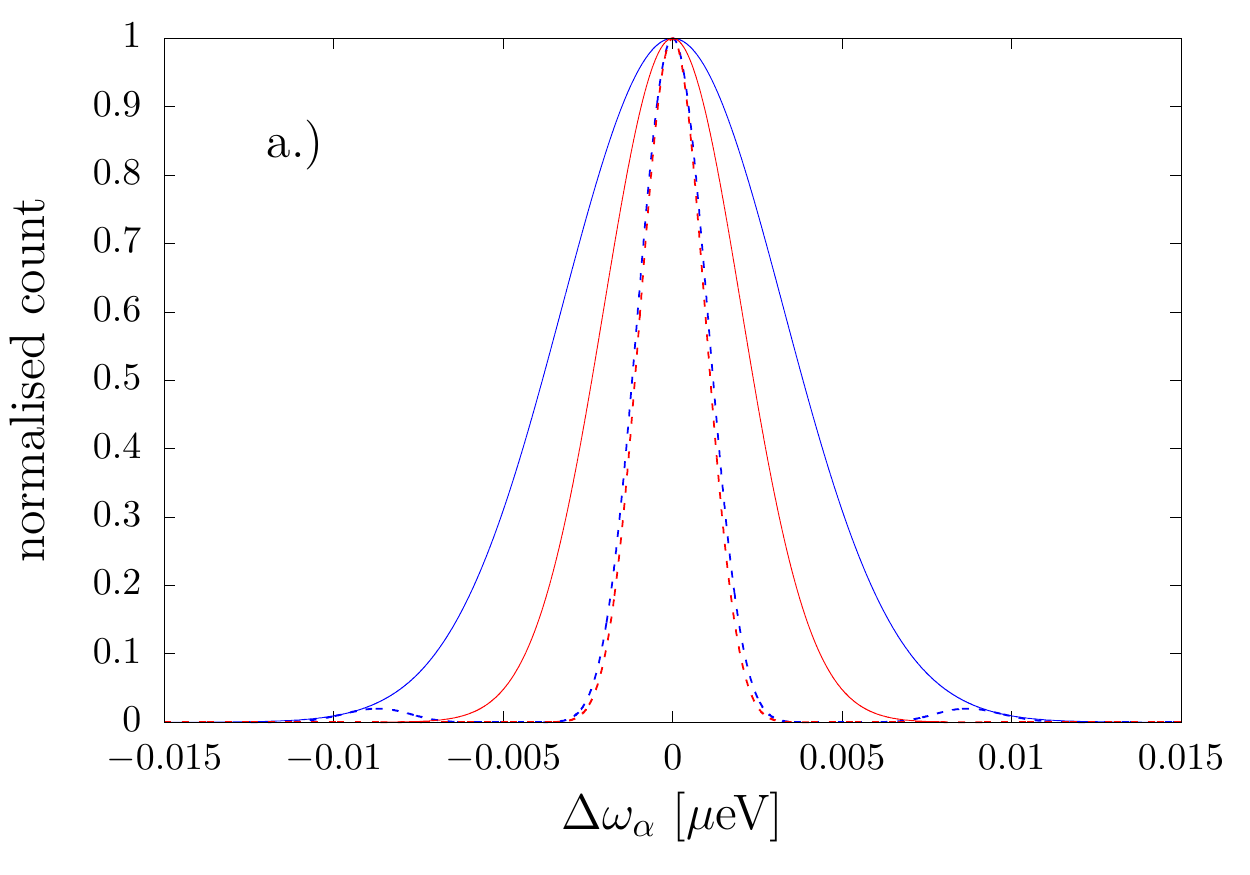}
}\hfill
\subfloat{\label{fig:dist-post-long}
		\includegraphics[width=\linewidth]{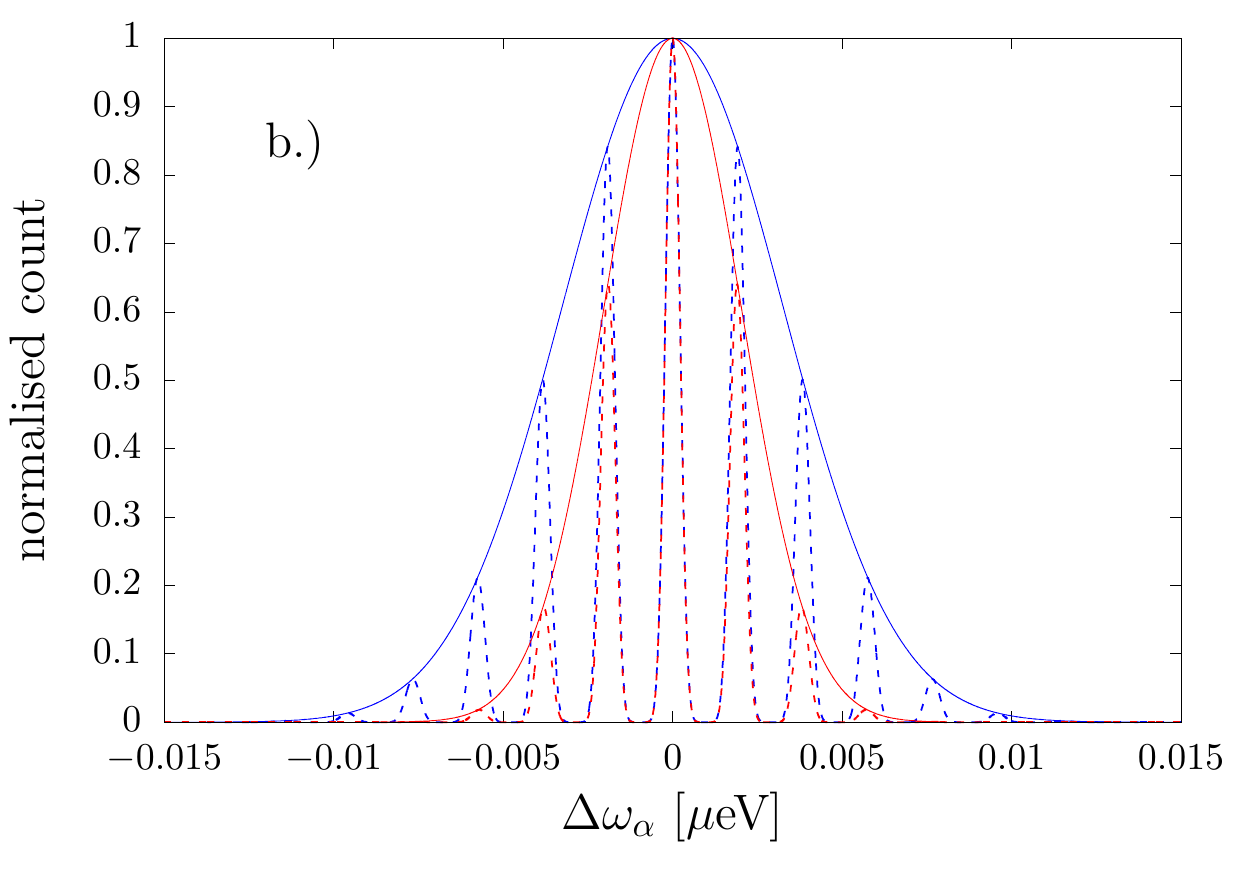}
}
	\caption{Coarse graining distributions of the effective frequency $\Delta\omega_\alpha$ with post selection Eq.~\eqref{eq:post-selected-distribution} for our choice of $\tau \! =\! 2.7$ $\mu$s. The similar distribution for longer $\tau \! = \! 13.5$ $\mu$s is exemplified in the lower panel. The number of measurements is $n=4$.}\label{fig:dist-post} % caption for whole figure
\end{figure}

Let us introduce now the notion of coarse-grained post-selected probability distribution $p_{M_j}(\Delta \omega)$, corresponding to the environment state obtained after a sequence $M_k$ of qubit measurements results was obtained. 
As discussed above, $M_0$ and $M_{2^n-1}$ sequences of identical results are the most probable when $\tau$ is longer than $T_{2}^{*}$, or equivalently than $1/\sigma$ of the initial distribution of $\Delta \omega$. We focus then on one of them, specifically on $M_0$, and on initially completely mixed state of environment, for which the coarse-grained distribution of $\Delta \omega$ is well approximated by a Gaussian with standard deviation $\sigma$. Using Eq.~(\ref{eq:pMj}) with $j\!=\! 0$ we obtain for the post-selected distribution:
	\begin{equation}\begin{split}
		n\cv{\Delta\omega} p_{M_0}\cv{\Delta\omega} &\sim \frac{e^{-\frac{\cv{\Delta\omega}^2}{2\sigma^2}}}{\sqrt{2\pi\sigma^2}}\cos^{2n}\cv{\frac{\tau\Delta\omega}{2}}.\label{eq:post-selected-distribution}%\\
%			&= \frac{1}{2^{2n}\sqrt{2\pi\sigma^2}}\sum_{k=0}^{2n}\cv{\begin{array}{c}2n\\ k \end{array}} e^{-\frac{\cv{\Delta\omega}^2}{2\sigma^2}} \cos\cvb{(n-k)\tau\Delta\omega}
		\end{split}
	\end{equation}
The $\cos^{2n}\cv{\frac{\tau\Delta\omega}{2}}$ represents a frequency comb with teeth of width $\approx \! 1/\sqrt{n}\tau$ (for large $n$) and spacing $\propto 1/\tau$. 
This is illustrated in Fig.~\ref{fig:dist-post}. In panel (a) the evolution time $\tau$ is comparable to $1/\sigma$ of the initial distribution $p(\Delta \omega)$ , and the distribution $p_{M_0}(\Delta \omega)$ obtained after getting $n\! =\! 4$ results of $\ket{+x}$ is characterized by much smaller rms of the central peak. For longer $\tau$, in panel (b) we see the appearance of multiple peaks in $p_{M_0}(\Delta \omega)$. The coherence decay obtained for qubit interating with such post-selected environments will be discussed in Sec.~\ref{sec:dephasing}.

%%%%%%%%%%%%%%%%%%%%%%%%%%%
%%% POST-SELECTED DEPHASING
%%%%%%%%%%%%%%%%%%%%%%%%%%%
\section{Dephasing after environmental state post-selection}  \label{sec:dephasing}
In Section \ref{sec:postselected_state} we have shown that the state of the environment, post-selected after obtaining one of the most probable sequence of measurements, is described by a distribution of $\Delta \omega$ frequencies that is characterized by diminished standard deviation - in other words the post-selected state is ``narrowed'' \cite{Coish_PRB04,Klauser_PRB06}. According to expression (\ref{eq:Wpost}), dephasing of the qubit affected by such a post-selected environment should be modified compared to dephasing of the qubit interacting with the environment described by $\hat{\rho}_{0}^{E}$ state. 

%\subsection{Results: exact and within Gaussian approximation}
Let us demonstrate this behaviour by inspecting the quantity $\langle\hat{\sigma}_x\cv{t}\rangle \! = \! \mathrm{Re}W(t)$. The imaginary part of $W(\tau)$ is zero for an initial distribution of $\Delta \omega$ being an even function of $\Delta \omega$, which is the case for high temperature of the environment. Using the Gaussian approximation for distribution of $\Delta \omega$, we obtain for $\langle\hat{\sigma}_x\cv{t}\rangle$ after $n$ measurements, all giving $\ket{+x}$ result, 
	\begin{equation}
		\langle\hat{\sigma}_x\cv{t}\rangle_n \approx \dfrac{\displaystyle\Re\cvb{\int_{-\infty}^\infty \cos^{2n}\cv{\frac{\tau\Delta\omega}{2}}e^{-\frac{\cv{\Delta\omega}^2}{2\sigma^2}}e^{i\Delta\omega t}d\Delta\omega}}{\displaystyle\int_{-\infty}^\infty \cos^{2n}\cv{\frac{\tau\Delta\omega}{2}}e^{-\frac{\cv{\Delta\omega}^2}{2\sigma^2}} d\Delta\omega}.
	\end{equation}
A straightforward calculation gives then
	\begin{equation}
			\langle\hat{\sigma}_x\cv{t}\rangle_n = \dfrac{\displaystyle\sum_{r=0}^{2n} \cv{\begin{array}{c}2n\\r\end{array}}e^{-\frac{1}{2}\sigma^2\cvb{t+ \cv{n-r}\tau}^2}}{\displaystyle\sum_{r=0}^{2n} \cv{\begin{array}{c}2n\\r\end{array}}e^{-\frac{1}{2}\sigma^2\cv{n-r}^2\tau^2}}.\label{eq:decoherence-Gaussian}
	\end{equation}
For $n=0$ this is a free induction decay that one can expected from the Gaussian bath, while in the finite $n$ case the decoherence function is a sum of Gaussian profiles with the same width $1/\sigma$ (controlled by the initial state of $E$) but different means (e.g. different values of $-\cv{n-r}\tau$) determined by the number of measurements $n$ and the duration $\tau$ of pre-measurement evolutions.

% FIGURE: n=0 vs n=4
\begin{figure}[!]
\subfloat{\label{fig:decoherence-no-field-0and4}
  \includegraphics[width=0.9\linewidth]{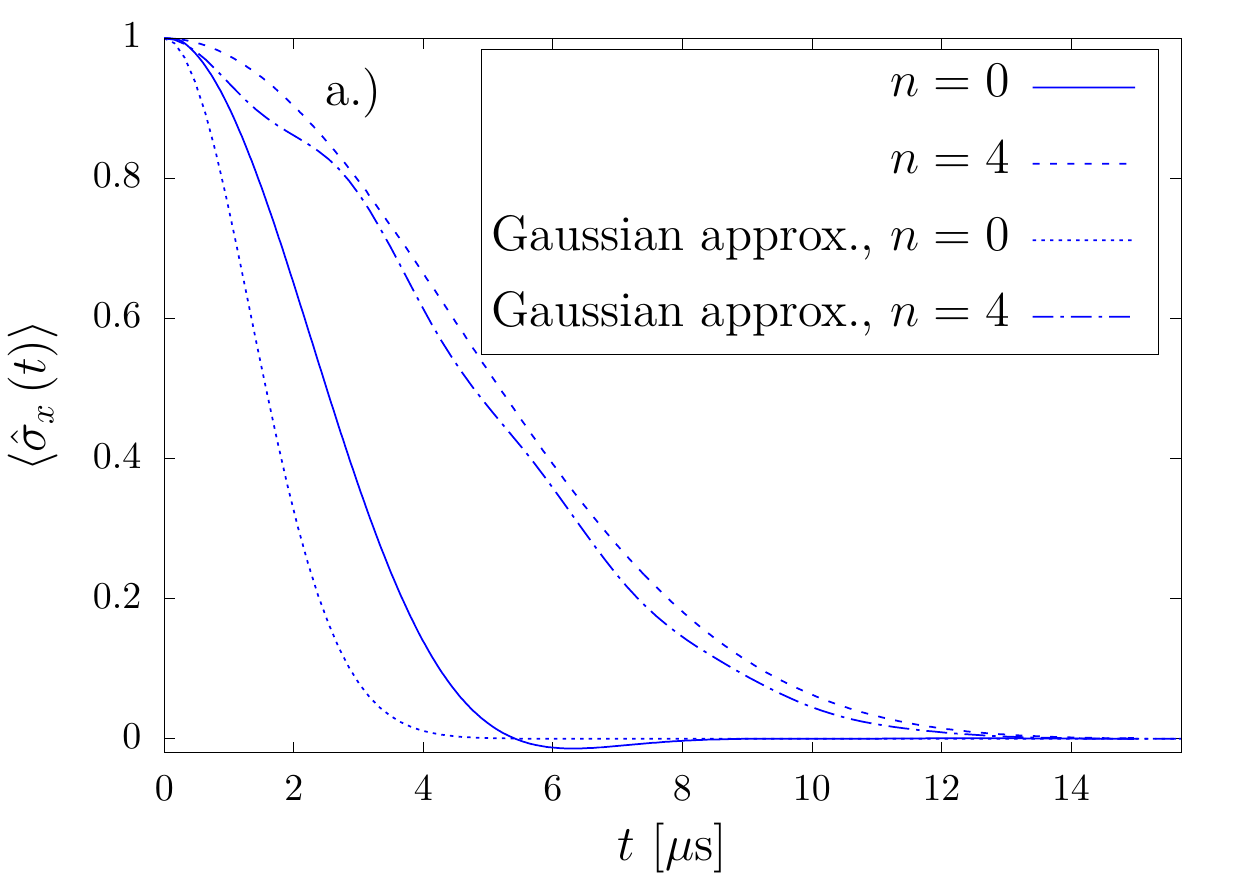}%
}\hfill
\subfloat{\label{fig:decoherence-strong-0and4}
  \includegraphics[width=0.9\linewidth]{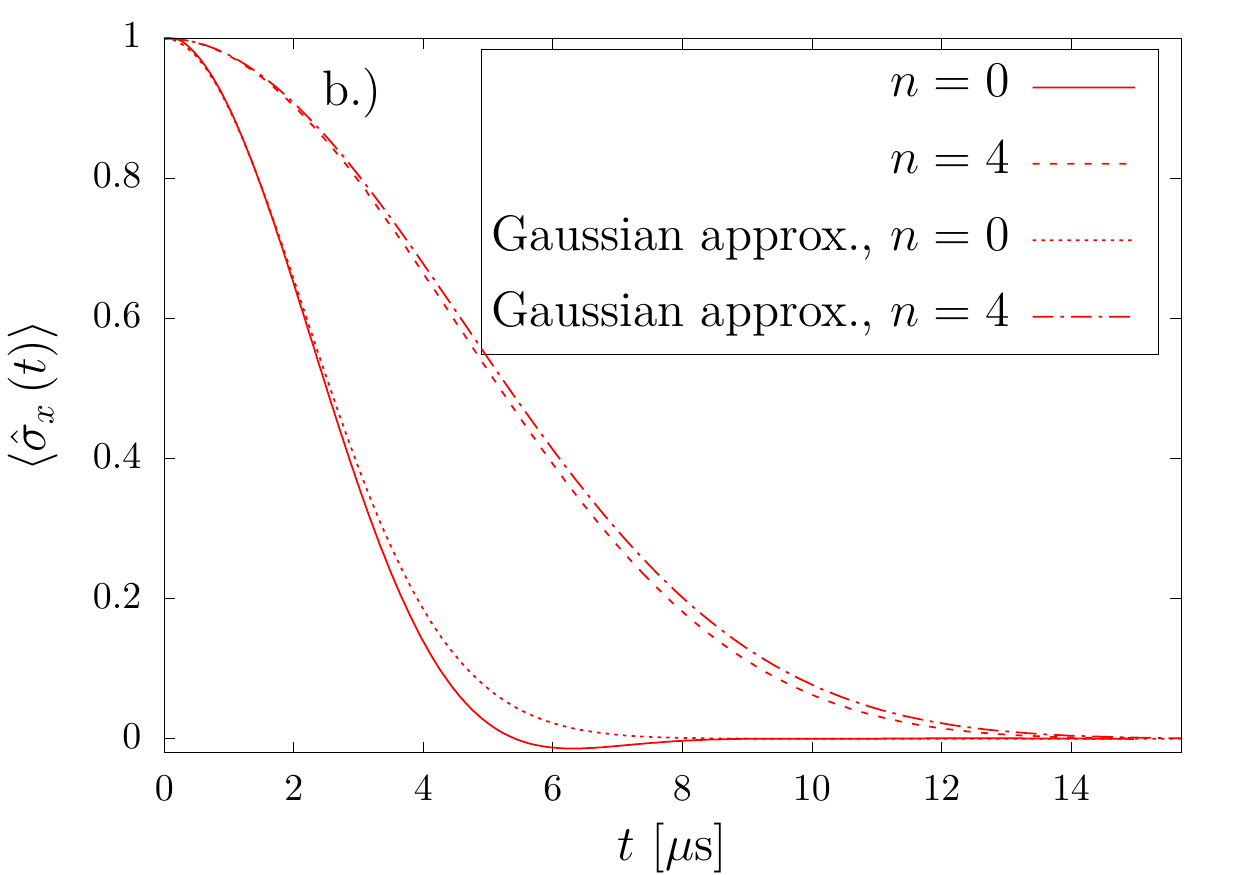}%
}
\caption{Signals correspond to the expectation value $\langle\hat{\sigma}_x\cv{t}\rangle$ with for the initial and the post-selected state of the environment from exact calculation and Gaussian approximation calculation at (a) zero field and (b) high field ($B\! =\! 100$ mT). } \label{fig:decoherence-0and4}
\end{figure}

% FIGURE: other post-selections:
\begin{figure}
	\begin{center}
		\includegraphics[width=\linewidth]{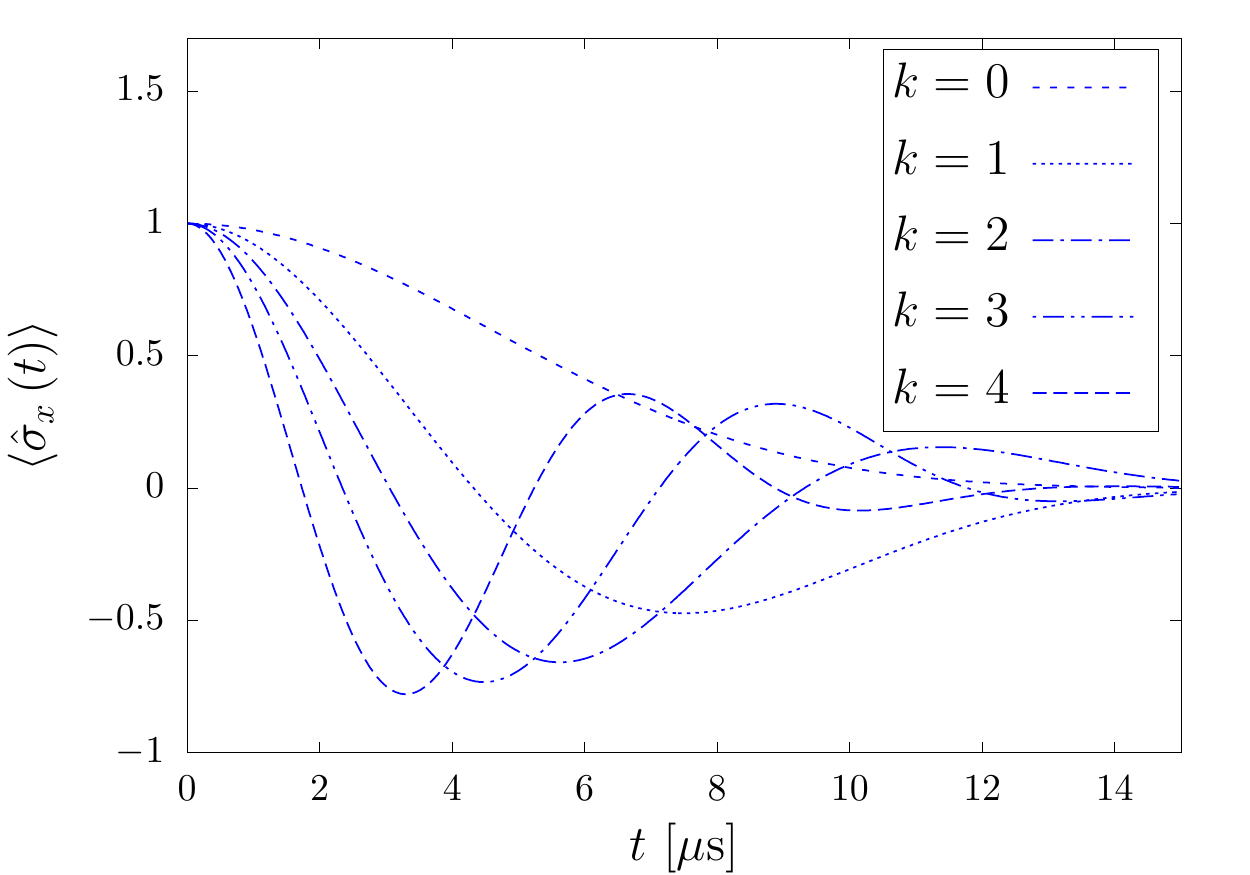}
		\caption{Signals in Eq. \eqref{eq:decoherence-Gaussian-Mj}, expected from all posible sequences $M_j,$ grouped according to their pathlength $k\cv{j}=k$ with $n=4,$ $\tau=2.7~\mu$s in the zero field case. }\label{fig:decoherence-zero-all-post}
	\end{center}
\end{figure}

The numerical calculations directly from the distribution over $\left\{\alpha\right\}$ and the corresponding results of Gaussian approximation calculation are shown in Fig. \ref{fig:decoherence-0and4}. One can see that both calculations are in good quantitative agreement already for $N\! =\! 20$..

For other sequences $M_j$ defining the post-selected environmental state, the decoherence function can be written in the from of combination over $M_0$ of shorter sequences by the same technique as in Eq.~\eqref{eq:pMj_inpM0}, giving
	\begin{equation}
		\begin{split}
			\langle&\hat{\sigma}_x\cv{t}\rangle_{M_j}= \dfrac{\displaystyle\sum_{r=0}^{k\cv{j}}\cv{-1}^{k\cv{j}+r}w_{n-r,r,k\cv{j}}\cv{t}}{\displaystyle\sum_{r=0}^{k\cv{j}}\cv{-1}^{k\cv{j}+r}w_{n-r,r,k\cv{j}}\cv{0}},\label{eq:decoherence-Gaussian-Mj}
		\end{split}
	\end{equation}
	where $w_{n,r,k}\cv{t} = \displaystyle\sum_{l=0}^{2n} \cv{\begin{array}{c}k\\r\end{array}}\cv{\begin{array}{c}2n\\l\end{array}}e^{-\frac{1}{2}\sigma^2\cvb{t+\cv{n-l}\tau}^2}.$ 
The appearance of alternating sum of different Gaussian functions  introduces an oscillation in $	\langle\hat{\sigma}_x\cv{t}\rangle_n$ signal. With increasing $k$, the value of $	\langle\hat{\sigma}_x\cv{t}\rangle_n$ at time $t\! =\! \tau$ should become progressively closer to $-1$, as we are post-selecting the environmental states that cause the qubit's initial $\ket{+x}$ to rotate towards $\ket{-x}$ state at this time delay. This is seen in Fig.~\ref{fig:decoherence-zero-all-post}. Clearly, if our aim is to protect for protect for an enhanced time the initial $\ket{+x}$ state of the qubit, we should focus on post-selection following observation of $M_{0}$ sequence. 

Let us also note that if we sum all the coherence signals from Fig.~\ref{fig:decoherence-zero-all-post} while weighing each by the probability of obtaining the given sequence of $n$ results, we will obtain the signal exactly the same as the $n\! =\! 0$ one from Fig.~\ref{fig:decoherence-0and4}. This follows from the fact that for the quasi-static environment approximation defined in Sec.~\ref{sec:QSE} and used to obtain all the results in this paper, performing $n$ of measurements and {\it not} post-selecting based on the obtained sequence of results, leaves the state of the environment unchanged, as follow from Eq.~\ref{eq:invariant}.

% FIGURE: color map
\begin{figure}[tb]
\subfloat{\label{fig:decoherence-no-field-n}
  \includegraphics[width=0.9\linewidth]{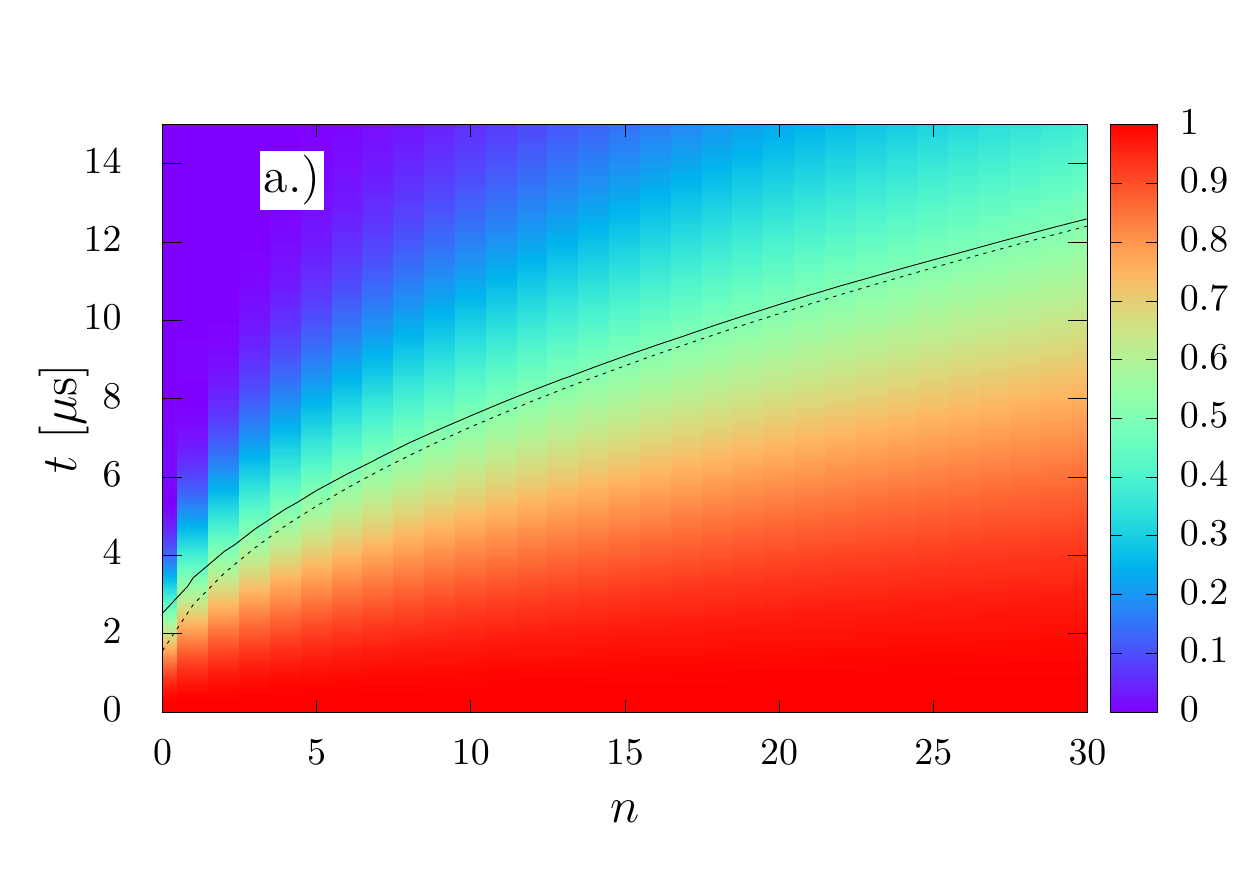}%
}\hfill
\subfloat{\label{fig:Gauss-no-field_30}
  \includegraphics[width=0.9\linewidth]{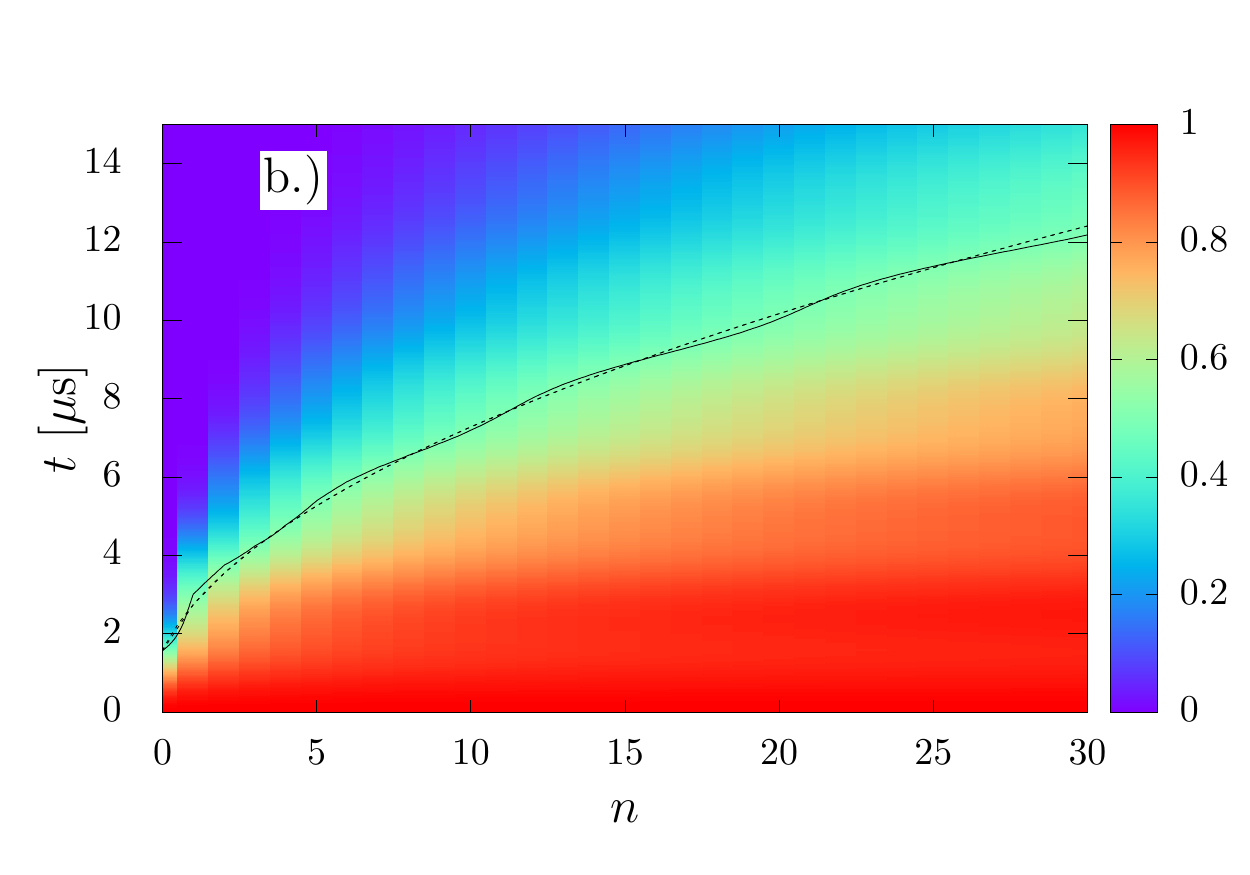}%
}\hfill
\caption{Signals correspond to the expectation value $\langle\hat{\sigma}_x\cv{t}\rangle$ in exact calculation (top panel) and calculation within Gaussian approximation from Eq.~\eqref{eq:decoherence-Gaussian} (bottom banel)  of $\langle\hat{\sigma}_x\cv{t}\rangle$ with $N=20$ and $\tau=2.7\mu$s, versus the number of  measurements $n$ for zero field case. The analogous figures for the high field case are very similar. The solid lines are  the exact iso-coherence lines corresponding to $W \! = 0.5$, while the dashed lines are obtained from Eq.~\eqref{eq:T_half}.}\label{fig:W_vs_n}
\end{figure}

Finally, let us focus on large-$n$ asymptotics, and discuss it within the Gaussian approximation.
For $n \! \gg \! 1$ we can replace summation over $r$ in Eq.~(\ref{eq:decoherence-Gaussian}) by integration, and approximate the binomial coefficient by a Gaussian, arriving at
\begin{align}
			\langle\hat{\sigma}_x\cv{t}\rangle_{\text{large }n}\cv{t} &\approx \dfrac{\displaystyle\int_{-\infty}^{\infty} e^{-\frac{\cv{x-n}^2}{n}} e^{-\frac{1}{2}\sigma^2\cvb{t+\cv{n-x}\tau}^2} dx}{\displaystyle\int_{-\infty}^{\infty}e^{-\frac{\cv{x-n}^2}{n}} e^{-\frac{1}{2}\sigma^2\cv{n-x}^2\tau^2} dx}\nonumber\\
			 &= \exp\cvb{-\frac{1}{2}f^2\cv{n,\tau}\sigma^2t^2}
		\end{align}
%	where $\sigma'^2 = \dfrac{2}{n\tau^2} + \sigma^2 = \dfrac{2+n\sigma^2\tau^2}{n\tau^2}$ and \[f^2\cv{n,\tau} = 1-\cv{\frac{\sigma}{\sigma'}}^2 = \dfrac{2}{2+n\sigma^2\tau^2}.\]
where $f^2\cv{n,\tau} \! =  \!2/(2+n\sigma^2\tau^2)$ appeared previously in expression (\ref{eq:pM0large}) for $\prob_{\text{large }n}\cv{M_0}$. 
%For $\tau=0$ we have $f^2(n,\tau) \! =\! 1$ and $W_{n}(t) \! =\! W_{0}(t)$, as the $\ket{+x}$ result of each measurement is guaranteed, and from the sequence of such measurements immediately following the qubit initializations we learn nothing about the state of E. 
With increasing $n$ and for finite $\tau$ the factor $f^2(n,\tau)$ decreases towards $0$ as $1/n$. Hence for both large $n$ and $N$ the decoherence function will be approximately constant (and very close to $1$) up to $t \propto \sqrt{n}$, and only for longer times $t$ it will exhibit Gaussian decay. 
The half-decay coherence time $T_{1/2}$, at which $W_{n}(1/2) \! =\! 1/2$, can be estimated as
		\begin{equation}
			T_{1/2} \approx\sqrt{\dfrac{\ln 2\cv{2+n\sigma^2\tau^2}}{\sigma^2}} \,\, ,	\label{eq:T_half}
		\end{equation}	
		so that $T_{1/2} \! \propto \sqrt{n}$ when $\sigma\tau \! \gg \! 1$ or when $\sqrt{n} \sigma\tau\! \gg \! 1$. The large degree of agreement between this prediction and the exact results for  $N\! =\! 20$ nuclei is shown in Fig.~\ref{fig:W_vs_n}.

%%%%%%%%%%%%%%%%%%%%%%%%%%%%%%%%%%%%%%%%%%%%%%%%%%%%%%%%%
\section{Discussion and Conclusion}
We have presented a simple theoretical approach to calculation of state of a quasi-static environment obtained after getting a particular sequence of results of projective measurements on a qubit that is precessing under the influence of this environment. We have pointed out that for such a quasi-static environment (the definition of which we have carefully discussed), the statistics of measurements is essentially classical: the initial state of the environment determines the probability density of having a qubit behave in a certain way (precess with a certain angular frequency), and a sequence of measurements on a qubit repeatedly re-initialized in the same state progressively diminishes our ignorance of which exact environmental state we are dealing with. For long qubit evolution times (longer than the inverse bandwidth of the initial environmental state), the most probable sequences are the ones in which the same result appears repeatedly. 
For $n$ measurements, they appear with probability $\propto 1/\sqrt{n}$. After recording such a sequence, the environment is in a ``narrowed'' state of diminished uncertainty in the field exerted on the qubit that leads to a slowed-down decay of fidelity of qubit state if the qubit is initialized again in the same initial state as the one used for the $n$-measurement sequence. The enhancement of fidelity decay time is by a factor $\propto \sqrt{n}$ compared to decay observed without any post-selection. If our goal is to make a particular superposition state of the qubit more resistant to environmental influence, such a simple post-selection procedure is an optimal strategy if we do not want to use feedback schemes \cite{Shulman_NC14,Cappellaro_PRA12} that are much harder to implement than simply repeating the same sequence of qubit intialization-evolution-measurement a few times. 
While the $\sqrt{n}$ factors in probability of obtaining one of the ``extremal'' sequences of measurement results and in enhancement of coherence time balance each other out, the possibility of obtaining a visible increase in free evolution coherence time with little overhead cost could be useful in some cases, e.g.~when a qubit state with enhanced fidelity in a finite time-window is needed for a specific purpose, and we do not have to fulfill strict criteria of having this qubit on-demand, i.e.~if we can tolerate a non-deterministic time delay associated with proper ``narrowing'' of qubit's environment.  

For all the relevant quantities (probabilities of various sequences, post-measurements states of environment), we have given both general formulas, and approximate ones, valid when the distribution of qubit energy shifts due to environment can be approximated by a Gaussian one. We have discussed to what degree such a Gaussian approximation applies to an NV center spin qubit interacting with a nuclear spin bath. For such a qubit we have presented exact numerical results, showing very good agreement with recent experiments \cite{RaoEAPreprint2018}, strongly suggesting that, according to expectations, that experiment is performed in the regime in which the environment is quasi-static. Let us remark that the statement from Ref.~\cite{RaoEAPreprint2018} that for $\tau \! \gg \! T_{2}^{*}$ the probabilities of measuring $\ket{+x}$ and $\ket{-x}$ states of the qubit are equal, is true only when considering {\it ensemble-averaged} probabilities, not when applied to a single sequence of measurements. In any given sequence of $n$ measurements performed on timescale on which the environment does not evolve due to its intrinsic dynamics, for $\tau \! \gg \! T_{2}^{*}$ the probabilities of obtaining $\ket{+x}$ and $\ket{-x}$ are typically visibly distinct, and this result can be obtained from a classical reasoning concerning the influence of the quasi-static environment on the qubit.

%all the 16 measurement results in a sequence of $n\!=\! 4$ measurements should occur with the same probability

\section*{Acknowledgements}
We thank Jan Krzywda, Damian Kwiatkowski, and Piotr Sza\'nkowski for discussions. 
This work is supported by funds of Polish National Science Center (NCN), Grant no.~2015/19/B/ST3/03152.

\bibliography{measurements_correlations_QSBA_revision.bbl}

\begin{thebibliography}{47}%
\makeatletter
\providecommand \@ifxundefined [1]{%
 \@ifx{#1\undefined}
}%
\providecommand \@ifnum [1]{%
 \ifnum #1\expandafter \@firstoftwo
 \else \expandafter \@secondoftwo
 \fi
}%
\providecommand \@ifx [1]{%
 \ifx #1\expandafter \@firstoftwo
 \else \expandafter \@secondoftwo
 \fi
}%
\providecommand \natexlab [1]{#1}%
\providecommand \enquote  [1]{``#1''}%
\providecommand \bibnamefont  [1]{#1}%
\providecommand \bibfnamefont [1]{#1}%
\providecommand \citenamefont [1]{#1}%
\providecommand \href@noop [0]{\@secondoftwo}%
\providecommand \href [0]{\begingroup \@sanitize@url \@href}%
\providecommand \@href[1]{\@@startlink{#1}\@@href}%
\providecommand \@@href[1]{\endgroup#1\@@endlink}%
\providecommand \@sanitize@url [0]{\catcode `\\12\catcode `\$12\catcode
  `\&12\catcode `\#12\catcode `\^12\catcode `\_12\catcode `\%12\relax}%
\providecommand \@@startlink[1]{}%
\providecommand \@@endlink[0]{}%
\providecommand \url  [0]{\begingroup\@sanitize@url \@url }%
\providecommand \@url [1]{\endgroup\@href {#1}{\urlprefix }}%
\providecommand \urlprefix  [0]{URL }%
\providecommand \Eprint [0]{\href }%
\providecommand \doibase [0]{http://dx.doi.org/}%
\providecommand \selectlanguage [0]{\@gobble}%
\providecommand \bibinfo  [0]{\@secondoftwo}%
\providecommand \bibfield  [0]{\@secondoftwo}%
\providecommand \translation [1]{[#1]}%
\providecommand \BibitemOpen [0]{}%
\providecommand \bibitemStop [0]{}%
\providecommand \bibitemNoStop [0]{.\EOS\space}%
\providecommand \EOS [0]{\spacefactor3000\relax}%
\providecommand \BibitemShut  [1]{\csname bibitem#1\endcsname}%
\let\auto@bib@innerbib\@empty
%</preamble>
\bibitem [{\citenamefont {{\.Z}urek}(2003)}]{Zurek_RMP03}%
  \BibitemOpen
  \bibfield  {author} {\bibinfo {author} {\bibfnamefont {Wojciech~Hubert}\
  \bibnamefont {{\.Z}urek}},\ }\bibfield  {title} {\enquote {\bibinfo {title}
  {Decoherence, einselection, and the quantum origins of the classical},}\
  }\href {\doibase 10.1103/RevModPhys.75.715} {\bibfield  {journal} {\bibinfo
  {journal} {Rev.\ Mod.\ Phys.}\ }\textbf {\bibinfo {volume} {75}},\ \bibinfo
  {pages} {715} (\bibinfo {year} {2003})}\BibitemShut {NoStop}%
\bibitem [{\citenamefont {Schlosshauer}(2007)}]{Schlosshauer_book}%
  \BibitemOpen
  \bibfield  {author} {\bibinfo {author} {\bibfnamefont {M.}~\bibnamefont
  {Schlosshauer}},\ }\href@noop {} {\emph {\bibinfo {title} {Decoherence and
  the Quantum-to-Classical Transition}}}\ (\bibinfo  {publisher} {Springer},\
  \bibinfo {address} {Berlin/Heidelberg},\ \bibinfo {year} {2007})\BibitemShut
  {NoStop}%
\bibitem [{\citenamefont {Wiseman}\ and\ \citenamefont
  {Milburn}(2010)}]{Wiseman}%
  \BibitemOpen
  \bibfield  {author} {\bibinfo {author} {\bibfnamefont {H.~M.}\ \bibnamefont
  {Wiseman}}\ and\ \bibinfo {author} {\bibfnamefont {G.~J.}\ \bibnamefont
  {Milburn}},\ }\href@noop {} {\emph {\bibinfo {title} {Quantum Measurement and
  Control}}}\ (\bibinfo  {publisher} {Cambridge University Press},\ \bibinfo
  {address} {Cambridge, England},\ \bibinfo {year} {2010})\BibitemShut
  {NoStop}%
\bibitem [{\citenamefont {Pfender}\ \emph {et~al.}(2019)\citenamefont
  {Pfender}, \citenamefont {Wang}, \citenamefont {Sumiya}, \citenamefont
  {Onoda}, \citenamefont {Yang}, \citenamefont {Dasari}, \citenamefont
  {Neumann}, \citenamefont {Pan}, \citenamefont {Isoya}, \citenamefont {Liu},\
  and\ \citenamefont {Wrachtrup}}]{Pfender_NC19}%
  \BibitemOpen
  \bibfield  {author} {\bibinfo {author} {\bibfnamefont {Matthias}\
  \bibnamefont {Pfender}}, \bibinfo {author} {\bibfnamefont {Ping}\
  \bibnamefont {Wang}}, \bibinfo {author} {\bibfnamefont {Hitoshi}\
  \bibnamefont {Sumiya}}, \bibinfo {author} {\bibfnamefont {Shinobu}\
  \bibnamefont {Onoda}}, \bibinfo {author} {\bibfnamefont {Wen}\ \bibnamefont
  {Yang}}, \bibinfo {author} {\bibfnamefont {Durga Bhaktavatsala~Rao}\
  \bibnamefont {Dasari}}, \bibinfo {author} {\bibfnamefont {Philipp}\
  \bibnamefont {Neumann}}, \bibinfo {author} {\bibfnamefont {Xin-Yu}\
  \bibnamefont {Pan}}, \bibinfo {author} {\bibfnamefont {Junichi}\ \bibnamefont
  {Isoya}}, \bibinfo {author} {\bibfnamefont {Ren-Bao}\ \bibnamefont {Liu}}, \
  and\ \bibinfo {author} {\bibfnamefont {J\"org}\ \bibnamefont {Wrachtrup}},\
  }\bibfield  {title} {\enquote {\bibinfo {title} {High-resolution spectroscopy
  of single nuclear spins via sequential weak measurements},}\ }\href {\doibase
  10.1038/s41467-019-08544-z} {\bibfield  {journal} {\bibinfo  {journal}
  {Nature Communications}\ }\textbf {\bibinfo {volume} {10}},\ \bibinfo {pages}
  {594} (\bibinfo {year} {2019})}\BibitemShut {NoStop}%
\bibitem [{\citenamefont {Ma}\ \emph {et~al.}(2018)\citenamefont {Ma},
  \citenamefont {Wang}, \citenamefont {Leong},\ and\ \citenamefont
  {Liu}}]{Ma_PRA18}%
  \BibitemOpen
  \bibfield  {author} {\bibinfo {author} {\bibfnamefont {Wen-Long}\
  \bibnamefont {Ma}}, \bibinfo {author} {\bibfnamefont {Ping}\ \bibnamefont
  {Wang}}, \bibinfo {author} {\bibfnamefont {Weng-Hang}\ \bibnamefont {Leong}},
  \ and\ \bibinfo {author} {\bibfnamefont {Ren-Bao}\ \bibnamefont {Liu}},\
  }\bibfield  {title} {\enquote {\bibinfo {title} {Phase transitions in
  sequential weak measurements},}\ }\href {\doibase 10.1103/PhysRevA.98.012117}
  {\bibfield  {journal} {\bibinfo  {journal} {Phys. Rev. A}\ }\textbf {\bibinfo
  {volume} {98}},\ \bibinfo {pages} {012117} (\bibinfo {year}
  {2018})}\BibitemShut {NoStop}%
\bibitem [{\citenamefont {Blok}\ \emph {et~al.}(2014)\citenamefont {Blok},
  \citenamefont {Bonato}, \citenamefont {Markham}, \citenamefont {Twitchen},
  \citenamefont {Dobrovitski},\ and\ \citenamefont {Hanson}}]{Blok_NP14}%
  \BibitemOpen
  \bibfield  {author} {\bibinfo {author} {\bibfnamefont {M.~S.}\ \bibnamefont
  {Blok}}, \bibinfo {author} {\bibfnamefont {C.}~\bibnamefont {Bonato}},
  \bibinfo {author} {\bibfnamefont {M.~L.}\ \bibnamefont {Markham}}, \bibinfo
  {author} {\bibfnamefont {D.~J.}\ \bibnamefont {Twitchen}}, \bibinfo {author}
  {\bibfnamefont {V.~V.}\ \bibnamefont {Dobrovitski}}, \ and\ \bibinfo {author}
  {\bibfnamefont {R.}~\bibnamefont {Hanson}},\ }\bibfield  {title} {\enquote
  {\bibinfo {title} {Manipulating a qubit through the backaction of sequential
  partial measurements and real-time feedback},}\ }\href {\doibase
  10.1038/nphys2881} {\bibfield  {journal} {\bibinfo  {journal} {Nat. Phys.}\
  }\textbf {\bibinfo {volume} {10}},\ \bibinfo {pages} {189} (\bibinfo {year}
  {2014})}\BibitemShut {NoStop}%
\bibitem [{\citenamefont {Muhonen}\ \emph {et~al.}(2018)\citenamefont
  {Muhonen}, \citenamefont {Dehollain}, \citenamefont {Laucht}, \citenamefont
  {Simmons}, \citenamefont {Kalra}, \citenamefont {Hudson}, \citenamefont
  {Dzurak}, \citenamefont {Morello}, \citenamefont {Jamieson}, \citenamefont
  {McCallum},\ and\ \citenamefont {Itoh}}]{Muhonen_PRB18}%
  \BibitemOpen
  \bibfield  {author} {\bibinfo {author} {\bibfnamefont {J.~T.}\ \bibnamefont
  {Muhonen}}, \bibinfo {author} {\bibfnamefont {J.~P.}\ \bibnamefont
  {Dehollain}}, \bibinfo {author} {\bibfnamefont {A.}~\bibnamefont {Laucht}},
  \bibinfo {author} {\bibfnamefont {S.}~\bibnamefont {Simmons}}, \bibinfo
  {author} {\bibfnamefont {R.}~\bibnamefont {Kalra}}, \bibinfo {author}
  {\bibfnamefont {F.~E.}\ \bibnamefont {Hudson}}, \bibinfo {author}
  {\bibfnamefont {A.~S.}\ \bibnamefont {Dzurak}}, \bibinfo {author}
  {\bibfnamefont {A.}~\bibnamefont {Morello}}, \bibinfo {author} {\bibfnamefont
  {D.~N.}\ \bibnamefont {Jamieson}}, \bibinfo {author} {\bibfnamefont {J.~C.}\
  \bibnamefont {McCallum}}, \ and\ \bibinfo {author} {\bibfnamefont {K.~M.}\
  \bibnamefont {Itoh}},\ }\bibfield  {title} {\enquote {\bibinfo {title}
  {Coherent control via weak measurements in $^{31}\mathrm{P}$ single-atom
  electron and nuclear spin qubits},}\ }\href {\doibase
  10.1103/PhysRevB.98.155201} {\bibfield  {journal} {\bibinfo  {journal} {Phys.
  Rev. B}\ }\textbf {\bibinfo {volume} {98}},\ \bibinfo {pages} {155201}
  (\bibinfo {year} {2018})}\BibitemShut {NoStop}%
\bibitem [{\citenamefont {Klauser}\ \emph {et~al.}(2006)\citenamefont
  {Klauser}, \citenamefont {Coish},\ and\ \citenamefont
  {Loss}}]{Klauser_PRB06}%
  \BibitemOpen
  \bibfield  {author} {\bibinfo {author} {\bibfnamefont {D.}~\bibnamefont
  {Klauser}}, \bibinfo {author} {\bibfnamefont {W.~A.}\ \bibnamefont {Coish}},
  \ and\ \bibinfo {author} {\bibfnamefont {Daniel}\ \bibnamefont {Loss}},\
  }\bibfield  {title} {\enquote {\bibinfo {title} {Nuclear spin state narrowing
  via gate-controlled rabi oscillations in a double quantum dot},}\ }\href
  {\doibase 10.1103/PhysRevB.73.205302} {\bibfield  {journal} {\bibinfo
  {journal} {Phys.\ Rev.\ B}\ }\textbf {\bibinfo {volume} {73}},\ \bibinfo
  {pages} {205302} (\bibinfo {year} {2006})}\BibitemShut {NoStop}%
\bibitem [{\citenamefont {Stepanenko}\ \emph {et~al.}(2006)\citenamefont
  {Stepanenko}, \citenamefont {Burkard}, \citenamefont {Giedke},\ and\
  \citenamefont {Imamo\u{g}lu}}]{Stepanenko_PRL06}%
  \BibitemOpen
  \bibfield  {author} {\bibinfo {author} {\bibfnamefont {Dimitrije}\
  \bibnamefont {Stepanenko}}, \bibinfo {author} {\bibfnamefont {Guido}\
  \bibnamefont {Burkard}}, \bibinfo {author} {\bibfnamefont {Geza}\
  \bibnamefont {Giedke}}, \ and\ \bibinfo {author} {\bibfnamefont {Atac}\
  \bibnamefont {Imamo\u{g}lu}},\ }\bibfield  {title} {\enquote {\bibinfo
  {title} {Enhancement of electron spin coherence by optical preparation of
  nuclear spins},}\ }\href {\doibase 10.1103/PhysRevLett.96.136401} {\bibfield
  {journal} {\bibinfo  {journal} {Phys.\ Rev.\ Lett.}\ }\textbf {\bibinfo
  {volume} {96}},\ \bibinfo {pages} {136401} (\bibinfo {year}
  {2006})}\BibitemShut {NoStop}%
\bibitem [{\citenamefont {Giedke}\ \emph {et~al.}(2006)\citenamefont {Giedke},
  \citenamefont {Taylor}, \citenamefont {D'Alessandro}, \citenamefont {Lukin},\
  and\ \citenamefont {Imamo\u{g}lu}}]{Giedke_PRA06}%
  \BibitemOpen
  \bibfield  {author} {\bibinfo {author} {\bibfnamefont {G.}~\bibnamefont
  {Giedke}}, \bibinfo {author} {\bibfnamefont {J.~M.}\ \bibnamefont {Taylor}},
  \bibinfo {author} {\bibfnamefont {D.}~\bibnamefont {D'Alessandro}}, \bibinfo
  {author} {\bibfnamefont {M.~D.}\ \bibnamefont {Lukin}}, \ and\ \bibinfo
  {author} {\bibfnamefont {A.}~\bibnamefont {Imamo\u{g}lu}},\ }\bibfield
  {title} {\enquote {\bibinfo {title} {Quantum measurement of a mesoscopic spin
  ensemble},}\ }\href {\doibase 10.1103/PhysRevA.74.032316} {\bibfield
  {journal} {\bibinfo  {journal} {Phys.\ Rev.\ A}\ }\textbf {\bibinfo {volume}
  {74}},\ \bibinfo {pages} {032316} (\bibinfo {year} {2006})}\BibitemShut
  {NoStop}%
\bibitem [{\citenamefont {Coish}\ and\ \citenamefont
  {Baugh}(2009)}]{Coish_PSSB09}%
  \BibitemOpen
  \bibfield  {author} {\bibinfo {author} {\bibfnamefont {W.~A.}\ \bibnamefont
  {Coish}}\ and\ \bibinfo {author} {\bibfnamefont {J.}~\bibnamefont {Baugh}},\
  }\bibfield  {title} {\enquote {\bibinfo {title} {Nuclear spins in
  nanostructures},}\ }\href {\doibase 10.1002/pssb.200945229} {\bibfield
  {journal} {\bibinfo  {journal} {Phys. Status Solidi B}\ }\textbf {\bibinfo
  {volume} {246}},\ \bibinfo {pages} {2203} (\bibinfo {year}
  {2009})}\BibitemShut {NoStop}%
\bibitem [{\citenamefont {Cywi{\'n}ski}(2011)}]{Cywinski_APPA11}%
  \BibitemOpen
  \bibfield  {author} {\bibinfo {author} {\bibfnamefont {{\L}ukasz}\
  \bibnamefont {Cywi{\'n}ski}},\ }\bibfield  {title} {\enquote {\bibinfo
  {title} {Dephasing of electron spin qubits due to their interaction with
  nuclei in quantum dots},}\ }\href {\doibase 10.12693/APhysPolA.119.576}
  {\bibfield  {journal} {\bibinfo  {journal} {Acta Phys.~Pol.~A}\ }\textbf
  {\bibinfo {volume} {119}},\ \bibinfo {pages} {576} (\bibinfo {year}
  {2011})}\BibitemShut {NoStop}%
\bibitem [{\citenamefont {Chekhovich}\ \emph {et~al.}(2013)\citenamefont
  {Chekhovich}, \citenamefont {Makhonin}, \citenamefont {Tartakovskii},
  \citenamefont {Yacoby}, \citenamefont {Bluhm}, \citenamefont {Nowack},\ and\
  \citenamefont {Vandersypen}}]{Chekhovich_NM13}%
  \BibitemOpen
  \bibfield  {author} {\bibinfo {author} {\bibfnamefont {E.~A.}\ \bibnamefont
  {Chekhovich}}, \bibinfo {author} {\bibfnamefont {M.~N.}\ \bibnamefont
  {Makhonin}}, \bibinfo {author} {\bibfnamefont {A.~I.}\ \bibnamefont
  {Tartakovskii}}, \bibinfo {author} {\bibfnamefont {A.}~\bibnamefont
  {Yacoby}}, \bibinfo {author} {\bibfnamefont {H.}~\bibnamefont {Bluhm}},
  \bibinfo {author} {\bibfnamefont {K.~C.}\ \bibnamefont {Nowack}}, \ and\
  \bibinfo {author} {\bibfnamefont {L.~M.~K.}\ \bibnamefont {Vandersypen}},\
  }\bibfield  {title} {\enquote {\bibinfo {title} {Nuclear spin effects in
  semiconductor quantum dots},}\ }\href {\doibase 10.1038/nmat3652} {\bibfield
  {journal} {\bibinfo  {journal} {Nature Materials}\ }\textbf {\bibinfo
  {volume} {12}},\ \bibinfo {pages} {494} (\bibinfo {year} {2013})}\BibitemShut
  {NoStop}%
\bibitem [{\citenamefont {Yang}\ \emph {et~al.}(2017)\citenamefont {Yang},
  \citenamefont {Ma},\ and\ \citenamefont {Liu}}]{Yang_RPP17}%
  \BibitemOpen
  \bibfield  {author} {\bibinfo {author} {\bibfnamefont {Wen}\ \bibnamefont
  {Yang}}, \bibinfo {author} {\bibfnamefont {Wen-Long}\ \bibnamefont {Ma}}, \
  and\ \bibinfo {author} {\bibfnamefont {Ren-Bao}\ \bibnamefont {Liu}},\
  }\bibfield  {title} {\enquote {\bibinfo {title} {Quantum many-body theory for
  electron spin decoherence in nanoscale nuclear spin baths},}\ }\href
  {\doibase 10.1088/0034-4885/80/1/016001} {\bibfield  {journal} {\bibinfo
  {journal} {Rep. Prog. Phys.}\ }\textbf {\bibinfo {volume} {80}},\ \bibinfo
  {pages} {016001} (\bibinfo {year} {2017})}\BibitemShut {NoStop}%
\bibitem [{\citenamefont {Paladino}\ \emph {et~al.}(2014)\citenamefont
  {Paladino}, \citenamefont {Galperin}, \citenamefont {Falci},\ and\
  \citenamefont {Altshuler}}]{Paladino_RMP14}%
  \BibitemOpen
  \bibfield  {author} {\bibinfo {author} {\bibfnamefont {E.}~\bibnamefont
  {Paladino}}, \bibinfo {author} {\bibfnamefont {Y.~M.}\ \bibnamefont
  {Galperin}}, \bibinfo {author} {\bibfnamefont {G.}~\bibnamefont {Falci}}, \
  and\ \bibinfo {author} {\bibfnamefont {B.~L.}\ \bibnamefont {Altshuler}},\
  }\bibfield  {title} {\enquote {\bibinfo {title} {$1/f$ noise: Implications
  for solid-state quantum information},}\ }\href {\doibase
  10.1103/RevModPhys.86.361} {\bibfield  {journal} {\bibinfo  {journal} {Rev.
  Mod. Phys.}\ }\textbf {\bibinfo {volume} {86}},\ \bibinfo {pages} {361}
  (\bibinfo {year} {2014})}\BibitemShut {NoStop}%
\bibitem [{\citenamefont {Sza\'nkowski}\ \emph {et~al.}(2017)\citenamefont
  {Sza\'nkowski}, \citenamefont {Ramon}, \citenamefont {Krzywda}, \citenamefont
  {Kwiatkowski},\ and\ \citenamefont {Cywi\'nski}}]{Szankowski_JPCM17}%
  \BibitemOpen
  \bibfield  {author} {\bibinfo {author} {\bibfnamefont {P.}~\bibnamefont
  {Sza\'nkowski}}, \bibinfo {author} {\bibfnamefont {G.}~\bibnamefont {Ramon}},
  \bibinfo {author} {\bibfnamefont {J.}~\bibnamefont {Krzywda}}, \bibinfo
  {author} {\bibfnamefont {D.}~\bibnamefont {Kwiatkowski}}, \ and\ \bibinfo
  {author} {\bibfnamefont {{\L}.}~\bibnamefont {Cywi\'nski}},\ }\bibfield
  {title} {\enquote {\bibinfo {title} {Environmental noise spectroscopy with
  qubits subjected to dynamical decoupling},}\ }\href {\doibase
  10.1088/1361-648X/aa7648} {\bibfield  {journal} {\bibinfo  {journal} {J.
  Phys.:Condens. Matter}\ }\textbf {\bibinfo {volume} {29}},\ \bibinfo {pages}
  {333001} (\bibinfo {year} {2017})}\BibitemShut {NoStop}%
\bibitem [{\citenamefont {Coish}\ and\ \citenamefont
  {Loss}(2004)}]{Coish_PRB04}%
  \BibitemOpen
  \bibfield  {author} {\bibinfo {author} {\bibfnamefont {W.~A.}\ \bibnamefont
  {Coish}}\ and\ \bibinfo {author} {\bibfnamefont {Daniel}\ \bibnamefont
  {Loss}},\ }\bibfield  {title} {\enquote {\bibinfo {title} {Hyperfine
  interaction in a quantum dot: Non-markovian electron spin dynamics},}\ }\href
  {\doibase 10.1103/PhysRevB.70.195340} {\bibfield  {journal} {\bibinfo
  {journal} {Phys.\ Rev.\ B}\ }\textbf {\bibinfo {volume} {70}},\ \bibinfo
  {eid} {195340} (\bibinfo {year} {2004})}\BibitemShut {NoStop}%
\bibitem [{\citenamefont {Cappellaro}(2012)}]{Cappellaro_PRA12}%
  \BibitemOpen
  \bibfield  {author} {\bibinfo {author} {\bibfnamefont {Paola}\ \bibnamefont
  {Cappellaro}},\ }\bibfield  {title} {\enquote {\bibinfo {title} {Spin-bath
  narrowing with adaptive parameter estimation},}\ }\href {\doibase
  10.1103/PhysRevA.85.030301} {\bibfield  {journal} {\bibinfo  {journal} {Phys.
  Rev. A}\ }\textbf {\bibinfo {volume} {85}},\ \bibinfo {pages} {030301(R)}
  (\bibinfo {year} {2012})}\BibitemShut {NoStop}%
\bibitem [{\citenamefont {Barthel}\ \emph {et~al.}(2009)\citenamefont
  {Barthel}, \citenamefont {Reilly}, \citenamefont {Marcus}, \citenamefont
  {Hanson},\ and\ \citenamefont {Gossard}}]{Barthel_PRL09}%
  \BibitemOpen
  \bibfield  {author} {\bibinfo {author} {\bibfnamefont {C.}~\bibnamefont
  {Barthel}}, \bibinfo {author} {\bibfnamefont {D.~J.}\ \bibnamefont {Reilly}},
  \bibinfo {author} {\bibfnamefont {C.~M.}\ \bibnamefont {Marcus}}, \bibinfo
  {author} {\bibfnamefont {M.~P.}\ \bibnamefont {Hanson}}, \ and\ \bibinfo
  {author} {\bibfnamefont {A.~C.}\ \bibnamefont {Gossard}},\ }\bibfield
  {title} {\enquote {\bibinfo {title} {Rapid single-shot measurement of a
  singlet-triplet qubit},}\ }\href {\doibase 10.1103/PhysRevLett.103.160503}
  {\bibfield  {journal} {\bibinfo  {journal} {Phys.\ Rev.\ Lett.}\ }\textbf
  {\bibinfo {volume} {103}},\ \bibinfo {pages} {160503} (\bibinfo {year}
  {2009})}\BibitemShut {NoStop}%
\bibitem [{\citenamefont {Delbecq}\ \emph {et~al.}(2016)\citenamefont
  {Delbecq}, \citenamefont {Nakajima}, \citenamefont {Stano}, \citenamefont
  {Otsuka}, \citenamefont {Amaha}, \citenamefont {Yoneda}, \citenamefont
  {Takeda}, \citenamefont {Allison}, \citenamefont {Ludwig}, \citenamefont
  {Wieck},\ and\ \citenamefont {Tarucha}}]{Delbecq_PRL16}%
  \BibitemOpen
  \bibfield  {author} {\bibinfo {author} {\bibfnamefont {M.~R.}\ \bibnamefont
  {Delbecq}}, \bibinfo {author} {\bibfnamefont {T.}~\bibnamefont {Nakajima}},
  \bibinfo {author} {\bibfnamefont {P.}~\bibnamefont {Stano}}, \bibinfo
  {author} {\bibfnamefont {T.}~\bibnamefont {Otsuka}}, \bibinfo {author}
  {\bibfnamefont {S.}~\bibnamefont {Amaha}}, \bibinfo {author} {\bibfnamefont
  {J.}~\bibnamefont {Yoneda}}, \bibinfo {author} {\bibfnamefont
  {K.}~\bibnamefont {Takeda}}, \bibinfo {author} {\bibfnamefont
  {G.}~\bibnamefont {Allison}}, \bibinfo {author} {\bibfnamefont
  {A.}~\bibnamefont {Ludwig}}, \bibinfo {author} {\bibfnamefont {A.~D.}\
  \bibnamefont {Wieck}}, \ and\ \bibinfo {author} {\bibfnamefont
  {S.}~\bibnamefont {Tarucha}},\ }\bibfield  {title} {\enquote {\bibinfo
  {title} {Quantum dephasing in a gated gaas triple quantum dot due to
  nonergodic noise},}\ }\href {\doibase 10.1103/PhysRevLett.116.046802}
  {\bibfield  {journal} {\bibinfo  {journal} {Phys. Rev. Lett.}\ }\textbf
  {\bibinfo {volume} {116}},\ \bibinfo {pages} {046802} (\bibinfo {year}
  {2016})}\BibitemShut {NoStop}%
\bibitem [{\citenamefont {Shulman}\ \emph {et~al.}(2014)\citenamefont
  {Shulman}, \citenamefont {Harvey}, \citenamefont {Nichol}, \citenamefont
  {Bartlett}, \citenamefont {Doherty}, \citenamefont {Umansky}, ,\ and\
  \citenamefont {Yacoby}}]{Shulman_NC14}%
  \BibitemOpen
  \bibfield  {author} {\bibinfo {author} {\bibfnamefont {M.~D.}\ \bibnamefont
  {Shulman}}, \bibinfo {author} {\bibfnamefont {S.~P.}\ \bibnamefont {Harvey}},
  \bibinfo {author} {\bibfnamefont {J.~M.}\ \bibnamefont {Nichol}}, \bibinfo
  {author} {\bibfnamefont {S.~D.}\ \bibnamefont {Bartlett}}, \bibinfo {author}
  {\bibfnamefont {A.~C.}\ \bibnamefont {Doherty}}, \bibinfo {author}
  {\bibfnamefont {V.}~\bibnamefont {Umansky}}, , \ and\ \bibinfo {author}
  {\bibfnamefont {A.}~\bibnamefont {Yacoby}},\ }\bibfield  {title} {\enquote
  {\bibinfo {title} {Suppressing qubit dephasing using real-time hamiltonian
  estimation},}\ }\href {\doibase 10.1038/ncomms6156} {\bibfield  {journal}
  {\bibinfo  {journal} {Nature Communications}\ }\textbf {\bibinfo {volume}
  {5}},\ \bibinfo {pages} {5156} (\bibinfo {year} {2014})}\BibitemShut
  {NoStop}%
\bibitem [{\citenamefont {Zhao}\ \emph {et~al.}(2012)\citenamefont {Zhao},
  \citenamefont {Ho},\ and\ \citenamefont {Liu}}]{Zhao_PRB12}%
  \BibitemOpen
  \bibfield  {author} {\bibinfo {author} {\bibfnamefont {Nan}\ \bibnamefont
  {Zhao}}, \bibinfo {author} {\bibfnamefont {Sai-Wah}\ \bibnamefont {Ho}}, \
  and\ \bibinfo {author} {\bibfnamefont {Ren-Bao}\ \bibnamefont {Liu}},\
  }\bibfield  {title} {\enquote {\bibinfo {title} {Decoherence and dynamical
  decoupling control of nitrogen vacancy center electron spins in nuclear spin
  baths},}\ }\href {\doibase 10.1103/PhysRevB.85.115303} {\bibfield  {journal}
  {\bibinfo  {journal} {Phys. Rev. B}\ }\textbf {\bibinfo {volume} {85}},\
  \bibinfo {pages} {115303} (\bibinfo {year} {2012})}\BibitemShut {NoStop}%
\bibitem [{\citenamefont {Robledo}\ \emph {et~al.}(2011)\citenamefont
  {Robledo}, \citenamefont {Childress}, \citenamefont {Bernien}, \citenamefont
  {Hensen}, \citenamefont {Alkemade},\ and\ \citenamefont
  {Hanson}}]{Robledo_Nature11}%
  \BibitemOpen
  \bibfield  {author} {\bibinfo {author} {\bibfnamefont {Lucio}\ \bibnamefont
  {Robledo}}, \bibinfo {author} {\bibfnamefont {Lilian}\ \bibnamefont
  {Childress}}, \bibinfo {author} {\bibfnamefont {Hannes}\ \bibnamefont
  {Bernien}}, \bibinfo {author} {\bibfnamefont {Bas}\ \bibnamefont {Hensen}},
  \bibinfo {author} {\bibfnamefont {Paul F.~A.}\ \bibnamefont {Alkemade}}, \
  and\ \bibinfo {author} {\bibfnamefont {Ronald}\ \bibnamefont {Hanson}},\
  }\bibfield  {title} {\enquote {\bibinfo {title} {High-fidelity projective
  read-out of a solid-state spin quantum register},}\ }\href {\doibase
  10.1038/nature10401} {\bibfield  {journal} {\bibinfo  {journal} {Nature}\
  }\textbf {\bibinfo {volume} {477}},\ \bibinfo {pages} {574} (\bibinfo {year}
  {2011})}\BibitemShut {NoStop}%
\bibitem [{\citenamefont {Rao}\ \emph {et~al.}(2018)\citenamefont {Rao},
  \citenamefont {Yang}, \citenamefont {Jesenski}, \citenamefont {Kaiser},\ and\
  \citenamefont {Wrachtrup}}]{RaoEAPreprint2018}%
  \BibitemOpen
  \bibfield  {author} {\bibinfo {author} {\bibfnamefont {D.~D.~Bhaktavatsala}\
  \bibnamefont {Rao}}, \bibinfo {author} {\bibfnamefont {Sen}\ \bibnamefont
  {Yang}}, \bibinfo {author} {\bibfnamefont {Stefan}\ \bibnamefont {Jesenski}},
  \bibinfo {author} {\bibfnamefont {Florian}\ \bibnamefont {Kaiser}}, \ and\
  \bibinfo {author} {\bibfnamefont {J{\"o}rg}\ \bibnamefont {Wrachtrup}},\
  }\href {http://arxiv.org/abs/1804.07111v1; http://arxiv.org/pdf/1804.07111v1}
  {\enquote {\bibinfo {title} {Non-classical measurement statistics induced by
  a coherent spin environment},}\ } (\bibinfo {year} {2018}),\ \Eprint
  {http://arxiv.org/abs/1804.07111v1} {arXiv:1804.07111v1 [quant-ph]}
  \BibitemShut {NoStop}%
\bibitem [{\citenamefont {Wang}\ \emph {et~al.}(2019)\citenamefont {Wang},
  \citenamefont {Chen}, \citenamefont {Peng}, \citenamefont {Wrachtrup},\ and\
  \citenamefont {Liu}}]{Wang_arXiv19}%
  \BibitemOpen
  \bibfield  {author} {\bibinfo {author} {\bibfnamefont {Ping}\ \bibnamefont
  {Wang}}, \bibinfo {author} {\bibfnamefont {Chong}\ \bibnamefont {Chen}},
  \bibinfo {author} {\bibfnamefont {Xinhua}\ \bibnamefont {Peng}}, \bibinfo
  {author} {\bibfnamefont {J{\"o}rg}\ \bibnamefont {Wrachtrup}}, \ and\
  \bibinfo {author} {\bibfnamefont {Ren-Bao}\ \bibnamefont {Liu}},\ }\bibfield
  {title} {\enquote {\bibinfo {title} {Characterization of arbitrary-order
  correlations in quantum baths by weak measurement},}\ }\href
  {https://arxiv.org/abs/1902.03606} {\bibfield  {journal} {\bibinfo  {journal}
  {arXiv:1902.03606}\ } (\bibinfo {year} {2019})}\BibitemShut {NoStop}%
\bibitem [{\citenamefont {Reilly}\ \emph {et~al.}(2008)\citenamefont {Reilly},
  \citenamefont {Taylor}, \citenamefont {Laird}, \citenamefont {Petta},
  \citenamefont {Marcus}, \citenamefont {Hanson},\ and\ \citenamefont
  {Gossard}}]{Reilly_PRL08}%
  \BibitemOpen
  \bibfield  {author} {\bibinfo {author} {\bibfnamefont {D.~J.}\ \bibnamefont
  {Reilly}}, \bibinfo {author} {\bibfnamefont {J.~M.}\ \bibnamefont {Taylor}},
  \bibinfo {author} {\bibfnamefont {E.~A.}\ \bibnamefont {Laird}}, \bibinfo
  {author} {\bibfnamefont {J.~R.}\ \bibnamefont {Petta}}, \bibinfo {author}
  {\bibfnamefont {C.~M.}\ \bibnamefont {Marcus}}, \bibinfo {author}
  {\bibfnamefont {M.~P.}\ \bibnamefont {Hanson}}, \ and\ \bibinfo {author}
  {\bibfnamefont {A.~C.}\ \bibnamefont {Gossard}},\ }\bibfield  {title}
  {\enquote {\bibinfo {title} {Measurement of temporal correlations of the
  overhauser field in a double quantum dot},}\ }\href {\doibase
  10.1103/PhysRevLett.101.236803} {\bibfield  {journal} {\bibinfo  {journal}
  {Phys.\ Rev.\ Lett.}\ }\textbf {\bibinfo {volume} {101}},\ \bibinfo {pages}
  {236803} (\bibinfo {year} {2008})}\BibitemShut {NoStop}%
\bibitem [{\citenamefont {Reilly}\ \emph {et~al.}(2010)\citenamefont {Reilly},
  \citenamefont {Taylor}, \citenamefont {Petta}, \citenamefont {Marcus},
  \citenamefont {Hanson},\ and\ \citenamefont {Gossard}}]{Reilly_PRL10}%
  \BibitemOpen
  \bibfield  {author} {\bibinfo {author} {\bibfnamefont {D.~J.}\ \bibnamefont
  {Reilly}}, \bibinfo {author} {\bibfnamefont {J.~M.}\ \bibnamefont {Taylor}},
  \bibinfo {author} {\bibfnamefont {J.~R.}\ \bibnamefont {Petta}}, \bibinfo
  {author} {\bibfnamefont {C.~M.}\ \bibnamefont {Marcus}}, \bibinfo {author}
  {\bibfnamefont {M.~P.}\ \bibnamefont {Hanson}}, \ and\ \bibinfo {author}
  {\bibfnamefont {A.~C.}\ \bibnamefont {Gossard}},\ }\bibfield  {title}
  {\enquote {\bibinfo {title} {Exchange control of nuclear spin diffusion in a
  double quantum dot},}\ }\href {\doibase 10.1103/PhysRevLett.104.236802}
  {\bibfield  {journal} {\bibinfo  {journal} {Phys.\ Rev.\ Lett.}\ }\textbf
  {\bibinfo {volume} {104}},\ \bibinfo {pages} {236802} (\bibinfo {year}
  {2010})}\BibitemShut {NoStop}%
\bibitem [{\citenamefont {Malinowski}\ \emph {et~al.}(2017)\citenamefont
  {Malinowski}, \citenamefont {Martins}, \citenamefont {Cywi{\'n}ski},
  \citenamefont {Rudner}, \citenamefont {Nissen}, \citenamefont {Fallahi},
  \citenamefont {Gardner}, \citenamefont {Manfra}, \citenamefont {Marcus},\
  and\ \citenamefont {Kuemmeth}}]{Malinowski_PRL17}%
  \BibitemOpen
  \bibfield  {author} {\bibinfo {author} {\bibfnamefont {Filip~K.}\
  \bibnamefont {Malinowski}}, \bibinfo {author} {\bibfnamefont {Frederico}\
  \bibnamefont {Martins}}, \bibinfo {author} {\bibfnamefont {{\L}ukasz}\
  \bibnamefont {Cywi{\'n}ski}}, \bibinfo {author} {\bibfnamefont {Mark~S.}\
  \bibnamefont {Rudner}}, \bibinfo {author} {\bibfnamefont {Peter~D.}\
  \bibnamefont {Nissen}}, \bibinfo {author} {\bibfnamefont {Saeed}\
  \bibnamefont {Fallahi}}, \bibinfo {author} {\bibfnamefont {Geoffrey~C.}\
  \bibnamefont {Gardner}}, \bibinfo {author} {\bibfnamefont {Michael~J.}\
  \bibnamefont {Manfra}}, \bibinfo {author} {\bibfnamefont {Charles~M.}\
  \bibnamefont {Marcus}}, \ and\ \bibinfo {author} {\bibfnamefont {Ferdinand}\
  \bibnamefont {Kuemmeth}},\ }\bibfield  {title} {\enquote {\bibinfo {title}
  {Spectrum of the nuclear environment for gaas spin qubits},}\ }\href
  {\doibase 10.1103/PhysRevLett.118.177702} {\bibfield  {journal} {\bibinfo
  {journal} {Phys.\ Rev.\ Lett.}\ }\textbf {\bibinfo {volume} {118}},\ \bibinfo
  {pages} {177702} (\bibinfo {year} {2017})}\BibitemShut {NoStop}%
\bibitem [{\citenamefont {Doherty}\ \emph {et~al.}(2013)\citenamefont
  {Doherty}, \citenamefont {Manson}, \citenamefont {Delaney}, \citenamefont
  {Jelezko}, \citenamefont {Wrachtrup},\ and\ \citenamefont
  {Hollenberg}}]{Doherty_PR13}%
  \BibitemOpen
  \bibfield  {author} {\bibinfo {author} {\bibfnamefont {Marcus~W.}\
  \bibnamefont {Doherty}}, \bibinfo {author} {\bibfnamefont {Neil~B.}\
  \bibnamefont {Manson}}, \bibinfo {author} {\bibfnamefont {Paul}\ \bibnamefont
  {Delaney}}, \bibinfo {author} {\bibfnamefont {Fedor}\ \bibnamefont
  {Jelezko}}, \bibinfo {author} {\bibfnamefont {J{\"o}rg}\ \bibnamefont
  {Wrachtrup}}, \ and\ \bibinfo {author} {\bibfnamefont {Lloyd~C.L.}\
  \bibnamefont {Hollenberg}},\ }\bibfield  {title} {\enquote {\bibinfo {title}
  {The nitrogen-vacancy colour centre in diamond},}\ }\href {\doibase
  10.1016/j.physrep.2013.02.001} {\bibfield  {journal} {\bibinfo  {journal}
  {Phys. Rep.}\ }\textbf {\bibinfo {volume} {528}},\ \bibinfo {pages} {1}
  (\bibinfo {year} {2013})}\BibitemShut {NoStop}%
\bibitem [{\citenamefont {Koppens}\ \emph {et~al.}(2008)\citenamefont
  {Koppens}, \citenamefont {Nowack},\ and\ \citenamefont
  {Vandersypen}}]{Koppens_PRL08}%
  \BibitemOpen
  \bibfield  {author} {\bibinfo {author} {\bibfnamefont {F.~H.~L.}\
  \bibnamefont {Koppens}}, \bibinfo {author} {\bibfnamefont {K.~C.}\
  \bibnamefont {Nowack}}, \ and\ \bibinfo {author} {\bibfnamefont {L.~M.~K.}\
  \bibnamefont {Vandersypen}},\ }\bibfield  {title} {\enquote {\bibinfo {title}
  {Spin echo of a single electron spin in a quantum dot},}\ }\href {\doibase
  10.1103/PhysRevLett.100.236802} {\bibfield  {journal} {\bibinfo  {journal}
  {Phys.\ Rev.\ Lett.}\ }\textbf {\bibinfo {volume} {100}},\ \bibinfo {pages}
  {236802} (\bibinfo {year} {2008})}\BibitemShut {NoStop}%
\bibitem [{\citenamefont {Bechtold}\ \emph {et~al.}(2015)\citenamefont
  {Bechtold}, \citenamefont {Rauch}, \citenamefont {Li}, \citenamefont
  {Simmet}, \citenamefont {Ardelt}, \citenamefont {Regler}, \citenamefont
  {M{\"u}ller}, \citenamefont {Sinitsyn},\ and\ \citenamefont
  {Finley}}]{Bechtold_NP15}%
  \BibitemOpen
  \bibfield  {author} {\bibinfo {author} {\bibfnamefont {Alexander}\
  \bibnamefont {Bechtold}}, \bibinfo {author} {\bibfnamefont {Dominik}\
  \bibnamefont {Rauch}}, \bibinfo {author} {\bibfnamefont {Fuxiang}\
  \bibnamefont {Li}}, \bibinfo {author} {\bibfnamefont {Tobias}\ \bibnamefont
  {Simmet}}, \bibinfo {author} {\bibfnamefont {Per-Lennart}\ \bibnamefont
  {Ardelt}}, \bibinfo {author} {\bibfnamefont {Armin}\ \bibnamefont {Regler}},
  \bibinfo {author} {\bibfnamefont {Kai}\ \bibnamefont {M{\"u}ller}}, \bibinfo
  {author} {\bibfnamefont {Nikolai~A.}\ \bibnamefont {Sinitsyn}}, \ and\
  \bibinfo {author} {\bibfnamefont {Jonathan~J.}\ \bibnamefont {Finley}},\
  }\bibfield  {title} {\enquote {\bibinfo {title} {Three-stage decoherence
  dynamics of an electron spin qubit in an optically active quantum dot},}\
  }\href {\doibase 10.1038/nphys3470} {\bibfield  {journal} {\bibinfo
  {journal} {Nat. Phys.}\ }\textbf {\bibinfo {volume} {11}},\ \bibinfo {pages}
  {1005} (\bibinfo {year} {2015})}\BibitemShut {NoStop}%
\bibitem [{\citenamefont {Bechtold}\ \emph {et~al.}(2016)\citenamefont
  {Bechtold}, \citenamefont {Li}, \citenamefont {M\"uller}, \citenamefont
  {Simmet}, \citenamefont {Ardelt}, \citenamefont {Finley},\ and\ \citenamefont
  {Sinitsyn}}]{Bechtold_PRL16}%
  \BibitemOpen
  \bibfield  {author} {\bibinfo {author} {\bibfnamefont {A.}~\bibnamefont
  {Bechtold}}, \bibinfo {author} {\bibfnamefont {F.}~\bibnamefont {Li}},
  \bibinfo {author} {\bibfnamefont {K.}~\bibnamefont {M\"uller}}, \bibinfo
  {author} {\bibfnamefont {T.}~\bibnamefont {Simmet}}, \bibinfo {author}
  {\bibfnamefont {P.-L.}\ \bibnamefont {Ardelt}}, \bibinfo {author}
  {\bibfnamefont {J.~J.}\ \bibnamefont {Finley}}, \ and\ \bibinfo {author}
  {\bibfnamefont {N.~A.}\ \bibnamefont {Sinitsyn}},\ }\bibfield  {title}
  {\enquote {\bibinfo {title} {Quantum effects in higher-order correlators of a
  quantum-dot spin qubit},}\ }\href {\doibase 10.1103/PhysRevLett.117.027402}
  {\bibfield  {journal} {\bibinfo  {journal} {Phys. Rev. Lett.}\ }\textbf
  {\bibinfo {volume} {117}},\ \bibinfo {pages} {027402} (\bibinfo {year}
  {2016})}\BibitemShut {NoStop}%
\bibitem [{\citenamefont {Kawakami}\ \emph {et~al.}(2016)\citenamefont
  {Kawakami}, \citenamefont {Jullien}, \citenamefont {Scarlino}, \citenamefont
  {Ward}, \citenamefont {Savage}, \citenamefont {Lagally}, \citenamefont
  {Dobrovitski}, \citenamefont {Friesen}, \citenamefont {Coppersmith},
  \citenamefont {Eriksson},\ and\ \citenamefont
  {Vandersypen}}]{Kawakami_PNAS16}%
  \BibitemOpen
  \bibfield  {author} {\bibinfo {author} {\bibfnamefont {Erika}\ \bibnamefont
  {Kawakami}}, \bibinfo {author} {\bibfnamefont {Thibaut}\ \bibnamefont
  {Jullien}}, \bibinfo {author} {\bibfnamefont {Pasquale}\ \bibnamefont
  {Scarlino}}, \bibinfo {author} {\bibfnamefont {Daniel~R.}\ \bibnamefont
  {Ward}}, \bibinfo {author} {\bibfnamefont {Donald~E.}\ \bibnamefont
  {Savage}}, \bibinfo {author} {\bibfnamefont {Max~G.}\ \bibnamefont
  {Lagally}}, \bibinfo {author} {\bibfnamefont {Viatcheslav~V.}\ \bibnamefont
  {Dobrovitski}}, \bibinfo {author} {\bibfnamefont {Mark}\ \bibnamefont
  {Friesen}}, \bibinfo {author} {\bibfnamefont {Susan~N.}\ \bibnamefont
  {Coppersmith}}, \bibinfo {author} {\bibfnamefont {Mark~A.}\ \bibnamefont
  {Eriksson}}, \ and\ \bibinfo {author} {\bibfnamefont {Lieven M.~K.}\
  \bibnamefont {Vandersypen}},\ }\bibfield  {title} {\enquote {\bibinfo {title}
  {Gate fidelity and coherence of an electron spin in an si/sige quantum dot
  with micromagnet},}\ }\href {\doibase 10.1073/pnas.1603251113} {\bibfield
  {journal} {\bibinfo  {journal} {PNAS}\ }\textbf {\bibinfo {volume} {113}},\
  \bibinfo {pages} {11738} (\bibinfo {year} {2016})}\BibitemShut {NoStop}%
\bibitem [{\citenamefont {Tyryshkin}\ \emph {et~al.}(2012)\citenamefont
  {Tyryshkin}, \citenamefont {Tojo}, \citenamefont {Morton}, \citenamefont
  {Riemann}, \citenamefont {Abrosimov}, \citenamefont {Becker}, \citenamefont
  {Pohl}, \citenamefont {Schenkel}, \citenamefont {Thewalt}, \citenamefont
  {Itoh},\ and\ \citenamefont {Lyon}}]{Tyryshkin_NM12}%
  \BibitemOpen
  \bibfield  {author} {\bibinfo {author} {\bibfnamefont {Alexei~M.}\
  \bibnamefont {Tyryshkin}}, \bibinfo {author} {\bibfnamefont {Shinichi}\
  \bibnamefont {Tojo}}, \bibinfo {author} {\bibfnamefont {John J.~L.}\
  \bibnamefont {Morton}}, \bibinfo {author} {\bibfnamefont {Helge}\
  \bibnamefont {Riemann}}, \bibinfo {author} {\bibfnamefont {Nikolai~V.}\
  \bibnamefont {Abrosimov}}, \bibinfo {author} {\bibfnamefont {Peter}\
  \bibnamefont {Becker}}, \bibinfo {author} {\bibfnamefont {Hans-Joachim}\
  \bibnamefont {Pohl}}, \bibinfo {author} {\bibfnamefont {Thomas}\ \bibnamefont
  {Schenkel}}, \bibinfo {author} {\bibfnamefont {Michael L.~W.}\ \bibnamefont
  {Thewalt}}, \bibinfo {author} {\bibfnamefont {Kohei~M.}\ \bibnamefont
  {Itoh}}, \ and\ \bibinfo {author} {\bibfnamefont {S.~A.}\ \bibnamefont
  {Lyon}},\ }\bibfield  {title} {\enquote {\bibinfo {title} {Electron spin
  coherence exceeding seconds in high purity silicon},}\ }\href {\doibase
  10.1038/nmat3182} {\bibfield  {journal} {\bibinfo  {journal} {Nat.
  Materials}\ }\textbf {\bibinfo {volume} {11}},\ \bibinfo {pages} {143}
  (\bibinfo {year} {2012})}\BibitemShut {NoStop}%
\bibitem [{\citenamefont {Pla}\ \emph {et~al.}(2013=2)\citenamefont {Pla},
  \citenamefont {Tan}, \citenamefont {Dehollain}, \citenamefont {Lim},
  \citenamefont {Morton}, \citenamefont {Jamieson}, \citenamefont {Dzurak},\
  and\ \citenamefont {Morello}}]{Pla_Nature12}%
  \BibitemOpen
  \bibfield  {author} {\bibinfo {author} {\bibfnamefont {Jarryd~J.}\
  \bibnamefont {Pla}}, \bibinfo {author} {\bibfnamefont {Kuan~Y.}\ \bibnamefont
  {Tan}}, \bibinfo {author} {\bibfnamefont {Juan~P.}\ \bibnamefont
  {Dehollain}}, \bibinfo {author} {\bibfnamefont {Wee~H.}\ \bibnamefont {Lim}},
  \bibinfo {author} {\bibfnamefont {John J.~L.}\ \bibnamefont {Morton}},
  \bibinfo {author} {\bibfnamefont {David~N.}\ \bibnamefont {Jamieson}},
  \bibinfo {author} {\bibfnamefont {Andrew~S.}\ \bibnamefont {Dzurak}}, \ and\
  \bibinfo {author} {\bibfnamefont {Andrea}\ \bibnamefont {Morello}},\
  }\bibfield  {title} {\enquote {\bibinfo {title} {A single-atom electron spin
  qubit in silicon},}\ }\href {\doibase 10.1038/nature11449} {\bibfield
  {journal} {\bibinfo  {journal} {Nature.}\ }\textbf {\bibinfo {volume}
  {489}},\ \bibinfo {pages} {541} (\bibinfo {year} {2013=2})}\BibitemShut
  {NoStop}%
\bibitem [{\citenamefont {Wolfowicz}\ \emph {et~al.}(2013)\citenamefont
  {Wolfowicz}, \citenamefont {Tyryshkin}, \citenamefont {George}, \citenamefont
  {Riemann}, \citenamefont {Abrosimov}, \citenamefont {Becker}, \citenamefont
  {Pohl}, \citenamefont {Thewalt}, \citenamefont {Lyon},\ and\ \citenamefont
  {Morton}}]{Wolfowicz_NN13}%
  \BibitemOpen
  \bibfield  {author} {\bibinfo {author} {\bibfnamefont {Gary}\ \bibnamefont
  {Wolfowicz}}, \bibinfo {author} {\bibfnamefont {Alexei~M.}\ \bibnamefont
  {Tyryshkin}}, \bibinfo {author} {\bibfnamefont {Richard~E.}\ \bibnamefont
  {George}}, \bibinfo {author} {\bibfnamefont {Helge}\ \bibnamefont {Riemann}},
  \bibinfo {author} {\bibfnamefont {Nikolai~V.}\ \bibnamefont {Abrosimov}},
  \bibinfo {author} {\bibfnamefont {Peter}\ \bibnamefont {Becker}}, \bibinfo
  {author} {\bibfnamefont {Hans-Joachim}\ \bibnamefont {Pohl}}, \bibinfo
  {author} {\bibfnamefont {Mike L.~W.}\ \bibnamefont {Thewalt}}, \bibinfo
  {author} {\bibfnamefont {Stephen~A.}\ \bibnamefont {Lyon}}, \ and\ \bibinfo
  {author} {\bibfnamefont {John J.~L.}\ \bibnamefont {Morton}},\ }\bibfield
  {title} {\enquote {\bibinfo {title} {Atomic clock transitions in
  silicon-based spin qubits},}\ }\href {\doibase 10.1038/nnano.2013.117}
  {\bibfield  {journal} {\bibinfo  {journal} {Nature Nanotechnology}\ }\textbf
  {\bibinfo {volume} {8}},\ \bibinfo {pages} {561} (\bibinfo {year}
  {2013})}\BibitemShut {NoStop}%
\bibitem [{\citenamefont {Rogers}\ \emph {et~al.}(2014)\citenamefont {Rogers},
  \citenamefont {Jahnke}, \citenamefont {Metsch}, \citenamefont {Sipahigil},
  \citenamefont {Binder}, \citenamefont {Teraji}, \citenamefont {Sumiya},
  \citenamefont {Isoya}, \citenamefont {Lukin}, \citenamefont {Hemmer},\ and\
  \citenamefont {Jelezko}}]{Rogers_PRL14}%
  \BibitemOpen
  \bibfield  {author} {\bibinfo {author} {\bibfnamefont {Lachlan~J.}\
  \bibnamefont {Rogers}}, \bibinfo {author} {\bibfnamefont {Kay~D.}\
  \bibnamefont {Jahnke}}, \bibinfo {author} {\bibfnamefont {Mathias~H.}\
  \bibnamefont {Metsch}}, \bibinfo {author} {\bibfnamefont {Alp}\ \bibnamefont
  {Sipahigil}}, \bibinfo {author} {\bibfnamefont {Jan~M.}\ \bibnamefont
  {Binder}}, \bibinfo {author} {\bibfnamefont {Tokuyuki}\ \bibnamefont
  {Teraji}}, \bibinfo {author} {\bibfnamefont {Hitoshi}\ \bibnamefont
  {Sumiya}}, \bibinfo {author} {\bibfnamefont {Junichi}\ \bibnamefont {Isoya}},
  \bibinfo {author} {\bibfnamefont {Mikhail~D.}\ \bibnamefont {Lukin}},
  \bibinfo {author} {\bibfnamefont {Philip}\ \bibnamefont {Hemmer}}, \ and\
  \bibinfo {author} {\bibfnamefont {Fedor}\ \bibnamefont {Jelezko}},\
  }\bibfield  {title} {\enquote {\bibinfo {title} {All-optical initialization,
  readout, and coherent preparation of single silicon-vacancy spins in
  diamond},}\ }\href {\doibase 10.1103/PhysRevLett.113.263602} {\bibfield
  {journal} {\bibinfo  {journal} {Phys. Rev. Lett.}\ }\textbf {\bibinfo
  {volume} {113}},\ \bibinfo {pages} {263602} (\bibinfo {year}
  {2014})}\BibitemShut {NoStop}%
\bibitem [{\citenamefont {Widmann}\ \emph {et~al.}(2015)\citenamefont
  {Widmann}, \citenamefont {Lee}, \citenamefont {Rendler}, \citenamefont {Son},
  \citenamefont {Fedder}, \citenamefont {Paik}, \citenamefont {Yang},
  \citenamefont {Zhao}, \citenamefont {Yang}, \citenamefont {Booker},
  \citenamefont {Denisenko}, \citenamefont {Jamali}, \citenamefont
  {Momenzadeh}, \citenamefont {Gerhardt}, \citenamefont {Ohshima},
  \citenamefont {Gali}, \citenamefont {Janz{\'e}n},\ and\ \citenamefont
  {Wrachtrup}}]{Widmann_NM15}%
  \BibitemOpen
  \bibfield  {author} {\bibinfo {author} {\bibfnamefont {Matthias}\
  \bibnamefont {Widmann}}, \bibinfo {author} {\bibfnamefont {Sang-Yun}\
  \bibnamefont {Lee}}, \bibinfo {author} {\bibfnamefont {Torsten}\ \bibnamefont
  {Rendler}}, \bibinfo {author} {\bibfnamefont {Nguyen~Tien}\ \bibnamefont
  {Son}}, \bibinfo {author} {\bibfnamefont {Helmut}\ \bibnamefont {Fedder}},
  \bibinfo {author} {\bibfnamefont {Seoyoung}\ \bibnamefont {Paik}}, \bibinfo
  {author} {\bibfnamefont {Li-Ping}\ \bibnamefont {Yang}}, \bibinfo {author}
  {\bibfnamefont {Nan}\ \bibnamefont {Zhao}}, \bibinfo {author} {\bibfnamefont
  {Sen}\ \bibnamefont {Yang}}, \bibinfo {author} {\bibfnamefont {Ian}\
  \bibnamefont {Booker}}, \bibinfo {author} {\bibfnamefont {Andrej}\
  \bibnamefont {Denisenko}}, \bibinfo {author} {\bibfnamefont {Mohammad}\
  \bibnamefont {Jamali}}, \bibinfo {author} {\bibfnamefont {S.~Ali}\
  \bibnamefont {Momenzadeh}}, \bibinfo {author} {\bibfnamefont {Ilja}\
  \bibnamefont {Gerhardt}}, \bibinfo {author} {\bibfnamefont {Takeshi}\
  \bibnamefont {Ohshima}}, \bibinfo {author} {\bibfnamefont {Adam}\
  \bibnamefont {Gali}}, \bibinfo {author} {\bibfnamefont {Erik}\ \bibnamefont
  {Janz{\'e}n}}, \ and\ \bibinfo {author} {\bibfnamefont {J{\"o}rg}\
  \bibnamefont {Wrachtrup}},\ }\bibfield  {title} {\enquote {\bibinfo {title}
  {Coherent control of single spins in silicon carbide at room temperature},}\
  }\href {\doibase 10.1038/nmat4145} {\bibfield  {journal} {\bibinfo  {journal}
  {Nature Materials}\ }\textbf {\bibinfo {volume} {14}},\ \bibinfo {pages}
  {164} (\bibinfo {year} {2015})}\BibitemShut {NoStop}%
\bibitem [{\citenamefont {Carter}\ \emph {et~al.}(2015)\citenamefont {Carter},
  \citenamefont {Soykal}, \citenamefont {Dev}, \citenamefont {Economou},\ and\
  \citenamefont {Glaser}}]{Carter_PRB15}%
  \BibitemOpen
  \bibfield  {author} {\bibinfo {author} {\bibfnamefont {S.~G.}\ \bibnamefont
  {Carter}}, \bibinfo {author} {\bibfnamefont {\"O.~O.}\ \bibnamefont
  {Soykal}}, \bibinfo {author} {\bibfnamefont {Pratibha}\ \bibnamefont {Dev}},
  \bibinfo {author} {\bibfnamefont {Sophia~E.}\ \bibnamefont {Economou}}, \
  and\ \bibinfo {author} {\bibfnamefont {E.~R.}\ \bibnamefont {Glaser}},\
  }\bibfield  {title} {\enquote {\bibinfo {title} {Spin coherence and echo
  modulation of the silicon vacancy in $4h\ensuremath{-}\mathrm{SiC}$ at room
  temperature},}\ }\href {\doibase 10.1103/PhysRevB.92.161202} {\bibfield
  {journal} {\bibinfo  {journal} {Phys. Rev. B}\ }\textbf {\bibinfo {volume}
  {92}},\ \bibinfo {pages} {161202(R)} (\bibinfo {year} {2015})}\BibitemShut
  {NoStop}%
\bibitem [{\citenamefont {Cywi{\'n}ski}\ \emph {et~al.}(2009)\citenamefont
  {Cywi{\'n}ski}, \citenamefont {Witzel},\ and\ \citenamefont {{Das
  Sarma}}}]{Cywinski_PRB09}%
  \BibitemOpen
  \bibfield  {author} {\bibinfo {author} {\bibfnamefont {{\L}ukasz}\
  \bibnamefont {Cywi{\'n}ski}}, \bibinfo {author} {\bibfnamefont {Wayne~M.}\
  \bibnamefont {Witzel}}, \ and\ \bibinfo {author} {\bibfnamefont
  {S.}~\bibnamefont {{Das Sarma}}},\ }\bibfield  {title} {\enquote {\bibinfo
  {title} {Pure quantum dephasing of a solid-state electron spin qubit in a
  large nuclear spin bath coupled by long-range hyperfine-mediated
  interaction},}\ }\href {\doibase 10.1103/PhysRevB.79.245314} {\bibfield
  {journal} {\bibinfo  {journal} {Phys.\ Rev.\ B}\ }\textbf {\bibinfo {volume}
  {79}},\ \bibinfo {pages} {245314} (\bibinfo {year} {2009})}\BibitemShut
  {NoStop}%
\bibitem [{\citenamefont {Neder}\ \emph {et~al.}(2011)\citenamefont {Neder},
  \citenamefont {Rudner}, \citenamefont {Bluhm}, \citenamefont {Foletti},
  \citenamefont {Halperin},\ and\ \citenamefont {Yacoby}}]{Neder_PRB11}%
  \BibitemOpen
  \bibfield  {author} {\bibinfo {author} {\bibfnamefont {Izhar}\ \bibnamefont
  {Neder}}, \bibinfo {author} {\bibfnamefont {Mark~S.}\ \bibnamefont {Rudner}},
  \bibinfo {author} {\bibfnamefont {Hendrik}\ \bibnamefont {Bluhm}}, \bibinfo
  {author} {\bibfnamefont {Sandra}\ \bibnamefont {Foletti}}, \bibinfo {author}
  {\bibfnamefont {Bertrand~I.}\ \bibnamefont {Halperin}}, \ and\ \bibinfo
  {author} {\bibfnamefont {Amir}\ \bibnamefont {Yacoby}},\ }\bibfield  {title}
  {\enquote {\bibinfo {title} {Semiclassical model for the dephasing of a
  two-electron spin qubit coupled to a coherently evolving nuclear spin
  bath},}\ }\href {\doibase 10.1103/PhysRevB.84.035441} {\bibfield  {journal}
  {\bibinfo  {journal} {Phys.\ Rev.\ B}\ }\textbf {\bibinfo {volume} {84}},\
  \bibinfo {pages} {035441} (\bibinfo {year} {2011})}\BibitemShut {NoStop}%
\bibitem [{\citenamefont {Bluhm}\ \emph {et~al.}(2010)\citenamefont {Bluhm},
  \citenamefont {Foletti}, \citenamefont {Neder}, \citenamefont {Rudner},
  \citenamefont {Mahalu}, \citenamefont {Umansky},\ and\ \citenamefont
  {Yacoby}}]{Bluhm_NP10}%
  \BibitemOpen
  \bibfield  {author} {\bibinfo {author} {\bibfnamefont {Hendrik}\ \bibnamefont
  {Bluhm}}, \bibinfo {author} {\bibfnamefont {Sandra}\ \bibnamefont {Foletti}},
  \bibinfo {author} {\bibfnamefont {Izhar}\ \bibnamefont {Neder}}, \bibinfo
  {author} {\bibfnamefont {Mark}\ \bibnamefont {Rudner}}, \bibinfo {author}
  {\bibfnamefont {Diana}\ \bibnamefont {Mahalu}}, \bibinfo {author}
  {\bibfnamefont {Vladimir}\ \bibnamefont {Umansky}}, \ and\ \bibinfo {author}
  {\bibfnamefont {Amir}\ \bibnamefont {Yacoby}},\ }\bibfield  {title} {\enquote
  {\bibinfo {title} {Long coherence of electron spins coupled to a nuclear spin
  bath},}\ }\href {\doibase 10.1038/nphys1856} {\bibfield  {journal} {\bibinfo
  {journal} {Nat. Phys.}\ }\textbf {\bibinfo {volume} {7}},\ \bibinfo {pages}
  {109} (\bibinfo {year} {2010})}\BibitemShut {NoStop}%
\bibitem [{\citenamefont {Witzel}\ \emph {et~al.}(2007)\citenamefont {Witzel},
  \citenamefont {Hu},\ and\ \citenamefont {{Das Sarma}}}]{Witzel_AHF_PRB07}%
  \BibitemOpen
  \bibfield  {author} {\bibinfo {author} {\bibfnamefont {W.~M.}\ \bibnamefont
  {Witzel}}, \bibinfo {author} {\bibfnamefont {Xuedong}\ \bibnamefont {Hu}}, \
  and\ \bibinfo {author} {\bibfnamefont {S.}~\bibnamefont {{Das Sarma}}},\
  }\bibfield  {title} {\enquote {\bibinfo {title} {Decoherence induced by
  anisotropic hyperfine interaction in si spin qubits},}\ }\href {\doibase
  10.1103/PhysRevB.76.035212} {\bibfield  {journal} {\bibinfo  {journal}
  {Phys.\ Rev.\ B}\ }\textbf {\bibinfo {volume} {76}},\ \bibinfo {pages}
  {035212} (\bibinfo {year} {2007})}\BibitemShut {NoStop}%
\bibitem [{\citenamefont {{Gurudev Dutt}}\ \emph {et~al.}(2007)\citenamefont
  {{Gurudev Dutt}}, \citenamefont {Childress}, \citenamefont {Jiang},
  \citenamefont {Togan}, \citenamefont {Maze}, \citenamefont {Jelezko},
  \citenamefont {Zibrov}, \citenamefont {Hemmer},\ and\ \citenamefont
  {Lukin}}]{Dutt_Science07}%
  \BibitemOpen
  \bibfield  {author} {\bibinfo {author} {\bibfnamefont {M.~V.}\ \bibnamefont
  {{Gurudev Dutt}}}, \bibinfo {author} {\bibfnamefont {L.}~\bibnamefont
  {Childress}}, \bibinfo {author} {\bibfnamefont {L.}~\bibnamefont {Jiang}},
  \bibinfo {author} {\bibfnamefont {E.}~\bibnamefont {Togan}}, \bibinfo
  {author} {\bibfnamefont {J.}~\bibnamefont {Maze}}, \bibinfo {author}
  {\bibfnamefont {F.}~\bibnamefont {Jelezko}}, \bibinfo {author} {\bibfnamefont
  {A.~S.}\ \bibnamefont {Zibrov}}, \bibinfo {author} {\bibfnamefont {P.~R.}\
  \bibnamefont {Hemmer}}, \ and\ \bibinfo {author} {\bibfnamefont {M.~D.}\
  \bibnamefont {Lukin}},\ }\bibfield  {title} {\enquote {\bibinfo {title}
  {Quantum register based on individual electronic and nuclear spin qubits in
  diamond},}\ }\href {\doibase 10.1126/science.1139831} {\bibfield  {journal}
  {\bibinfo  {journal} {Science}\ }\textbf {\bibinfo {volume} {316}},\ \bibinfo
  {pages} {1312} (\bibinfo {year} {2007})}\BibitemShut {NoStop}%
\bibitem [{\citenamefont {Jiang}\ \emph {et~al.}(2009)\citenamefont {Jiang},
  \citenamefont {Hodges}, \citenamefont {Maze}, \citenamefont {Maurer},
  \citenamefont {Taylor}, \citenamefont {Cory}, \citenamefont {Hemmer},
  \citenamefont {Walsworth}, \citenamefont {Yacoby}, \citenamefont {Zibrov},\
  and\ \citenamefont {Lukin}}]{Jiang_Science09}%
  \BibitemOpen
  \bibfield  {author} {\bibinfo {author} {\bibfnamefont {L.}~\bibnamefont
  {Jiang}}, \bibinfo {author} {\bibfnamefont {J.~S.}\ \bibnamefont {Hodges}},
  \bibinfo {author} {\bibfnamefont {J.~R.}\ \bibnamefont {Maze}}, \bibinfo
  {author} {\bibfnamefont {P.}~\bibnamefont {Maurer}}, \bibinfo {author}
  {\bibfnamefont {J.~M.}\ \bibnamefont {Taylor}}, \bibinfo {author}
  {\bibfnamefont {D.~G.}\ \bibnamefont {Cory}}, \bibinfo {author}
  {\bibfnamefont {P.~R.}\ \bibnamefont {Hemmer}}, \bibinfo {author}
  {\bibfnamefont {R.~L.}\ \bibnamefont {Walsworth}}, \bibinfo {author}
  {\bibfnamefont {A.}~\bibnamefont {Yacoby}}, \bibinfo {author} {\bibfnamefont
  {A.~S.}\ \bibnamefont {Zibrov}}, \ and\ \bibinfo {author} {\bibfnamefont
  {M.~D.}\ \bibnamefont {Lukin}},\ }\bibfield  {title} {\enquote {\bibinfo
  {title} {Repetitive readout of a single electronic spin via quantum logic
  with nuclear spin ancillae},}\ }\href {\doibase 10.1126/science.1176496}
  {\bibfield  {journal} {\bibinfo  {journal} {Science}\ }\textbf {\bibinfo
  {volume} {326}},\ \bibinfo {pages} {267} (\bibinfo {year}
  {2009})}\BibitemShut {NoStop}%
\bibitem [{\citenamefont {Taminiau}\ \emph {et~al.}(2014)\citenamefont
  {Taminiau}, \citenamefont {Cramer}, \citenamefont {{van der Sar}},
  \citenamefont {Dobrovitski},\ and\ \citenamefont {Hanson}}]{Taminiau_NN14}%
  \BibitemOpen
  \bibfield  {author} {\bibinfo {author} {\bibfnamefont {T.~H.}\ \bibnamefont
  {Taminiau}}, \bibinfo {author} {\bibfnamefont {J.}~\bibnamefont {Cramer}},
  \bibinfo {author} {\bibfnamefont {T.}~\bibnamefont {{van der Sar}}}, \bibinfo
  {author} {\bibfnamefont {V.~V.}\ \bibnamefont {Dobrovitski}}, \ and\ \bibinfo
  {author} {\bibfnamefont {R.}~\bibnamefont {Hanson}},\ }\bibfield  {title}
  {\enquote {\bibinfo {title} {Universal control and error correction in
  multi-qubit spin registers in diamond},}\ }\href {\doibase
  10.1038/nnano.2014.2} {\bibfield  {journal} {\bibinfo  {journal} {Nature
  Nanotechnology}\ }\textbf {\bibinfo {volume} {9}},\ \bibinfo {pages} {171}
  (\bibinfo {year} {2014})}\BibitemShut {NoStop}%
\bibitem [{\citenamefont {Waldherr}\ \emph {et~al.}(2014)\citenamefont
  {Waldherr}, \citenamefont {Wang}, \citenamefont {Zaiser}, \citenamefont
  {Jamali}, \citenamefont {Schulte-Herbr\"uggen}, \citenamefont {Abe},
  \citenamefont {Ohshima}, \citenamefont {Isoya}, \citenamefont {Du},
  \citenamefont {Neumann},\ and\ \citenamefont
  {Wrachtrup}}]{Waldherr_Nature14}%
  \BibitemOpen
  \bibfield  {author} {\bibinfo {author} {\bibfnamefont {G.}~\bibnamefont
  {Waldherr}}, \bibinfo {author} {\bibfnamefont {Y.}~\bibnamefont {Wang}},
  \bibinfo {author} {\bibfnamefont {S.}~\bibnamefont {Zaiser}}, \bibinfo
  {author} {\bibfnamefont {M.}~\bibnamefont {Jamali}}, \bibinfo {author}
  {\bibfnamefont {T.}~\bibnamefont {Schulte-Herbr\"uggen}}, \bibinfo {author}
  {\bibfnamefont {H.}~\bibnamefont {Abe}}, \bibinfo {author} {\bibfnamefont
  {T.}~\bibnamefont {Ohshima}}, \bibinfo {author} {\bibfnamefont
  {J.}~\bibnamefont {Isoya}}, \bibinfo {author} {\bibfnamefont {J.~F.}\
  \bibnamefont {Du}}, \bibinfo {author} {\bibfnamefont {P.}~\bibnamefont
  {Neumann}}, \ and\ \bibinfo {author} {\bibfnamefont {J.}~\bibnamefont
  {Wrachtrup}},\ }\bibfield  {title} {\enquote {\bibinfo {title} {Quantum error
  correction in a solid-state hybrid spin register},}\ }\href {\doibase
  10.1038/nature12919} {\bibfield  {journal} {\bibinfo  {journal} {Nature}\
  }\textbf {\bibinfo {volume} {506}},\ \bibinfo {pages} {204} (\bibinfo {year}
  {2014})}\BibitemShut {NoStop}%
\end{thebibliography}%
%\bibliography{../../../../../refs_quant,../../../../../refs_entanglement,../../../../../refs_ddns_magnetometry,../../../../../refs_decoherence,../../../../../refs_Si,multiplemeasurement,../../../../../refs_measurement}

\end{document}